\documentclass{article}

\usepackage{CJKutf8}
\usepackage{longcat_style}
\usepackage{adjustbox}
\usepackage[utf8]{inputenc} %
\usepackage[T1]{fontenc}    %
\usepackage{newunicodechar}
\usepackage{hyperref}       %
\usepackage{xcolor}
\usepackage[normalem]{ulem} %
\hypersetup{
    colorlinks=true,      %
    linkcolor=blue,      %
    urlcolor=blue,       %
    citecolor=blue,      %
    linkbordercolor=blue, %
    urlbordercolor=blue,
    citebordercolor=blue,
    pdfborderstyle={/S/U/W 1}, %
}

\usepackage{float}
\usepackage{url}            %
\usepackage{booktabs}       %
\usepackage{amsfonts}       %
\usepackage{nicefrac}       %
\usepackage{microtype}      %
\usepackage{lipsum}		%
\usepackage{graphicx}
\usepackage{natbib}
\usepackage{doi}
\usepackage{amsmath}
\usepackage{amssymb} %
\usepackage{xspace}
\usepackage{enumitem}
\usepackage{multirow}
\usepackage{subcaption} 
\usepackage{makecell}
\usepackage{hyperref, cleveref}
\usepackage{pifont}
\usepackage[inkscapelatex=false]{svg}

\setlist[itemize]{leftmargin=*}
\setlist[enumerate]{leftmargin=*}
\setlist[description]{leftmargin=*}

\newcommand{\longcat}{LongCat-Flash-Omni\xspace}

\definecolor{midnightgreen}{rgb}{0.0, 0.29, 0.33}

\title{\longcat Technical Report}

\author{ Meituan LongCat Team \\
	\texttt{longcat-team@meituan.com} \\
}

\renewcommand{\headeright}{\raisebox{-0.2\height}{\includegraphics[height=1.8em]{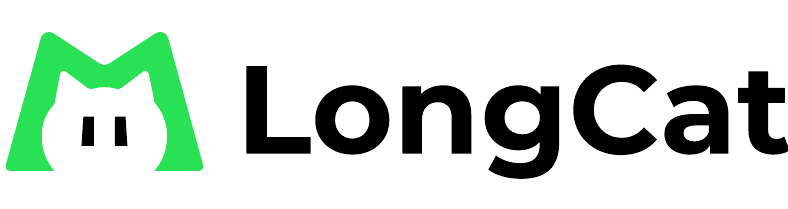}}}
\renewcommand{\shorttitle}{\longcat Technical Report}

\begin{document}
\maketitle

\begin{abstract}
We introduce \longcat, a state-of-the-art open-source omni-modal model with 560 billion parameters, excelling at real-time audio-visual interaction.
By adopting a curriculum-inspired progressive training strategy that transitions from simpler to increasingly complex modality sequence modeling tasks, \longcat attains comprehensive multimodal capabilities while maintaining strong unimodal capability.
Building upon LongCat-Flash, which adopts a high-performance Shortcut-connected Mixture-of-Experts (MoE) architecture with zero-computation experts, \longcat integrates efficient multimodal perception and speech reconstruction modules. Despite its immense size of 560B parameters (with 27B activated), \longcat achieves low-latency real-time audio-visual interaction.
For training infrastructure, we developed a modality-decoupled parallelism scheme specifically designed to manage the data and model heterogeneity inherent in large-scale multimodal training. This innovative approach demonstrates exceptional efficiency by sustaining over 90\% of the throughput achieved by text-only training. Extensive evaluations show that \longcat achieves state-of-the-art performance on omni-modal benchmarks among open-source models. Furthermore, it delivers highly competitive results across a wide range of modality-specific tasks, including text, image, and video understanding, as well as audio understanding and generation.
We provide a comprehensive overview of the model architecture design, training procedures, and data strategies, and open-source the model to foster future research and development in the community.

\textbf{LongCat Chat}: \href{https://longcat.ai}{https://longcat.ai} \\
\textbf{Hugging Face}: \href{https://huggingface.co/meituan-longcat}{https://huggingface.co/meituan-longcat/LongCat-Flash-Omni}\\
\textbf{GitHub}: \href{https://github.com/meituan-longcat}{https://github.com/meituan-longcat/LongCat-Flash-Omni}

\end{abstract}

\begin{figure}[h!]
    \centering
    \includegraphics[width=0.9\linewidth]{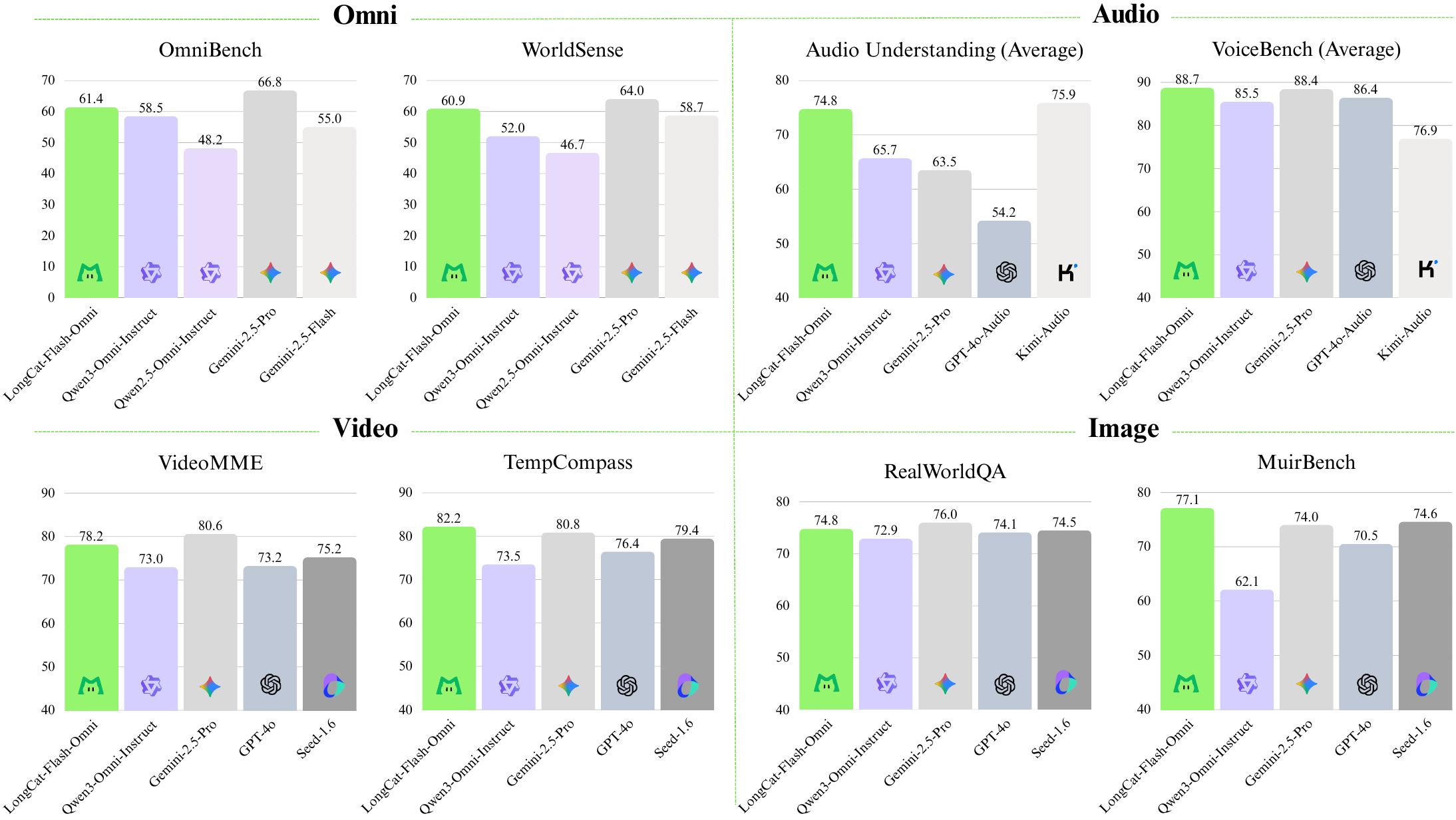}
    \caption{Benchmark performance of \longcat.}
    \label{fig: benchmark_overview}
\end{figure}

\clearpage
\tableofcontents
\clearpage

\section{Introduction} 

Humans are inherently omni-modal beings, capable of efficiently perceiving and integrating diverse forms of information, including visual and auditory inputs, to accomplish a wide range of challenging tasks. The seamless combination and transmission of multiple modalities significantly enhance the effectiveness and efficiency of human communication and interaction. In pursuit of Artificial General Intelligence (AGI), the field of large language models (LLMs) is now rapidly evolving toward the integration of richer multimodal capabilities and more efficient human-AI interaction.
Recent pioneers such as Gemini-2.5 \citep{comanici2025gemini} and GPT-4o \citep{gpt4o} have integrated text, audio, image, and video processing within a single model, enabling efficient audio-visual interaction. Following these advances, research on omni-modal models has attracted broad attention, with many subsequent efforts proposed in the community \citep{xu2025qwen3, guo2025m2, li2025baichuan, liu2025ola, fu2025vita}.

Training an omni-modal model that possesses both strong offline multimodal understanding and real-time audio-visual interaction capabilities is highly challenging. The difficulties mainly arise in the following aspects: (1) \textit{Cross-modal heterogeneity}: The substantial differences among modalities require exploring effective unified representations and fusion strategies to enable synergy across modalities, ensuring that the performance of any single modality does not degrade compared to their unimodality counterparts of similar scale. (2) \textit{Unified offline and streaming capabilities}: Integrating offline multimodal understanding with streaming audio-visual interaction presents a significant challenge. Streaming interaction scenarios require distinct capabilities not typically found in offline processing, such as the perception of relative time, precise synchronization of audio-visual information, and efficient management of multi-turn interaction contexts. (3) \textit{Real-time interaction}: Achieving real-time audio-visual interaction presents inherent difficulties, including the necessity of supporting both streaming audio and video input, as well as streaming speech output. The stringent low-latency requirement further imposes strict constraints on computational efficiency, thus placing high demands on both model architecture design and deployment infrastructure. (4) \textit{Training Efficiency}: The heterogeneity within model and data poses great challenges for the design of distributed strategies.

In this report, we attempt to address the aforementioned challenges. 
To overcome the first challenge, we carefully design a multi-stage large-scale pretraining pipeline. Based on an early-stage text pretraining foundation model, we progressively incorporate audio and visual data into the large-scale pretraining process. By adopting a balanced multimodal data mixture and effective ealy-fusion strategy, the model achieves deeply integrated comprehension across modalities while maintaining strong unimodal performance. 

To address the second challenge of balancing offline multimodal understanding with real-time audio-visual interaction, we introduce a human-in-the-loop strategy to construct high-quality interaction data, with careful consideration for long-term memory and multi-turn dialogue handling. In addition, we derive vision-speech question-answering data from existing vision-text corpora, enabling natural speech output and facilitating the transfer of strong offline multimodal understanding capabilities into interactive scenarios.

To address the third challenge of achieving low-latency audio-visual interaction in a large-scale model, we dedicate substantial effort to the design of all modules in \longcat.
We adopt the ScMoE architecture with zero-computation experts from LongCat-Flash~\citep{meituan2025longcat_flash_chat,meituan2025longcat_flash_chat_thinking} as the LLM backbone. To handle streaming input, we employ efficient audio and video encoders for feature extraction and introduce a synchronized chunk-wise interleaving strategy for real-time processing. For efficient audio reconstruction, we adopted a multi-codebook audio detokenization scheme with a coarser temporal resolution. This approach significantly improves decoding efficiency while preserving reconstruction quality.
Furthermore, we designed an efficient streaming pipeline for model serving to minimize end-to-end server-side latency. As a result, despite having up to 560B parameters (with 27B activated), \longcat achieves millisecond-level response latency in real-time interactive scenarios.

To address the fourth challenge of training efficiency, we devote substantial effort to large-scale omni-modal distributed training. We propose a modality-decoupled parallelism (MDP) strategy. This approach enables independent optimization of both performance and memory usage for the LLM, vision encoder, and audio encoder. For the LLM component, we systematically optimize the distributed training configuration for computational efficiency and apply multiple memory-reduction techniques to ensure robust system stability throughout training. Experimental results demonstrate the effectiveness of this strategy, our system sustains more than 90\% of the throughput achieved during pure text training.

Comprehensive evaluations demonstrate that our model delivers strong and consistent performance across both omni-modal benchmarks and real-time interactive tasks. \longcat achieves state-of-the-art results on omni-modal benchmarks such as Omni-Bench and WorldSense, while also exhibiting highly competitive performance on a wide range of unimodal tasks, including text, image, and video understanding, as well as speech comprehension and generation, establishing itself as the most powerful omni-modal model in the open-source community.
Furthermore, extensive subjective evaluations confirm that \longcat supports high-quality audio-visual interaction with low latency.

The key features of \longcat are summarized as follows:
\begin{itemize}
    \item \textbf{SOTA and Unified Omni-Modal Model:} \longcat achieves state-of-the-art cross-modal comprehension performance among open-source models. It seamlessly integrates powerful offline multimodal understanding with real-time audio-visual interaction within a single all-in-one framework. 
    \item \textbf{Large-Scale with Real-time Audio-Visual Interaction:} By leveraging an efficient LLM backbone, carefully designed lightweight modality encoders and decoder, and a chunk-wise audio-visual feature interleaving mechanism, \longcat achieves low-latency, high-quality audio-visual processing and streaming speech generation. It supports a context window of up to 128K tokens, enabling advanced capabilities in long-term memory, multi-turn dialogue, and temporal reasoning across multiple modalities.
    \item \textbf{Effective Early-Fusion Training:} The model adopts an innovative multi-stage pretraining pipeline that progressively incorporates text, audio, and visual modalities under a balanced data strategy and early-fusion training paradigm, ensuring strong omni-modal performance without degradation in any single modality.
    \item \textbf{Efficient Training Infrastructure:} Inspired by the concept of modality decoupling, we propose a modality-decoupled parallelism training scheme that significantly enhances the efficiency of large-scale and highly challenging multimodal training.
    \item \textbf{Open-Source Contribution:} We provide a comprehensive overview of the training methodology and data strategies behind \longcat, and release the model to accelerate future research and innovation in omni-modal intelligence.
\end{itemize}

The remainder of this paper is organized as follows. Section~\ref{sec2:archi} presents the model architecture of \longcat. Sections~\ref{sec:pre-train} and \ref{sec4:post-train} describe the pretraining and post-training datasets and pipelines, respectively. Sections~\ref{sec6:infra} and \ref{sec7:infer} introduce training infrastructures and inference deployment. Section~\ref{sec5:eval} reports the experimental results. Section~\ref{sec8:conclusion} draws a conclusion of this report.

\section{Architecture}
\label{sec2:archi}


\begin{figure}
    \centering
    \includegraphics[width=0.7 \linewidth]{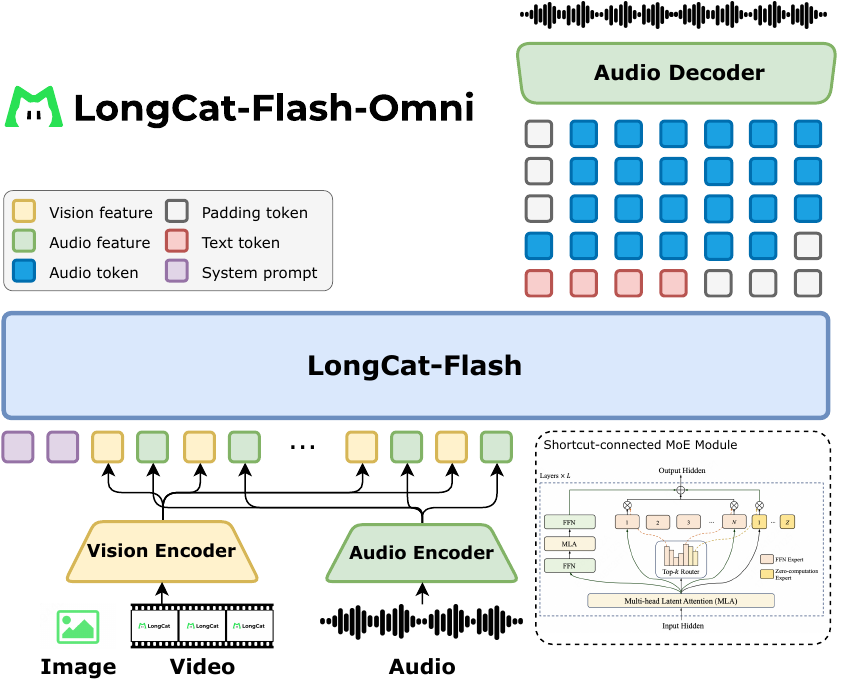}
    \caption{An overview of the LongCat-Flash-Omni model architecture. The model is fully end-to-end and unifies multimodal understanding and generation across text, image, video, and audio within a single large language model framework. An vision encoder and an audio encoder are used to obtain vision features and audio features, respectively, which are then projected into a shared latent token space and fed into the LongCat-Flash LLM backbone. The LLM decoder directly generates multi-codebook speech tokens, parallel to generated text tokens, which are then converted to audio waveforms by an audio decoder. Shortcut-connected MoE (ScMoE) with zero-computation experts module proposed in LongCat-Flash is employed to achieve efficient multimodal fusion. Vision and audio features are chunk-wisely interleaved to support streaming audio-visual input.}
    \label{fig:architecture}
\end{figure}

As illustrated in Figure~\ref{fig:architecture}, \longcat is a fully end-to-end omni-modal model. It can receive various modalities as input, including text, audio, image, video, and their any arbitrary combination, and is able to directly generates speech tokens from the LLM backbone. 
\longcat employs a vision encoder and an audio encoder as multimodal perceivers. An LLM processes the multimodal inputs and generates text and audio tokens. An audio decoder reconstructs the waveform from the speech tokens generated by the LLM, enabling natural speech interaction.
All modules are carefully designed to support efficient streaming inference. The audio encoder, vision encoder, and audio decoder are lightweight components, each with approximately 600M parameters. The large-scale LLM backbone uses the novel and efficient architectural design proposed in the LongCat LLM family~\citep{meituan2025longcat_flash_chat,meituan2025longcat_flash_chat_thinking}.


In the reminder of this section, we first introduce each structural component of \longcat, including the vision encoder that supports inputs with arbitrary aspect ratios and native-resolution feature encoding, the audio encoder, decoder, and tokenizer, and the LLM backbone. Then we elaborate on the video processing strategy and the architectural components that enable low-latency, real-time audio-visual interaction.

\subsection{Vision Encoder} \label{sec:vision_encoder} 
The vision encoder serves as a critical component in multimodal language models. To effectively encode visual inputs such as images and videos, LongCat-Flash-Omni incorporates a well-designed Vision Transformer (ViT), referred to as LongCat-ViT~\citep{qiao2025univitar}. LongCat-ViT achieves high performance across multimodal tasks, natively supports inputs with various resolutions and aspect ratios, while providing unified encoding capabilities for both image and video data.

\textbf{Architecture Design}
LongCat-ViT is a Transformer-based encoder that retains the core structure of the conventional Vision Transformer while integrating several key enhancements: a unified patchification module for native image and video inputs, 2D rotary position embeddings (2D-RoPE)~\citep{su2024roformer}, SwiGLU activation function, RMSNorm layer, a LayerScale module, and Query-Key normalization. Together, these refinements yield a more robust and efficient architecture compared to conventional designs.
To improve computational efficiency in video frame encoding during real-time interaction while maintaining model performance, we adopt a relatively lightweight model configuration.
The detailed hyper-parameters of the model are provided in Table~\ref{tab:vit_arch}.
In line with common practice, a two-layer multilayer perceptron (MLP) with pre-normalization is employed as the vision-language projector to align visual and textual representations. 
Moreover, a 2$\times$ pixel-unshuffle operation is applied along the spatial dimension to mitigate the quadratic computational complexity associated with high-resolution inputs.
\begin{table}[ht]
  \caption{Detailed architectural configuration for LongCat-ViT.}
  \centering
  \renewcommand{\arraystretch}{1.1}
  \scalebox{0.92}{
  \begin{tabular}{cccccccc}
    \toprule
    Patch Size & Pos Embed & Hidden Size & Intermediate Size & Num Layers & Num Heads & Parameters (M)  \\
    \midrule
    14 & 2D-RoPE& 1280 & 5184 & 32 & 16 & 637 \\
    \bottomrule
    \end{tabular}
}
   \label{tab:vit_arch}
\end{table}

\textbf{Native Resolution Encoding}
Conventional ViT models widely adopted in the community (e.g., CLIP~\citep{radford2021clip}, SigLIP~\citep{zhai2023sigmoid}) typically resize input images to a fixed resolution, which can result in substantial information loss, particularly for images with extreme aspect ratios or high native resolutions.
To mitigate this limitation, LongCat-ViT encodes visual inputs at their native resolutions, circumventing the limitations of fixed-resolution ViT models. This preserves the spatial and contextual information inherent in visual data, thereby enhancing the model’s capability to comprehend and reason over complex visual inputs.
For each image or video frame, if the number of patches falls within a predefined range (576-5832 during training), only minimal resizing is applied to ensure both dimensions are divisible by 112. Otherwise, the image is rescaled to fit within this range while preserving its aspect ratio.

\textbf{Contrastive Vision-Language Pretraining} 
LongCat-ViT adopts a progressive training scheme that integrates two complementary adaptation strategies: (1) progressive resolution adaptation, which leverages curriculum learning by transitioning from fixed low-resolution (e.g., 224) pretraining to native-resolution fine-tuning; and (2) progressive visual modality adaptation, which postpones the incorporation of video data until the final training stage to reduce computational overhead. To further facilitate convergence in the early phases, feature distillation from a frozen pretrained vision model is introduced as an auxiliary objective, and the weight of this objective gradually reduces in later stages. The model is trained from scratch on a total of 14.6 billion samples during the contrastive pretraining phase.

\subsection{Audio Tokenizer, Encoder, and Decoder} 
\label{sec:audio}

We input audio in different formats to the LLM backbone of \longcat across training phases. Specifically, during pre-training stages 1-4, an \textit{audio tokenizer} is employed to convert raw speech into four-codebook discrete tokens, enabling consistent next-token prediction and improved training efficiency.
However, we observe that such discretization hinders the model’s ability to capture fine-grained acoustic details. Therefore, from pre-training stage 5 (Section~\ref{subsec:stage-5}) we incorporate an \textit{audio encoder} to convert raw speech into continuous audio features before input to the LLM.
For speech generation, to align with the inherent next token prediction paradigm, the LLM consistently outputs the four-codebook discrete tokens, which are subsequently converted back into a waveform using an \textit{audio decoder}.

\begin{figure}[t]
  \centering
  \begin{minipage}[b]{0.44\textwidth}
    \centering
    \includegraphics[width=\textwidth]{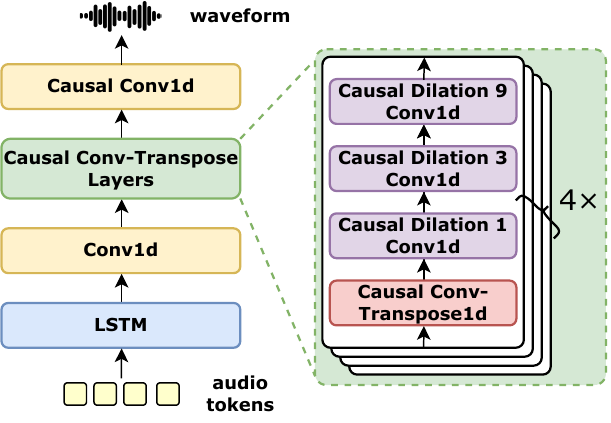}
    \caption{Architecture of the audio decoder.}
    \label{fig:Architecture of Code2Wav}
  \end{minipage} 
  \hfill
  \begin{minipage}[b]{0.54\textwidth}
    \centering
    \includegraphics[width=\textwidth]{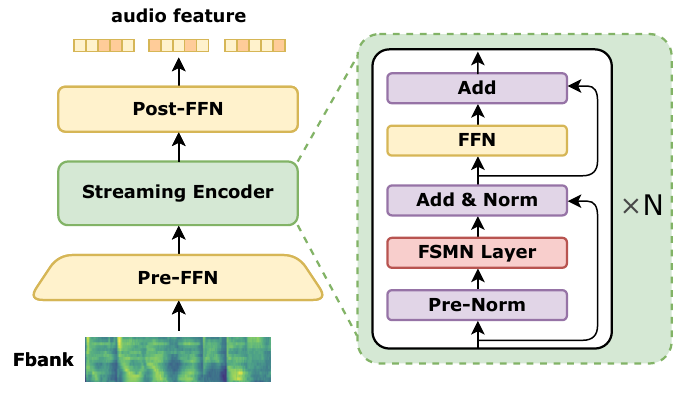}
    \caption{Architecture of the audio encoder.}
    \label{fig:audio_encoder}
  \end{minipage}
\end{figure}

\textbf{Audio Tokenizer and Decoder} We adopt LongCat-Audio-Codec~\citep{longcat-codec2025zhao} as our audio \textit{tokenizer} and \textit{decoder}, owing to its robust semantic modeling, flexible acoustic feature extraction, and low-latency streaming synthesis capabilities. The tokenizer discretizes audio waveforms into four codebooks at a frame rate of 16.67 Hz, with one codebook representing semantic information and the other three capturing acoustic details.
To achieve low-latency inference in real-time interaction scenarios, unlike conventional waveform reconstruction methods that rely on diffusion- or flow-matching-based code2mel models followed by vocoders, we directly employ the decoder from LongCat-Audio-Codec as the audio decoder to reconstruct waveform from tokens. 
It supports streaming audio decoding with a look-ahead of only three frames. 
As illustrated in Figure~\ref{fig:Architecture of Code2Wav}, the audio decoder consists of LSTM layers, convolutional blocks, and causal transposed convolution layers, and is trained under a generative adversarial network (GAN) framework.

\textbf{Audio Encoder}
\label{audio_encoder}
To optimize response latency and accommodate speech inputs of arbitrary duration, the audio encoder is designed using a streaming architecture.
As illustrated in Figure~\ref{fig:audio_encoder}, the audio encoder takes 80-dimensional Fbank features as input. The architecture incorporates a Pre-FFN module that reduces the audio sequence length by a factor of eight through frame splicing downsampling technique, with each frame representing an 80ms time window.
The core process is performed by a streaming encoder that maintains a Transformer-like structure while incorporating several key modifications: (1) a Pre-Norm configuration for enhanced training stability, and (2) replacing standard self-attention modules with FSMN layers \citep{zhang2018deep} to enable efficient feature processing within constrained context windows. 
To balance latency and performance, we implement a hybrid approach where only the final six layers incorporate a one-frame look-ahead mechanism, while maintaining strict causality in the preceding layers.
The architecture concludes with a post-FFN module for additional feature refinement. 
The audio encoder is trained under supervised learning using speech recognition data using the CTC loss \citep{graves2006connectionist}.

\subsection{LLM Backbone} 
LongCat-Flash-Omni is built upon LongCat-Flash~\citep{meituan2025longcat_flash_chat}, a 560-billion-parameter Mixture-of-Experts (MoE) language model.
LongCat-Flash adopts Multi-head Latent Attention (MLA)~\citep{liu2024deepseek}, shortcut-connected MoE~\citep{cai2024shortcut} and zero-computation experts, performing variable computation per token by activating 18.6B-31.3B parameters (27B on average), thereby unifying efficiency, performance and sparsity.
These characteristics are retained and extended to multimodal understanding and audio-visual interaction by \longcat.

\subsection{Video Strategy and Streaming Audio-Visual Interaction} 
\label{sec:audio-visual}

LongCat-Flash-Omni is designed to seamlessly integrate robust offline multimodal understanding with low-latency audio-visual interaction. 
The audio and visual streams are independently processed by an audio encoder and a vision encoder, respectively. Their extracted features are subsequently time-aligned and chunked into synchronized segments, which are interleaved and fed into the LLM decoder for multimodal understanding.
Here we elaborate the video strategy adopted by \longcat and how audio-visual input is processed to support streaming interaction.

\subsubsection{Video Strategy}
\label{subsec:video-strategy}
Efficient video processing remains a significant challenge due to the substantial variability in video properties, including durations ranging from a few seconds to several hours and resolutions spanning a wide spectrum. To address these challenges, we adopt a series of strategies to effectively balance between model performance and computational efficiency.

\textbf{Dynamic Video Frame Sampling} We adopt a default sampling rate of 2 frames per second (2 FPS), while also enable dynamic adjustments according to video duration.
During training, shorter videos are sampled at a higher frame rate to capture denser temporal information, ensuring at least 16 frames for effective information utilization. In contrast, for excessively long videos, frames will be sampled uniformly based on a maximum frame number constraint. This frame cap further facilitates training by regulating memory consumption and preserving computational efficiency.

\textbf{Textual Timestamps}
LongCat-Flash-Omni introduces timestamps preceding each input video frame to strengthen the model’s temporal awareness and improve its ability to recognize specific time points.
The timestamps are input as pure text that are naturally aligned with the textual space. Similar to~\citep{chen2024timemarker}, when sampling a video frame at second $i$, we prepend the text ``$Second\{i\}$'' to its corresponding visual tokens before inputting them into the LLM. The resulting input sequence is structured as ``$Second\{i\} || \mathbf{V}_i || Second\{j\} || \mathbf{V}_j || \dots$'', where $Second\{i\}$ is the textual timestamp, $\mathbf{V}_i$ represents the visual tokens of the video frame at second $i$, and $||$ indicates concatenation.

\textbf{Hierarchical Token Compress in Video Inputs} We compress the video inputs with three successive steps:
First, we rescale the each frame according to a predefined upper limit on the number of patches as described in Section~\ref{sec:vision_encoder}.
Then we apply a 3D convolution with temporal stride of 2 before feeding the video into the visual encoder, compressing the $N$ input frames to $N/2$ in temporal size.
Finally, after the video is processed by the visual projector into visual tokens, if the number of visual tokens exceeds a predefined limit, we further perform an interpolative downsample on the visual tokens.

\subsubsection{Streaming Audio-Visual Interaction}
\label{subsec:audio-visual-design}

The streaming audio-visual interaction mechanism is a core component of LongCat-Flash-Omni, enabling real-time integration of video and speech signals to support interactive communication. The proposed audio-visual interaction framework is characterized by two key aspects:

\textbf{Streaming Audio-Visual Feature Interleaving} Unlike offline audio-visual understanding tasks, where audio and visual features can be concatenated at the sequence level, real-time audio-visual interaction requires features from the audio and video streams to be prefilled into the LLM backbone as early as possible to minimize response latency once the user query is received.
To achieve this, we design a temporally-synchronized, chunk-wise audio-visual feature interleaving mechanism. 
Audio-visual feature chunks are structured as ``<|timestamp|>:<|video-tokens|><|audio-start-token|><|audio-tokens|><|timestamp|>:<|video-tokens|><|audio-tokens|>...<|audio-end-token|>'', where the timestamp is represented in textual form, as described in Section~\ref{subsec:video-strategy}.


\textbf{Sparse-Dense Sampling Strategy} We design a sparse-dense sampling strategy to optimally balance computational cost and information loss during turn-taking interactions between the user and the model.
Specifically, we employ a chunk size of 1 second during the information input period to preserve as much audio-visual information as possible, using a denser video sampling rate of 2 FPS. During the model response period, video frames are buffered with a sparser sampling rate (i.e., chunk size of 2 seconds, 0.5 FPS) and prepended to the next user turn. This design effectively balances visual information retention during the model response period and computational overhead, enabling high-quality audio-visual interaction, a core capability that distinguishes it from other omni-modal models in the community.

\section{Pre-Training}
\label{sec:pre-train}


This section describes the data curation process (Section~\ref{sec:pretrain_data_curation}) and training strategy (Section~\ref{sec:train_stages}) of \longcat.

\subsection{Data Curation} \label{sec:pretrain_data_curation}
We collect a large-scale and diverse multimodal corpus with over 2.5 trillion tokens for pre-training. This pre-training corpus consists of the following components: audio data (Section~\ref{sec:pretrain_audio_data}), generic image-text data (Section~\ref{sec:pretrain_generic_image_text_data}), OCR, grounding and GUI data (Section~\ref{sec:pretrain_ocr_grounding_and_gui_data}), STEM data (Section~\ref{sec:pretrain_stem_data}), multi-image data (Section~\ref{sec:pretrain_multi_image_data}), video data (Section~\ref{sec:pretrain_video_data}), and long-context multimodal data (Section~\ref{sec:pretrain_long_context_multimodal_data}).

\subsubsection{Audio Data} \label{sec:pretrain_audio_data} 
\textbf{Speech-Text Interleaved Data}
Our data consists of tens of millions of hours of audio with rich diversity.
We employ the following pipeline to extract high-quality speech audios with consistent topics:
First, we use a VAD system to split a long audio into speaking segments and remove non-speaking regions. Next, we use two proprietary ASR models for cross-validation, filtering out segments with significant discrepancies on transcription results. A multilingual speech aligner~\citep{pratap2024scaling} is then applied for forced alignment, yielding precise transcription timestamps.
We further refine the dataset by computing the ratio between speech duration and text length for each segment, and discard the segments whose ratios fall outside the $0.5$-th to $99.5$-th percentile range.
Finally, we merge adjacent speaking segments separated by silences shorter than 10 seconds to form the training samples.

To build speech-text interleaved training data, we first split each training sample into fragments using punctuation marks as delimiters: $(A_1, T_1), (A_2, T_2), \dots, (A_n, T_n)$, where $A_i$ and $T_i$ represent the $i$-th audio fragment and its transcribed text, respectively. We then randomly mask either audio or text component of some fragments, resulting in the final training input such as $(T_1, A_2+T_2, T_3, A_4+T_4, \dots)$ or $(A_1, A_2+T_2, A_3, A_4+T_4, \dots)$.

\textbf{Audio Understanding Data} We curate an audio understanding dataset that encompasses a wide range of tasks, including audio captioning, semantic audio understanding, paralinguistic analysis, acoustic scene and event detection, audio question answering, and music understanding.
The dataset comprises a combination of open-source datasets and in-house proprietary datasets.

We apply text translation as data augmentation for speech recognition datasets, various speech models for pseudo-label generation and label quality filtering, particularly for paralinguistic understanding and audio captioning datasets, and a diverse set of prompts for each task to enhance instruction variability.

\subsubsection{Generic Image-Text Data} \label{sec:pretrain_generic_image_text_data}
\textbf{Image Caption Data}
High-quality image captions are crucial for aligning visual representations with the language model’s knowledge space. To this end, we build a large-scale image-caption dataset by first applying a multi-stage cleaning pipeline across text, image, and image-text pair levels. Low-quality samples, such as those with extremely short text, abnormal resolutions, or poor image quality, are removed using elaborately-refined heuristic rules. At the pair level, we further filter samples using a SigLIP similarity threshold, discarding those below a strict minimum to preserve downstream diversity. We then improve the dataset through re-captioning to generate dense, fine-grained captions with multiple open-source vision-language models while incorporating world knowledge from original annotations. Additional filtering, including mixed-language detection, repetition removal, and truncation handling is applied to help mitigate hallucinations and ensure the final image-text pairs are accurate, informative, and contextually rich.

To broaden semantic coverage and avoid overfitting to dominant patterns, we significantly enhance data diversity and balance. We cluster image-text pairs based on joint image and text embeddings and resample from each cluster to ensure representation of long-tail content, while simultaneously removing low-quality samples frequently concentrated in specific clusters. Experimental results confirm that preserving diversity at this stage is more beneficial than overly strict quality filtering. Furthermore, to address the pronounced long-tail distribution inherent in multimodal datasets, we adopt a concept-based resampling strategy inspired by MetaCLIP~\citep{xu2023demystifying}. By expanding the vocabulary with a 200K-scale Chinese lexicon and resampling for broader concept coverage, we achieve a more balanced distribution across semantic categories. 

\textbf{Interleaved Image-Text Data}
Interleaved image-text data provides broader visual-textual coverage and improves few-shot performance, but it is often noisy and has uneven quality. To address this, we construct a high-quality dataset through a two-stage pipeline of filtering and diversity sampling from open-source data. In the filtering stage, we remove samples with noisy tokens, sensitive content, and overly complex samples, and discard corrupted or low-resolution images. We then improve image-text alignment by pairing each image with its most relevant text segment by SigLIP similarity scores, and discard pairs with low semantic alignment.
To sample a diverse and evenly distributed subset, we apply density-based pruning and semantic clustering similar to~\citep{abbas2024effective}. Samples are scored by knowledge content, quality, and similarity, then evenly sampled based on the cluster they belong to and their distances to the cluster centroid. 
This process reduces the raw dataset by approximately 74\%, while maintaining diversity and quality for multimodal pretraining.

To enhance domain knowledge and reasoning capabilities, we further curate an in-house, high-quality image-text interleaved dataset derived from video content containing educational materials.
We use an automated pipeline to select only instructional video segments and remove other parts.
Textual information is extracted using ASR and OCR, then refined by an LLM to improve accuracy and consistency. Videos are segmented into meaningful scenes, and key frames are extracted and aligned with the refined text based on temporal and semantic correspondence. The resulting dataset comprises structured, context-rich sequences of synchronized visual and textual information, significantly strengthening the model’s capacity for academic reasoning and multimodal understanding.

\subsubsection{OCR, Grounding and GUI Data} \label{sec:pretrain_ocr_grounding_and_gui_data}

\textbf{OCR Data}
We curate richly annotated training samples encompassing various content types, including PDFs, papers and web pages. These efforts further enrich our document parsing dataset, enabling the model to learn both content extraction and structural understanding, and resulting in more robust and accurate document parsing.
We further include diverse OCR datasets covering scene text, structured documents, handwriting, and mathematical expressions.
We also synthesize multi-page and region-level OCR samples from scene and document OCR data, with each sample comprising 2 to 6 pages. 
For OCR-related VQA, we incorporate a diverse range of datasets for visual question answering, covering domains of text-centric VQA, document VQA, table VQA, and chart VQA.

\textbf{Grounding Data}
To enhance the model’s grounding capabilities, we adopt several widely used open-source object detection datasets \citep{krishna2017visualgenome,mao2016refcocoplusg,lin2014microsoftcoco,shao2019objects365}. We perform data quality validation to filter out incorrect and redundant samples. Using these high-quality open-source datasets, we construct two types of question-answering data: localization data and region-captioning data. We uniformly adopt relative coordinates, which are normalized to the range of 0-1000, and process the data into JSON format. We also incorporate PixMo dataset~\citep{deitke2024molmo} to enhance the model’s counting ability.

\textbf{GUI Data}
Graphical User Interface (GUI) data contains rich information on visual understanding and task planning, enabling automated interactions on both mobile and desktop platforms. To this end, we utilize diverse types of GUI-related data to enhance the model’s perception, grounding, and planning capabilities \citep{zeng2025uitron}.
For GUI perception, we utilize a large number of image-text pairs from screenshots of various PC and mobile applications, covering tasks such as OCR, VQA, and captioning with GUI images. For GUI grounding, we construct instruction-answer pairs based on visual screenshots from diverse platforms and devices. For each screenshot we create multiple pairs, accounting for different UI elements and interaction possibilities.
For GUI planning, we collect a diverse set of navigational paths with rich context from both mobile and desktop. The planning data comprises key components such as screenshot observations, summarizations, reasoning traces, and corresponding actions. To enhance the model’s reasoning capability, we integrate multiple inference settings that involve different combinations of action, summarization, and reasoning outputs. 
Furthermore, to capture the underlying logic of GUI dynamics, we incorporate contextual scenarios that enable the model to predict changes across different screenshots.

\subsubsection{STEM Data} \label{sec:pretrain_stem_data}
To strengthen the model’s foundational capabilities in scientific reasoning and problem-solving, we construct a large-scale, multimodal STEM (Science, Technology, Engineering, and Mathematics) dataset.
The collected data are meticulously processed and structured into both multiple-choice and open-ended generative question-answer formats.
We implement a rigorous multi-stage filtering pipeline to ensure factual accuracy, eliminate ambiguity, and standardize formatting.
This effort culminates in a high-fidelity pre-training dataset comprising 15 million image-text pairs with substantial diversity in both subject matter and academic level, and covering a wide range of disciplines from K12 education to advanced university studies.
This foundational dataset is essential for enabling the model to achieve deep conceptual understanding and robust reasoning capabilities across a broad spectrum of scientific and technical domains.

\subsubsection{Multi-Image Data} \label{sec:pretrain_multi_image_data} 

To enhance the model's fine-grained image understanding, we constructed a taxonomy with various coarse-grained and fine-grained capabilities, such as ``emotion recognition'', ``vehicle identification'', and ``time calculation with clocks''.
Based on this taxonomy, we employed diverse strategies to collect multi-image question answering data, including open-source datasets with careful data selection and augmentation, and synthesizing images using dedicated tools.


\subsubsection{Video Data} \label{sec:pretrain_video_data} 
Our video dataset is primarily sourced from a broad range of publicly available corpora, covering diverse task types such as video classification, temporal grounding, detection, captioning, and question answering. We design a comprehensive data processing pipeline centered on rigorous quality filtering and targeted enhancement, resulting in a refined, high-quality dataset tailored for large-scale pre-training.

Additionally, we curate an in-house dataset from publicly available video content with three components:

\textbf{High-quality video caption data} Our video captioning process begins by dividing videos into coherent scenes using a scene detection algorithm similar to Koala~\citep{wang2025koala}. To ensure a wide range of content, we then cluster and sample these scenes. Finally, we employ a proprietary model to generate detailed captions that describe the sequence of events and the overall context of the video.

\textbf{Temporally grounded video QA data} We construct a high-quality temporally grounded dataset by applying rule-based transformations to convert annotations from tasks such as temporal action detection, segmentation, video summarization, and temporal sentence grounding into question-answer (QA) pairs. Data complexity is further enriched through a proprietary model that generates more challenging and in-depth QA pairs.

\textbf{Action recognition video QA data} To strengthen the model’s action recognition capability, we transform several publicly available video recognition datasets~\citep{carreira2019short,Goyal_2017_ICCV} into multiple-choice and open-ended video QA pairs.

\subsubsection{Long-Context Multimodal Data} \label{sec:pretrain_long_context_multimodal_data} 

To address the challenges in long-context multimodal scenarios, we construct a specialized long-context multi-modal dataset. The dataset contains two main parts: (1) carefully filtered samples from open-sourced data and in-house pre-training data, with a focus on videos exceeding three minutes in length, and (2) in-house long-context synthesized data.

Our in-house long-context dataset includes image-text interleaved data and long-video QA data.
Image-text interleaved data is constructed by concatenating single-image data with the same topic or rendering some text segments within long text into images. This data can help the model to improve in-context learning and information retrieval from long context (i.e., ``a needle in a haystack'').
Long video QA data is constructed by the following pipeline: (1) First we segment videos into multiple clips and generates rich descriptive captions for each clip. (2) For consecutive clips from the same video, we analyze their scene continuity from their captions. If the clips are identified as continuous, we concatenate and refine their captions into a coherent longer-form video caption. (3) Finally, we construct temporally grounded and detailed QA pairs to form our long-video QA dataset.
This data can help the model to enhance long-sequence cross-modal modeling and memory retention.

\begin{figure}
    \centering
    \includegraphics[width=1.0\linewidth]{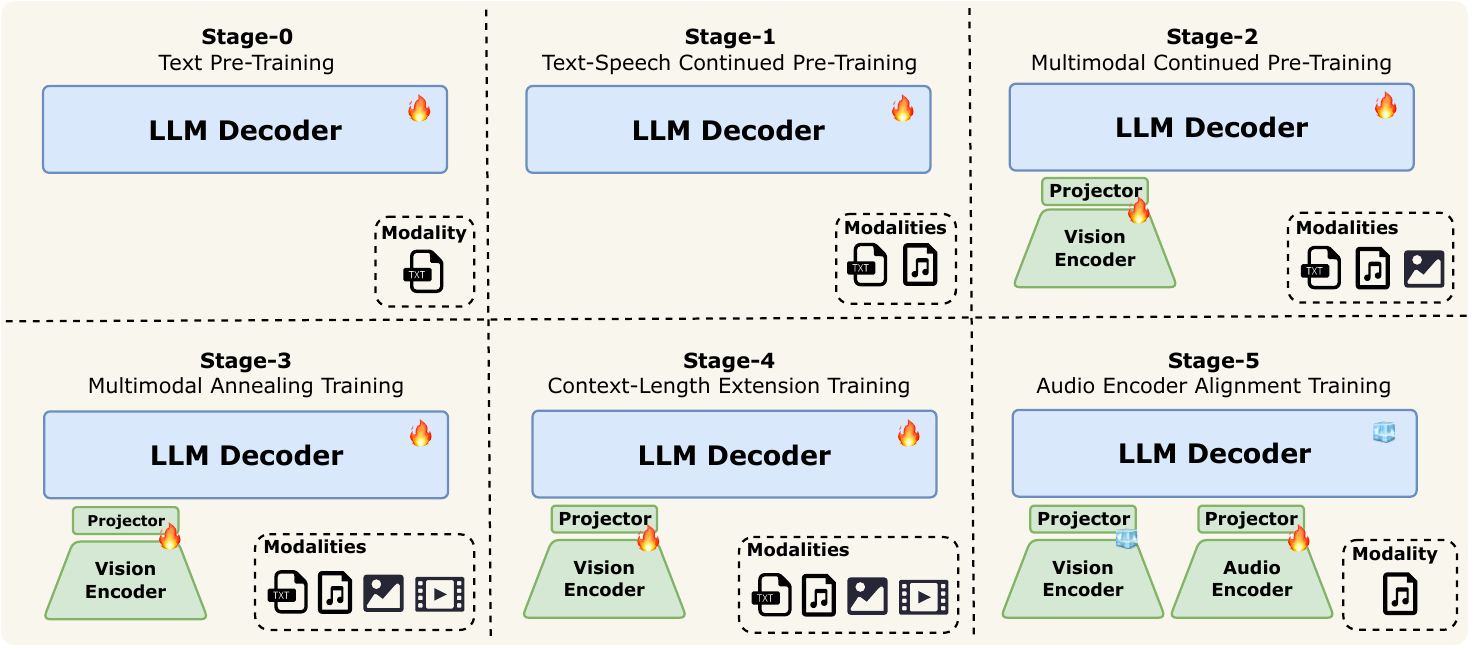}
    \caption{Pre-training stages.}
    \label{fig:pretraining}
\end{figure}

\subsection{Training Strategy} \label{sec:train_stages}

One of the most fundamental challenges in training omni-modal models lies in the significant heterogeneity of data distributions across modalities. Each modality exhibits distinct structural properties and representational characteristics.
Text, for instance, is a highly compressed and abstract symbolic representation of human knowledge, and LLMs have demonstrated remarkable success in modeling large-scale text sequences. Speech, on the other hand, is the acoustic manifestation of human thoughts and concepts, a sequential signal like text, but enriched with paralinguistic information such as speaker timbre, emotion, prosody, and accent. However, its semantic density is much lower than that of text: while a typical speech tokenizer operating at 12.5 Hz must generate roughly 12.5 tokens per second, humans only speak about 3-4 text tokens per second. This mismatch makes sequence modeling for speech inherently more challenging than for text.

Visual information introduces a fundamentally different modality. Unlike the sequential nature of text and speech, images encode spatial structures, requiring the model to reason over spatial relationships. Video data further increases the complexity by incorporating both spatial and temporal dynamics, making it a much more complicated modality to model effectively. This requires careful consideration of effective frame sampling strategies, token compression methods, and temporal modeling approaches, as well as the ability to handle long-range sequence dependencies.

Guided by these observations, we adopt a curriculum-inspired, progressive training strategy that gradually transitions from simpler to more complex sequence modeling tasks, as illustrated in Figure~\ref{fig:pretraining}.
We begin with large-scale text pretraining (\textbf{Stage-0}), leveraging the maturity and stability of LLMs as a strong initialization for subsequent multimodal learning. Building on this foundation, we introduce speech data, which is structurally closer to text, to align acoustic representations with the language model’s feature space and effectively integrate paralinguistic information (\textbf{Stage-1}). Once speech-text alignment is established, we incorporate large-scale image-caption pairs and vision-language interleaved corpora (\textbf{Stage-2}) for vision-language alignment, enriching the model’s visual knowledge. We then introduce the most complex video data to enable spatial-temporal reasoning (\textbf{Stage-3}), meanwhile integrating higher-quality and more diverse image datasets to enhance vision comprehension. 
To further support long-context reasoning and multi-turn interactions, we extend the model’s context window from 8K to 128K tokens (\textbf{Stage-4}). 
Finally, to mitigate information loss in audio inputs represented by discrete speech tokens, we introduce an audio encoder alignment stage (\textbf{Stage-5}) that enables the model to directly process continuous audio features, thereby enhancing fidelity in downstream speech tasks.

The following part of this section presents details of each pre-training stage.


\subsubsection{Stage-0 Text Pre-Training}
\label{subsec:stage-0}
To establish a robust text foundation, the text pre-training stage follows the same procedure as the initial phase of LongCat-Flash~\citep{meituan2025longcat_flash_chat}. The model is trained on approximately 16 trillion tokens drawn from a high-quality and diverse text corpus, where a constant learning rate is applied.
Throughout training, the proportion of high-quality reasoning data (e.g., STEM and code) is progressively increased to strengthen the model’s reasoning and programming capabilities, thereby providing a solid foundation for subsequent multimodal learning.

\subsubsection{Stage-1 Text-Speech Continued Pre-Training}
\label{subsec:stage-1}

\begin{figure}[h]
  \centering
  \includegraphics[width=0.8\textwidth]{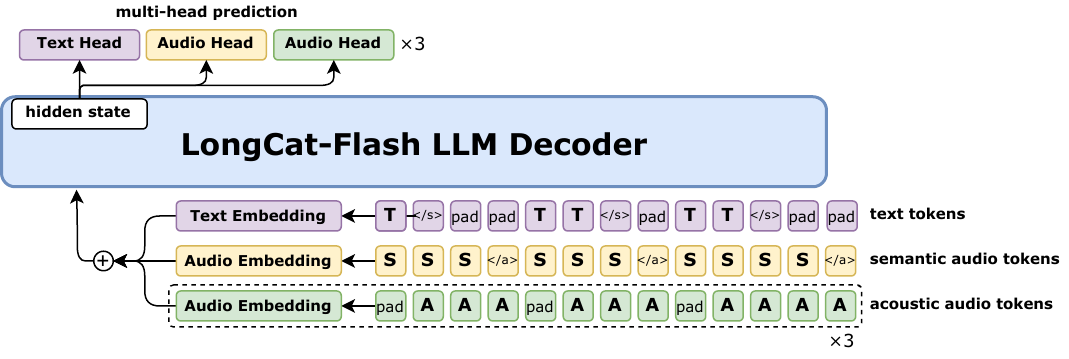}
  \caption{Schematic diagram of pre-training stage-1, where large-scale speech and text data is employed for the training process.}
  \label{fig:audio_data_pattern}
\end{figure}

Building upon the text foundation model obtained from stage-0, we continue pre-training using a mixture of text data, speech-text interleaved data, and ASR data. All speech samples are discretized into four-codebook token sequences. During training, we jointly optimize multiple objectives: (1) pure text next-token prediction (NTP) to preserve strong text understanding and generation capabilities; (2) text-speech interleaved NTP to align speech and text representations within a unified sequence modeling framework; and (3) ASR-style tasks to build basic speech perception capability.
As shown in Figure~\ref{fig:audio_data_pattern}, text and audio embeddings are fused before being fed into the LLM decoder. We introduce four audio prediction heads, enabling \longcat to directly generate audio tokens. The model simultaneously predicts text tokens, semantic tokens, and acoustic tokens, with a one-step temporal offset between semantic and acoustic token predictions. Empirical observations indicate that ASR tasks contribute minimally to modality alignment, prompting us to include only a small proportion of ASR data in the training process.
The overall training objective is defined as follows:
\[
\mathcal{L}_{\text{total}} 
= a\, \mathcal{L}_{\text{pure-text}} 
+ b\, \mathcal{L}_{\text{audio}} 
+ c\, \mathcal{L}_{\text{audio-text}} 
+ d\, \mathcal{L}_{\text{first-audio}}
\]
where $a$, $b$, $c$ and $d$ are loss weights. $\mathcal{L}_\text{pure-text}$ denotes the loss for pure text data, while $\mathcal{L}_\text{audio}$ and $\mathcal{L}_\text{audio-text}$ correspond to the audio loss and the text loss terms in the text-speech data. $\mathcal{L}_\text{first-audio}$ is an additional loss term applied to the semantic audio token. After carefully hyperparameter tuning and extensive preliminary experiments, we set \(a = 1.75\), \(b = 0.25\), \(c = 1.5\), and \(d = 0.1\), which achieves the best balance between preserving text capabilities and enhancing audio performance.

This training stage uses approximately 5.1 trillion tokens, 
with the ratio of text tokens and audio tokens is 2:1. A slight learning rate decay is applied.

\subsubsection{Stage-2 Multimodal Continued Pre-Training}
\label{sec:pretrain-stage2}

Based on the model from stage-1, we further incorporate large-scale image-text data into the pre-training procedure, including image caption data, and interleaved image-text data. We use a vision transformer (ViT) with weights initialized from the LongCat-ViT model introduced in Section~\ref{sec2:archi} to obtain visual features from images, and employ a randomly initialized vision projector to align the visual feature with the latent space of the LLM backbone.
We maintain the text-to-audio data ratio the same as in stage-1, i.e., 2:1, and make the text-to-vision data ratio also 2:1. In total, this pre-training stage consumes more than 3 trillion tokens.
The parameters of the ViT module and projector are jointly trained with the LLM decoder parameters throughout the stage-2 training, with a nearly constant learning rate. 
We reuse the loss weights in stage-1 and set the additional vision-related loss weight at 0.25 during this training process.

\subsubsection{Stage-3 Multimodal Annealing Training}
Following the pre-training stage, the model undergoes a multimodal annealing phase, where the model's training continues with a curated higher-quality data under an annealed learning rate to achieve superior performance.
We further incorporate video data, including video captioning and QA datasets, as well as a broader range of image-related data such as OCR, grounding, GUI, multi-image, and STEM datasets. We maintain the stage-2 data ratio to ensure cross-modal stability, using a text:vision:speech token ratio of 2:1:1, consuming 0.33 trillion tokens in total.

The effectiveness of the model is highly dependent on the quality and composition of the training data. Therefore, the domain mixture plays a crucial role in determining the final performance. Vision data, in particular, exhibits greater heterogeneity in content distribution, learning difficulty, and scale, thereby requiring more careful control over data composition.

We adopt a perplexity (PPL)-gap-based signal to automatically guide data sampling allocation. The corpus is first segmented into distinct subsets according to semantics and tasks. During training, we monitor PPL convergence for each subset: if a subset’s convergence lags behind the expected reference level, its sampling weight will be dynamically increased. 
For each subset, we build a corresponding validation set and use an off-the-shelf vision-language model to compute its per-sample PPL as the expected reference level~\citep{DBLP:conf/nips/Xie0DDLLLL0Y23, DBLP:conf/iclr/MichelHN22}.
To prevent high-value samples from being diluted by aggregate statistics, samples for which the PPL convergence is consistent with downstream performance will be isolated and relabeled into independent data subsets.
With the original data partitions largely preserved while the high-value samples organized more finely, this approach significantly enhances data efficiency and downstream task performance.

\subsubsection{Stage-4 Context-Length Extension Training} 

To augment the model's proficiency in capturing extended sequential relationships across various modalities, we gradually extend the context length to 32K and then 128K tokens, enabling the model to process and reason over more complex tasks such as long-term memory modeling and multi-turn real-time interaction.
We scale the context length from 8K to 32K tokens using 100B training tokens and adjust RoPE's base frequency~\citep{su2024roformer} from 1M to 5M to preserve positional encoding quality. The context length is then further expanded to 128K tokens with an additional 20B tokens of training, requiring a proportional increase of the RoPE's base frequency to 10M to maintain stable attention across the extended sequence length.

A key challenge in long-sequence multimodal modeling lies in preserving fine-grained visual information across multi-image compositions or extended video sequences. 
To address this, integrating our native resolution encoding strategy achieves dual optimization: it maintains critical visual details at their original fidelity while intelligently managing token allocation efficiency. This approach ensures the preservation of essential spatial relationships in images and temporal coherence in videos throughout extended sequences. Building upon the stage-3 vision-related data, we further augment the training mixture by incorporating 25\% additional long-context multimodal data, as detailed in Section~\ref{sec:pretrain_long_context_multimodal_data}. 

For textual and speech modalities, we utilize the foundation language model's high-quality text corpus~\citep{meituan2025longcat_flash_chat}, alongside carefully curated speech data from our multilingual audio repository (over 10 million hours of processed Chinese/English recordings). Notably, we maintain the same 2:1:1 text-to-vision-to-speech ratio as established in the pre-training phase, ensuring modality balance during context extension. This comprehensive approach ensures consistent performance across all modalities while scaling to extended contexts.

\subsubsection{Stage-5 Audio Encoder Alignment Training}
\label{subsec:stage-5}

In this stage, we maintain the LLM parameters in a frozen state while exclusively training the audio encoder. 
This approach serves dual objectives: (1) preserving the LLM's established multimodal processing capabilities, and (2) enhancing audio comprehension while aligning continuous speech inputs with the LLM's semantic space. 
We investigated the necessity of separate audio projector module pre-training for semantic space alignment, but empirical results showed negligible performance differences compared to direct end-to-end audio encoder training. 
To accelerate convergence, we initialize the audio encoder (excluding the projector) from the audio encoder trained with speech recognition introduced in Section~\ref{sec:audio}, while randomly initializing the projector module parameters.

During training, inputs are formatted as follows: ``Task Prompt + Speech Input + LLM Response'', where the loss is computed exclusively on the ``LLM Response'' segment. 
To improve task generalization, we design multiple prompt variations for each task and randomly select prompts during training.

\section{Post-Training} 
\label{sec4:post-train}

The post-training stage is a critical phase that transforms the pre-trained foundation model into a task-adaptive, human-aligned system with strong instruction-following, multimodal reasoning, and interactive capabilities. While the pre-training phase enhances multimodal perception and world knowledge, post-training focuses on refining response alignment, controllability, and fidelity through a combination of supervised and preference-based optimization.
This stage consists of two components: (1) Supervised Fine-Tuning (SFT), which equips the model with multimodal instruction-following, reasoning, and spoken interaction capabilities through high-quality and diverse instruction data, and (2) Reinforcement Learning (RL), which further enhances the model’s behavioral alignment, coherence, and consistency through Direct Preference Optimization (DPO)~\citep{rafailov2023direct}.
Together, these procedures ensure that the final model not only demonstrates robust omni-modal understanding and reasoning capabilities but also generates responses that are semantically accurate, perceptually natural, and contextually coherent across both offline and online interaction scenarios.

\subsection{Supervised Fine-Tuning}


The SFT stage focuses on two complementary objectives. First, it enhances the model’s multimodal instruction-following and reasoning capabilities by leveraging large-scale, high-quality, and diverse instruction data. Second, it strengthens the model’s multimodal interaction abilities by utilizing curated interaction data, including speech-to-speech conversational data and audio-visual interaction data.
In the following parts, we describe how we collect high-quality and diverse instruction data, then elaborate on the construction of multimodal interaction datasets, and finally present the SFT training recipes that unify and optimize these datasets.

\subsubsection{High-quality and Diverse Instruction Data Curation}
\label{sec:sft_data_curation}

\textbf{Image-Text SFT Data}
We curate a high-quality and diverse image-text SFT dataset, covering a wide spectrum of vision-language tasks, including: (1) Fundamental skills: image captioning and visual question answering, provided in both single-image and multi-image settings to capture diverse contextual dependencies. (2) Specialized skills: document and chart comprehension, OCR, visual grounding, agentic task execution, and STEM-related visual reasoning.

To ensure high data quality and strong alignment with human preferences, we employ an LLM-as-a-judge evaluation framework. In this pipeline, a strong multimodal language model evaluates candidate responses against reference answers and produces explanatory rationales. Samples labeled as inconsistent, low-quality, or semantically inaccurate are automatically filtered out, and we finally have approximately 3 million carefully curated datasets with high factual accuracy, stylistic consistency, and comprehensive task diversity.

\textbf{Video-Text SFT Data}
We construct a large-scale video data pool of approximately 3 million videos sourced from diverse proprietary and open-source datasets, covering tasks such as general video understanding, classification, reasoning, grounding, temporal localization, segmentation, and highlight detection. From each source, representative subsets are annotated with capability tags using multimodal language models, following a predefined taxonomy encompassing visual perception, temporal and causal reasoning, and domain-specific knowledge.

This taxonomy, consisting of 48 subcategories (e.g., entity attributes, counting, relationship comparison, OCR, anomaly detection, spatial relations, and camera motion), guides targeted sampling to enrich underrepresented capabilities and analyze coverage gaps. For capabilities with insufficient data, we augment annotations using proprietary expert models, and for particularly challenging tasks—such as action counting, relationship comparison, event localization, and attribute change—we incorporate manually annotated samples.

Through iterative sampling, augmentation, and quality refinement, we curate a high-quality video-text SFT dataset comprising approximately 700K samples, with balanced coverage of temporal reasoning, causal inference, and complex real-world visual understanding.

\textbf{Audio Understanding Data}
We re-utilize a subset of the comprehensive audio understanding dataset from the previous stage (Audio Encoder Alignment Training) for the SFT phase. This sampled data specifically targets tasks such as ASR, Audio-to-Speech Translation (AST), paralinguistic understanding, and audio-conditioned captioning and question answering. The purpose of integrating this kind of data is to improve the semantic alignment between the continuous audio representations generated by the audio encoder and the language model's semantic space.





\textbf{Vision-Speech QA Data} 
Fluent, speech-based question answering grounded in visual inputs represents a core capability of omni-modal models.     We designed a new vision-speech QA dataset to enhance this ability. It pairs visual inputs (images or videos) with spoken prompts and requires the model to output its answers in speech. To create this data, we sourced text-based QA pairs from existing SFT datasets, selecting only those suitable for TTS synthesis. The retained samples are then rewritten using an LLM to enhance fluency, naturalness, and stylistic consistency in spoken form. Finally, all text responses are converted into high-fidelity speech using a TTS engine.

\textbf{Audio-Visual Understanding Data} 
The joint audio-visual understanding capability fundamentally distinguishes an omni model from conventional vision-only or speech-only multimodal language models. To develop this capability, we curate an in-house time-synchronized audio-visual dataset by prompting a strong multimodal language model to generate high-quality text QA pairs that are involved with both audio and visual content in the video. 
Specifically, the videos are sourced from multiple domains, including multimodal data containing rich music and audio event contents.
All samples are organized in an interleaved chunk format, which enables the model to better perceive and reason over temporally aligned audio-visual content, thereby strengthening its capacity for multimodal understanding.

\subsubsection{Multimodal Interaction Data Curation}
 \noindent
To equip \longcat with real-time spoken and audio-visual interaction capabilities, we construct two types of specialized interaction datasets: (1) Speech-to-Speech Interaction Data, designed to enhance the naturalness and expressiveness of voice-based dialogues, and (2) Audio-Visual Interaction Data, aimed at improving the model’s ability to manage multi-turn, context-dependent audio-visual conversations.

\textbf{Speech-to-Speech Interaction Data} We construct a large-scale voice dialogue dataset using a two-stage approach. (1) Text-based dataset adaptation: we filter out contents like formulas, code, markdown and rewrite responses with an LLM to produce speech-friendly conversational language. (2) Voice-oriented dialogue generation: we prompt an LLM to generate diverse topics and create new multi-turn dialogues.
To enhance expressiveness and naturalness, professional voice actors record dialogues across a range of emotions, speaking styles, and major Chinese dialects (e.g., Sichuan, Beijing, Northeast). These recordings are then used to fine-tune a specialized TTS engine, ensuring consistent tone, high fidelity, and natural prosody. All dialogues are ultimately synthesized into high-quality speech using this engine.

\textbf{Audio-Visual Interaction Data}
Audio-visual speech interaction data plays a crucial role in endowing the omni model with the ability to simultaneously process audio and vision information and perform real-time voice response. Unlike offline video understanding or general audio-visual understanding, online interaction is characterized by bi-directional and multi-turn dialogues, which involve immediate feedback, dynamic barge-in, contextual memorization, logical progression, and co-reference resolution. However, collecting large-scale real-world audio-visual interaction data is resource-intensive and impractical, and synthesizing complex and reasonable interaction data is difficult for existing models that tend to generate hallucination. To address these challenges, we develop a semi-automated data production pipeline that leverages model-driven automation for initial generation, followed by a human-in-the-loop stage for verification and refinement:

\begin{itemize}
    \item \textbf{Model-Driven Automation} 
To comprehensively cover diverse interaction scenarios, we first establish an ability taxonomy encompassing six major dimensions: memorization, understanding, analysis, creation, application, and entertainment. Guided by this taxonomy, we collect a mixture of public and in-house videos as the source data to ensure balanced coverage across abilities. For each video, we first perform scene segmentation using PySceneDetect \citep{pyscenedetect} to obtain a sequence of clips. For every clip, a powerful multimodal language model generates multiple rounds of progressively deep and context-aware QA pairs. In each round, subsequent queries build upon preceding conversations with logical progression, referential dependency, or anaphoric linkage to earlier turns. Such conversations encourage the model to reason over extended dialogue contexts, maintain entity and reference consistency in a natural, human-like audio-visual interaction setting. We then apply an automatic verification pipeline, employing an LLM-as-a-judge framework to evaluate and discard low-quality or inconsistent QA pairs. The remaining qualified samples are then composed into multi-turn dialogue, enabling the model to learn both intra-scene conversational continuity and inter-scene contextual transitions, thereby better reflecting real-world audio-visual interactions.

To further enhance long-term memory and contextual reasoning in extended dialogue settings, we construct a long-context multi-turn video interaction dataset. In this dataset, dialogue sequences are reordered so that certain queries appear at temporal positions distant from their corresponding visual segments, encouraging the model to retain and retrieve information over extended temporal spans. After reordering, all QA sequences are refined with a strong multimodal language model to resolve anaphoric references, correct temporal inconsistencies, and improve contextual coherence, resulting in natural and logically consistent multi-turn conversations.

\item \textbf{Human-in-the-loop} 
Through manual verification, we observe that the generated QA data frequently exhibits several types of errors: factual inconsistency, response insufficiency, referential ambiguity, linguistic infelicity, and semantic irrelevance.
It is difficult to reuse a multimodal LLM to automatically correct these errors, as it may introduce new errors. Therefore, we employ human annotation for quality refinement.
For factual inconsistency, annotators are required to edit or replace any statement that contradicts the video/audio content. For response insufficiency, we ask annotators to augment the original answer with the missing information, ensuring all sub-questions are answered. For referential ambiguity, we ask annotators to replace ambiguous or incorrect pronouns with specific nouns or corrected pronouns. As a result, we obtain a high-quality interaction dataset that contains meaningful and natural conversations that align with verifiable truth.

Finally, we apply a TTS engine to convert the textual QA pairs into speech and integrate them into the videos to obtain the final audio-visual speech interaction dataset. The resulting data aligns the model’s response style with that of a helpful, interactive assistant and enhances the naturalness of its conversational behavior.

\end{itemize}

\subsubsection{SFT Training Recipe}

During SFT, we freeze the audio encoder while updating all other modules. The frozen audio encoder retains robust acoustic representations learned from large-scale audio pre-training (i.e., pre-training stage-5 as introduced in Section~\ref{subsec:stage-5}), whereas the unfrozen layers allow the model to learn fine-grained multimodal alignment and instruction-following behavior. Empirically, we find this selective fine-tuning strategy stabilizes convergence and avoids catastrophic forgetting of low-level auditory features. Apart from multi-modal SFT data, we also incorporate pure text SFT data used in developing LongCat-Flash, which covers domains ranging from reasoning, mathematics, coding, tool use, and general-purpose dialogue.

We adopt the AdamW optimizer with $\beta_1=0.9$, $\beta_2=0.95$, and a weight decay of 0.1. The learning rate follows a linear warm-up over the first 4\% of steps, followed by cosine decay from a peak value of $1\times10^{-5}$ to zero. The SFT stage consists of a single training phase conducted for one epoch. The training is performed with a batch size of 1024.

\subsection{Reinforcement Learning} 
To further enhance the model’s multimodal capabilities and improve human alignment, we employ the Direct Preference Optimization (DPO) \citep{rafailov2023direct} during the reinforcement learning stage. The \longcat model supports multimodal inputs and enables parallel streaming outputs of both text and speech. However, most existing DPO variants are designed for text-only outputs or optimize text and speech separately. We argue that such decoupled optimization is suboptimal for maintaining coherence between textual and speech responses.

To address this limitation, we extend the DPO to jointly optimize text and speech outputs to improve both alignment and stability across modalities. Specifically, since the \longcat includes one text head and multiple audio heads, we modify the DPO objective to optimize all heads simultaneously. The overall loss is defined as:


\[
\mathcal{L}_{\text{DPO}} = 
\alpha \, \mathcal{L}_{\text{DPO}} (\text{text}_{chosen},\text{text}_{rejected}) 
+ \beta \sum_{i=1}^{N} \mathcal{L}_{\text{DPO}}(\text{audio}^{i}_{chosen},\text{audio}^{i}_{rejected})
\]

where $N$ denotes the number of audio heads. $\mathcal{L}_{\text{DPO}} (\text{text}_{chosen},\text{text}_{rejected})$ focuses on the semantic quality of the text head, while each audio head $\mathcal{L}_{\text{DPO}}(\text{audio}^{i}_{chosen},\text{audio}^{i}_{rejected})$ emphasizes both the linguistic and pronunciation stability of the corresponding speech output. This joint optimization strategy leads to more coherent and natural multimodal responses, further enhancing the expressiveness and human-aligned communication capabilities.

\subsubsection{Data Construction} 
Our DPO training data consists of two components: general DPO data and model-generated DPO data. The general data covers samples focusing on safety, helpfulness, and response style. The model-generated data is sampled from an SFT checkpoint and used to refine broader multimodal capabilities through preference comparisons among model-produced responses. 
To comprehensively enhance the capability, alignment, and safety of \longcat across all modalities, we utilize all prompt types from the SFT datasets described in Section~\ref{sec:sft_data_curation} during DPO training. 
For each prompt, the model generates 6 rollouts to ensure diverse candidate responses for preference pair construction. 
To ensure data quality, we employ a hybrid evaluation strategy that combines manual annotation with automatic scoring from a robust multimodal language model, serving as a preference evaluator to assign quality scores and identify distinct and informative preference pairs. This process provides reliable supervisory signals for DPO training, enabling effective refinement of \longcat’s multimodal alignment, behavioral stability, and overall coherence.

\subsubsection{Training Details}
We train for one epoch with a batch size of 256, sampling each batch to achieve a well-balanced mix of modalities. The learning rate follows a cosine decay schedule, with a warm-up fraction of 0.03, gradually decreasing from $1\times10^{-6}$ to 0. To balance preference learning between the text head and the speech head, we set their loss weight ratio by setting $\alpha : \beta = 1 : 1$. To mitigate drift from the SFT model, we incorporate a Kullback-Leibler (KL) divergence regularizer with a weighting factor of 0.1.

\section{Training Infrastructures} 
\label{sec6:infra}
Our core design principles are largely inspired by the training infrastructure used in developing LongCat-Flash, with a particular emphasis on maximizing training efficiency while strictly ensuring numerical consistency. To guarantee numerical consistency, we enforce determinism, minimize errors, and maintain error interpretability, ensuring that every training run is both deterministic and reproducible. For efficiency, we decouple the components of the LLM, vision encoder, and audio encoder, enabling independent optimization of their performance and memory usage. Experimental results demonstrate the effectiveness of our approach: in multimodal settings, our system maintains over 90\% of the throughput of text-only training.

\subsection{Multimodal Decoupling Framework}

\begin{figure}[t]
  \centering
  \begin{minipage}[c]{0.58\textwidth}
    \centering
    \captionof{table}{Computation distribution per micro-batch across different modalities during the SFT stage.}
    \begin{tabular}{l|cccc}
      \toprule
      \multirow{2}{*}{Module} & \multicolumn{4}{c}{Computational Cost (TFLOPs)} \\
      & min & max & mean & std \\
      \midrule
      Audio Encoder & 0.01 & 109.96 & 3.29 & 7.94 \\
      Vision Encoder & 0.08 & 400.37 & 89.85 & 61.02 \\
      LLM Decoder & 1920.57 & 4667.74 & 3531.32 & 1111.11 \\
      \bottomrule
    \end{tabular}
    \label{tab:modal_cost_dit}
  \end{minipage}
  \hfill%
  \begin{minipage}[c]{0.38\textwidth}
    \centering
    \includegraphics[width=\textwidth]{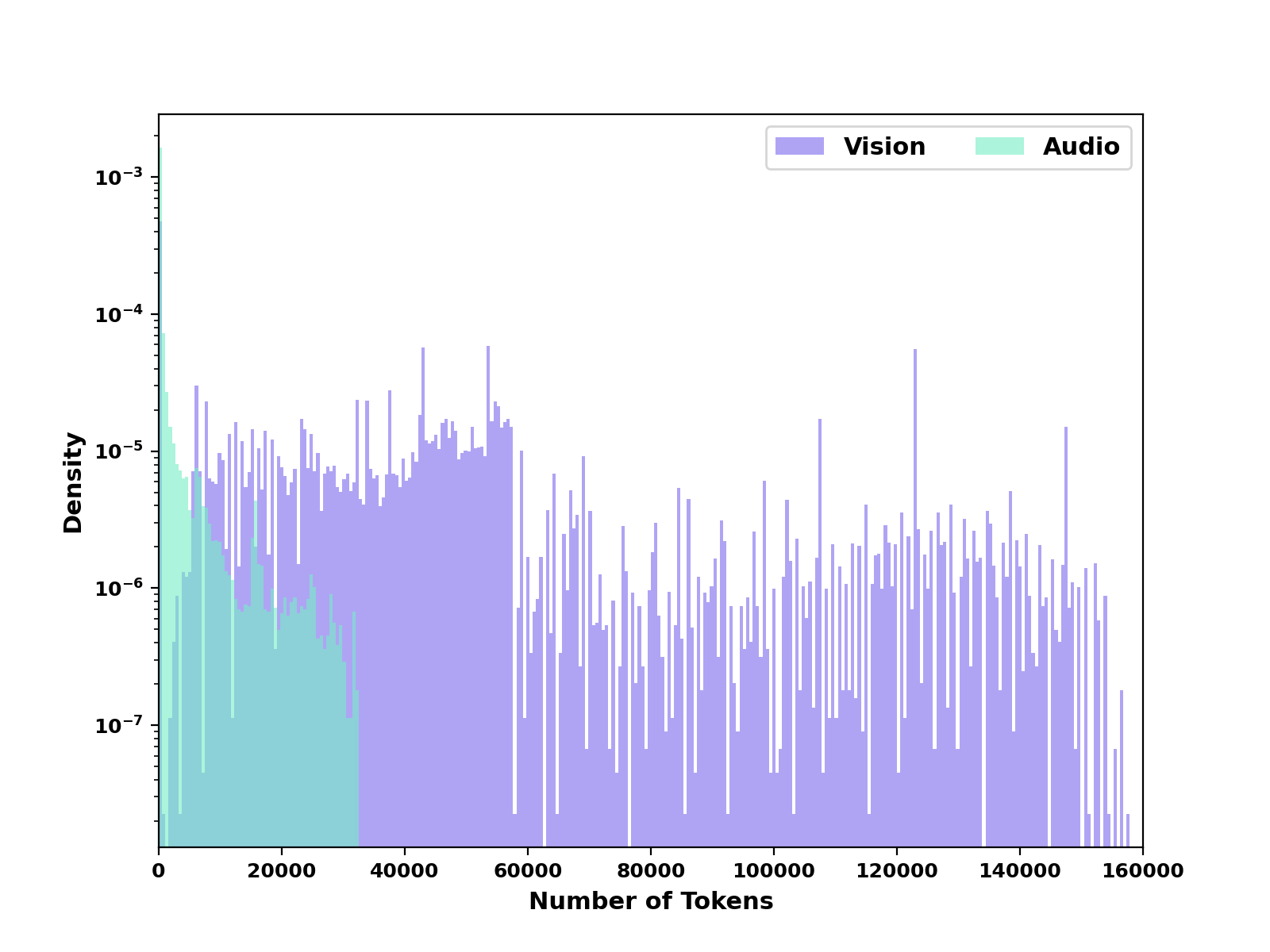}
    \caption{The distribution of sequence token lengths across different modalities during the SFT stage.}
    \label{fig:data_distrubtion}
  \end{minipage}
\end{figure}

In multimodal scenarios, optimizing training performance is challenging due to the heterogeneity of both data and models. The challenge of data heterogeneity arises because the training data for \longcat includes speech, vision, and text, which show significant and dynamic differences in token distribution (Figure~\ref{fig:data_distrubtion}). 
Model heterogeneity is evident in the substantially different computational workloads across the three core components of \longcat: the vision encoder, the audio encoder, and the LLM decoder (Table~\ref{tab:modal_cost_dit}).
To address the aforementioned challenges, there are two primary optimization approaches. The first relies on Fully Sharded Data Parallelism (FSDP) (e.g., OrchMLLM~\citep{zheng2025orchestrate}, veOmni~\citep{ma2025veomni}), which reduces static memory consumption through parameter sharding, avoids pipeline parallelism (PP) bubbles caused by model heterogeneity, and mitigates data heterogeneity via data-parallel (DP) group balancing. However, for \longcat, the total number of model parameters is too large for this approach to be practical.

The second approach is to decouple multimodal models across different parallel dimensions. For example, DistTrain~\citep{zhang2025disttrain} mitigates model and data heterogeneity by separating model partitioning strategies and sorting data; PipeWeaver~\citep{xue2025pipeweaver} reduces PP bubbles caused by heterogeneity through fine-grained data partitioning and dynamic pipeline scheduling; and Optimus~\citep{feng2025optimus} separates the parallelization strategies for encoders and LLMs, scheduling encoder computations into the LLM’s idle periods to eliminate bubbles caused by model heterogeneity.
Building on the Optimus approach, we develop modality-decoupled parallelism (MDP), a simple yet effective multimodal training strategy. The core idea is to completely decouple the modality encoders and the LLM backbone at the distributed level, enabling independent scheduling and more efficient utilization of computational resources.

\subsubsection{Modality-Decoupled Parallelism}

\begin{figure}[t]
    \centering
    \includegraphics[width=\linewidth]{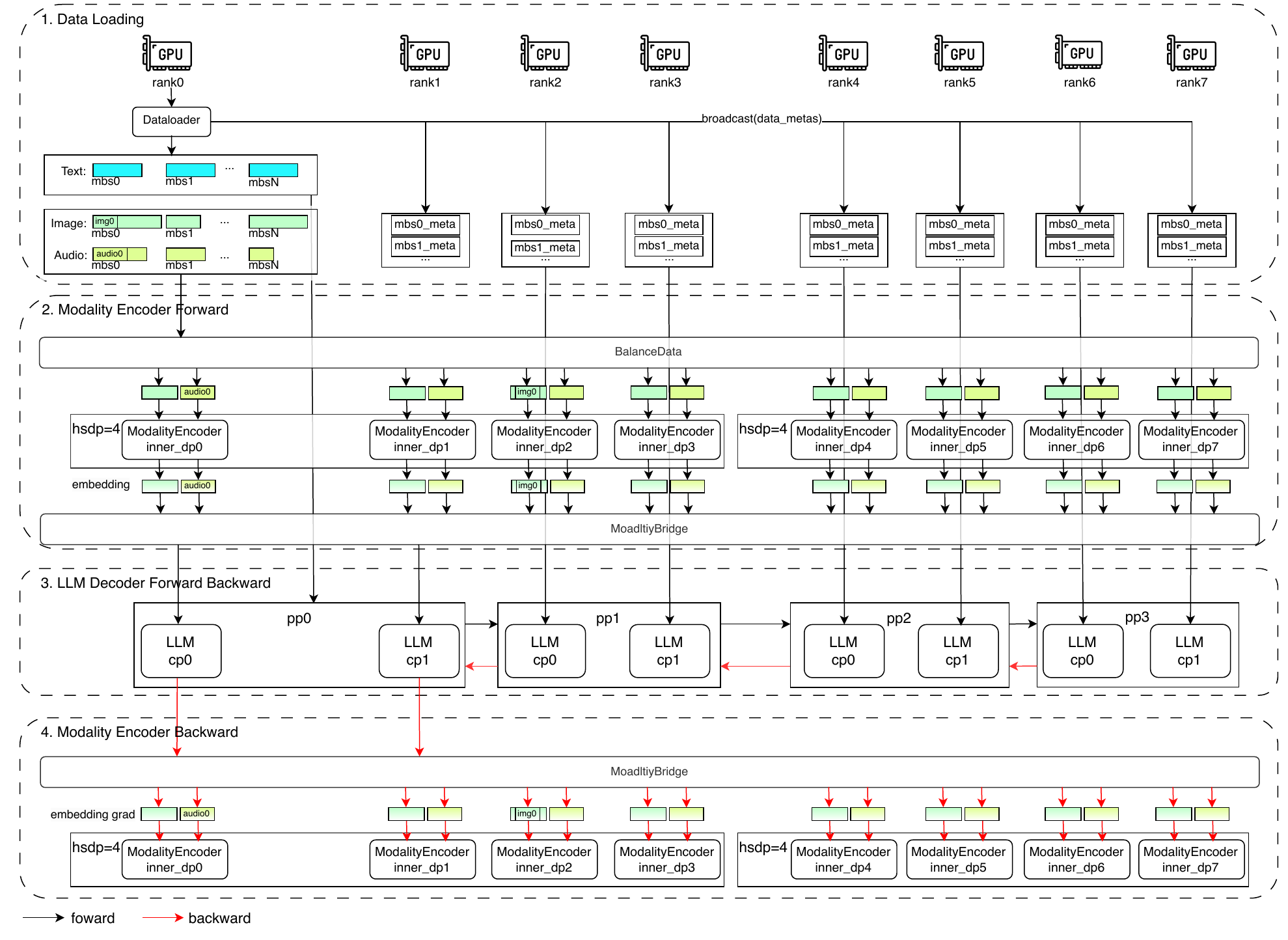}
    \caption{\textbf{Overview of modality-decoupled parallelism (MDP).} The modality encoders and the LLM backbone are fully decoupled at the distributed level, enabling independent scheduling and improved computational efficiency.}
    \label{fig:MDP}
\end{figure}

In our implementation, we co-locate the modality encoders and the LLM decoder. The modality encoders utilize Hybrid Sharding Data Parallelism (HSDP)\citep{zhao2023pytorchfsdpexperiencesscaling} to reduce static memory, and full activation recomputation is employed to reduce activation memory usage. The LLM decoder adopts a combined distributed strategy including pipeline parallelism (PP), ZeRO-1 data parallelism (DP), context parallelism (CP), and expert parallelism (EP). 
To simplify data mapping between modality encoders and the LLM decoder, we introduce an \textit{InnerDP} parallelism strategy, which further partitions modality data across all microbatches.
The DP rank of modality encoders corresponds one-to-one with the DP rank of the LLM decoder, and the size of the InnerDP dimension is the product of the PP and CP of the LLM decoder (i.e., $d_{inner\_dp}=d_{lm\_cp} \times d_{lm\_pp}$, $d_{world\_size}=d_{dp} \times d_{inner\_dp}$). As shown in Figure~\ref{fig:MDP}, the MDP execution timeline consists of four phases:

\textbf{Data Loading}: At the start of each training iteration, the rank with $inner\_dp=0$ fetches all micro-batches. Then it broadcasts the metadata of the data to the other ranks within the DP group. This mitigates I/O overhead and memory pressure. In this phase, all micro-batches are sorted by text data sequence length to ensure workload across DP groups remain similar, thereby reducing EP bubbles caused by workload imbalances.

\textbf{Modality Encoder Forward}: The BalanceData module first distributes the modality data from the $inner\_dp=0$ rank to the other $inner\_dp$ ranks. Then, the modality encoders on each rank compute the corresponding vision and audio embeddings. Finally, the ModalityBridge module aggregates these embeddings at the $inner\_dp=0$ rank, which are then passed as input to the LLM decoder.

\textbf{LLM Decoder Forward and Backward}: In this phase, modality embeddings are partitioned on the CP rank and fed into the LLM decoder for its forward and backward passes. The gradients of the vision and audio embeddings are then returned to the backward phase for the modality encoders. This design ensures the training efficiency of the LLM decoder, and also enables isolated optimization for modality encoders.

\textbf{Modality Encoder Backward}: The ModalityBridge module redistributes the modality embedding gradients from the LLM decoder to the other $inner\_dp$ ranks, and then the backward pass of the modality encoders is executed.

\subsubsection{ModalityBridge}

\begin{figure}
    \centering
    \includegraphics[width=1.0\linewidth]{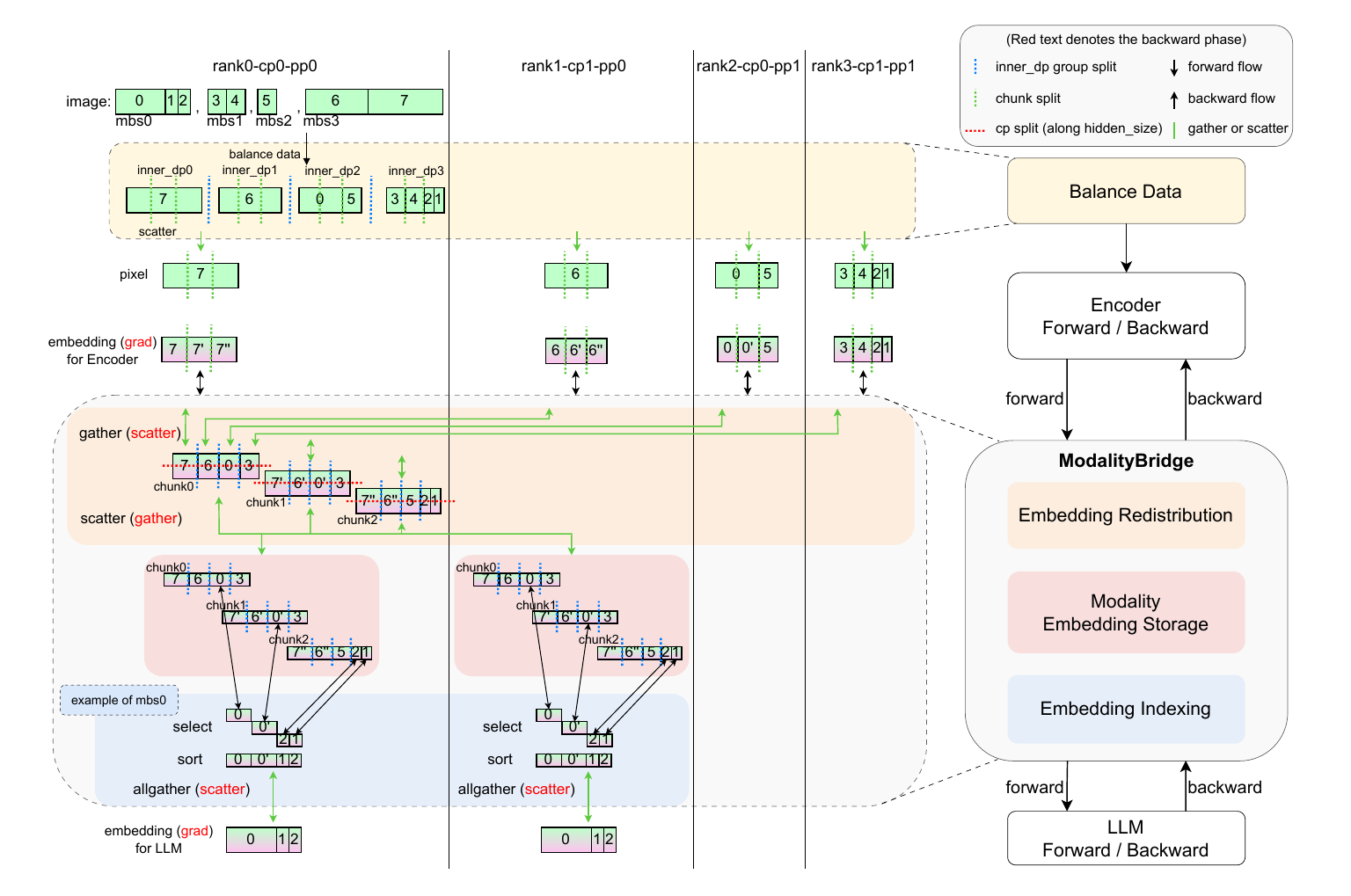}
    \caption{\textbf{Overview of the chunk-based ModalityBridge}. The example configuration with num\_chunk = 3, processing 4 microbatches of 8 images each across 4 GPUs (DP = 1, CP = 2, PP = 2)}
    \label{fig:chunk_balance}
\end{figure}

In MDP, the ModalityBridge serves as the communication layer between the multimodal encoders and the LLM decoder. 
It is responsible for transforming data organization formats to resolve discrepancies arising from the different parallelism strategies employed by the modality encoders and the LLM decoder. This process specifically involves handling modality embeddings during the forward phase and gradients during the backward phase.
However, when processing longer context length, significant memory pressure emerges as the $inner\_dp=0$ rank must read all micro-batches and perform gather and scatter operations on the modality embeddings and their gradients.
To address this challenge, we adopt chunk-based processing within ModalityBridge, which effectively alleviates the memory bottleneck while maintaining bitwise numerical consistency.
As illustrated in Figure~\ref{fig:chunk_balance}, this module comprises three core components:

\textbf{Embedding Redistribution}:
Employs a two-stage chunking approach for data gather and scatter, decomposing the complete data transformation into \(num\_chunk\) iterations. Each iteration consists of two stages:  
\begin{enumerate}
    \item \textbf{Aggregation:} Inner-DP embeddings are aggregated to the \(inner\_dp = 0\) rank.
    \item \textbf{CP partitioning:} The aggregated data is split along the \(hidden\_size\) dimension to ensure uniform embedding distribution and balanced memory usage, followed by scattering to the respective CP ranks to further reduce per-device memory consumption.
\end{enumerate}

During each chunking iteration, the overall memory footprint exhibits a rise-then-fall pattern, significantly lowering the peak memory usage.



\textbf{Modality Embedding Storage}: Stores all gathered-then-scattered chunk data while maintaining global offset information for subsequent indexing.

\textbf{Embedding Indexing}: Provides micro-batch-level embedding retrieval and gradient backpropagation during the LLM forward and backward phases.

Based on these three components, we accomplish the transformation of data formats across different stages. In the forward phase, the transformation is performed according to the two-stage chunking approach, while in the backward phase, \textit{Embedding Redistribution} executes the reverse process: gradients are first gathered via CP gather and then scattered to the corresponding \(inner\_dp\) ranks according to the global offset information for backward computation. Meanwhile, the chunking scheme reduces the peak memory usage to \(1/num\_chunk\) of the original, significantly alleviating memory pressure while ensuring bitwise numerical alignment.

\subsection{Performance Tuning}
Our overarching strategy is twofold: first, select a distributed configuration by profiling core operator efficiency; second, meet the configuration's memory budget via targeted memory optimizations. On the communication side, the shortcut architecture enables overlap between EP communication and computation within each micro-batch. In CPU-bound regimes, we employ fused operators to reduce kernel launch overhead and improve end-to-end throughput.
\subsubsection{Optimal Distributed Configuration}
\begin{figure}
    \centering
    \includegraphics[width=0.8\linewidth]{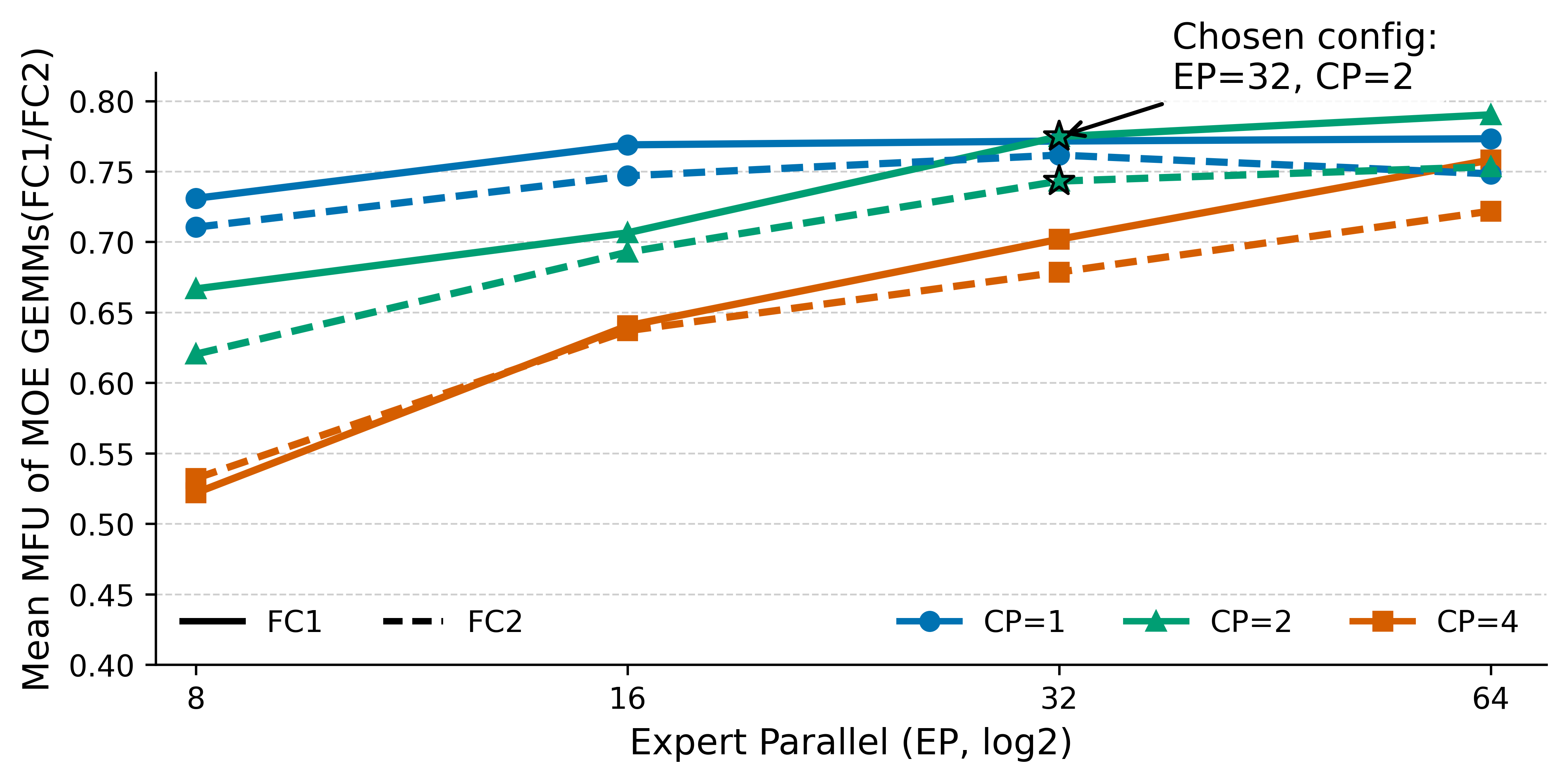}
    \caption{Benchmark results of MoE GEMMs under different CP/EP configurations with an 8K sequence length.}
    \label{fig:mfu-ops}
\end{figure}

In large-scale LLM pretraining, the efficiency of core operators is primarily constrained by the sequence length; increasing sequence length typically improves Model FLOPs Utilization (MFU). Accordingly, our system design increases the effective sequence length per EP rank to approach the operators' high-efficiency regime. Hardware-specific benchmarking results further indicate that reducing CP substantially improves the efficiency of the core operator, as shown in Figure~\ref{fig:mfu-ops}.
\subsubsection{Communication Optimizations}
Our training infra contains multiple compute-communication overlaps, mainly including shortcut-based EP overlap and point-to-point (P2P) overlap. We tune the number of streaming multiprocessors (SMs) assigned to communication and compute kernels for each scenario to maximize hardware utilization, and we use distinct P2P groups/streams so that inter-stage pipeline-parallel (PP) communication does not interfere across stages.

\subsubsection{Kernel Optimizations}
To optimize the training efficiency of MoE models, we maintain computational determinism while implementing kernel optimization and fusion along critical computational paths. Our technical contributions encompass the following areas:

\textbf{Optimized Grouped GEMM} 
Our optimization strategies for Grouped GEMM achieve approximately 75\% MFU in the evaluated scenarios. 
(1) \textbf{Dynamic SwapAB:} The WGMMA instruction is evaluated with the $N$ dimension varying from 8 to 256, while the $M$ dimension is fixed at 64. 
We design an optimized kernel that automatically performs SwapAB operations to maximize WGMMA performance. 
(2) \textbf{Configurable SM Usage:} The Grouped GEMM kernel allows configurable SM counts to avoid resource contention when overlapping with dispatch or combined communication operations. 
(3) \textbf{Fine-Tuning:} \textit{TileShape}, \textit{PipelineStage}, and \textit{ClusterShape} are carefully tuned for the target workload to maximize hardware utilization and prevent register spilling. 
(4) \textbf{Scheduling and Pipeline Strategy:} A swizzled block mapping schedule is adopted to significantly improve L2 cache hit rates. 
In addition, a ping-pong mechanism is introduced during the backward pass to overlap WGMMA computation with the loading and storing of weight gradients.

\textbf{Fused Permute} 
An efficient fused permute kernel is employed to rearrange tokens for expert alignment, while also integrating several critical functionalities, including metadata computation for the backward pass and token dropping within the CP group.

\textbf{Fused RoPE } 
We fuse RoPE into the MLA prologue kernel, eliminating the overhead of intermediate data writing and reloading, achieving a $3\times$ speedup compared to the baseline.

\textbf{Deterministic FA} 
We employ a semaphore-based synchronization scheme that ensures deterministic QK reduction in the backward phase and outperforms the default deterministic implementation, which requires an additional temporary workspace. 
This approach achieves approximately $0.8\times$ the performance of the non-deterministic version.

\subsection{Memory Optimization Strategies}
Without optimization, the required memory for our target configuration is approximately 137~GB per device (Table~\ref{tab:mem-usage-naive}). 
Given 80~GB devices and accounting for EP imbalance peaks, we constrain the theoretical memory footprint to approximately 72~GB.

\begin{table}[htbp]
  \centering
    \caption{Naïve memory usage breakdown by component.}
    \label{tab:mem-usage-naive}
    \begin{tabular}{@{}l|rr@{}}
      \toprule
      \textbf{Component} & \textbf{Static (ZeRO-1, GB)} & \textbf{Dynamic (GB)} \\
      \midrule
      LLM                & 16.8  & 103.24 \\
      NCCL               & --    & 12.93  \\
      DeepEP             & --    & 0.45  \\
      Modality Encoders  & 3.6   & --     \\
      \midrule
      \textbf{Total} & \multicolumn{2}{c}{\textbf{137.02\,GB} (peak at non-pp0 stage)} \\
      \bottomrule
    \end{tabular}
\end{table}

We combine the techniques in Table~\ref{tab:mem-opts}: (i) a V-shaped PP schedule to bound concurrently alive activation micro-batches \citep{qi2024pipelineparallelismcontrollablememory}; (ii) selective recomputation for low-FLOP, high-activation operators such as SwiGLU and LayerNorm \citep{deepseekai2025deepseekv3technicalreport}; (iii) memory-efficient permute, which moves aggregation of routing probabilities and hidden states into SwiGLU to reduce MoE unpermute memory \citep{megatron-lm}; (iv) fine-grained SM budgeting for NCCL communicators to reduce NCCL memory; and (v) Hybrid-Sharding Data Parallel (HSDP) for the modality encoder to further reduce its static footprint \citep{paszke2019pytorch}.
To accommodate potential EP load imbalance, we implement \emph{dynamic expert recomputation}: when any rank is assigned too many tokens, we dynamically recompute its down-projection, freeing all MoE memory except the down-projection inputs on trigger. This prevents training crashes from expert imbalance while incurring only modest overhead.

\begin{table}[htbp]
  \centering
  \caption{Memory footprint under different optimization settings.}
  \label{tab:mem-opts}
  \begin{tabular}{@{}l|cccc@{}}
    \toprule
    \textbf{Setting} &
    \begin{tabular}[c]{@{}c@{}}\textbf{Static}\end{tabular} &
    \textbf{Dynamic(GB)} &
    \textbf{Others(GB)} &
    \textbf{Total(GB)} \\
    \midrule
    Baseline                         & 20.4  & 103.2 & 13.4 & 137.0 \\
    +HSDP for Modality Encoder       & 17.5  & 103.2 & 13.4 & 134.1  \\
    +Vhalf                           & 17.5  & 58.6  & 13.4 & 89.4   \\
    +Select Recompute                & 17.5  & 49.2  & 13.4 & 80.1   \\
    +Memory Efficient Permute        & 17.5  & 42.8  & 13.4 & 73.6   \\
    +NCCL Memory Optimization        & 17.5  & 42.8  & 8.9  & \textbf{69.1}   \\
    \bottomrule
  \end{tabular}
\end{table}

\subsection{Numerical Consistency}
Large language model pretraining is extremely resource-intensive, often requiring weeks and incurring millions in computational costs. Ensuring the correctness of the training framework is essential to avoid costly retraining due to implementation errors. 

In our framework, we enforce deterministic implementations for all computation, communication, and data flow, ensuring that every experiment is reproducible and maintains bitwise-aligned loss values.

For new features, we prioritize bit-aligned implementations to ensure numerical consistency. For example, features like \verb|fused_permute| and \verb|fused_swiglu| ensure bitwise-aligned loss whether they are enabled or not. 
Similarly, kernels like Grouped GEMM and Router GEMM employ a non Split-K implementation to preserve alignment across varying batch sizes or distributed settings.

For features where bit alignment is not feasible, we systematically analyze all sources of numerical deviation and mitigate their impact by benchmarking against the gold reference implementation during verification. For example, we identified and aligned the accumulation order in DeepEP and All2All EP strategies, thereby achieving bit-level consistency in loss values and verifying implementation correctness.

\section{Inference and Deployment} 
\label{sec7:infer}

\subsection{Decoupled Framework}
We propose a decoupled multimodal inference framework that separates modality-specific encoders/decoders and the LLM for optimized deployment. Each module is deployed on dedicated hardware and accelerators tailored to its computational characteristics, mitigating cross-modal resource contention. The optimizations of LLM deployment follow LongCat-Flash, including Prefill-Decode (PD) Disaggregation and Single Batch Overlap for ScMoE to improve inference efficiency. 
Compared with conventional hybrid deployment, this separation achieves lower latency and higher throughput despite minor communication overhead.

\subsection{Asynchronous Streaming Pipeline}
\begin{figure}[t]
    \centering
    \includegraphics[width=1.0\linewidth]{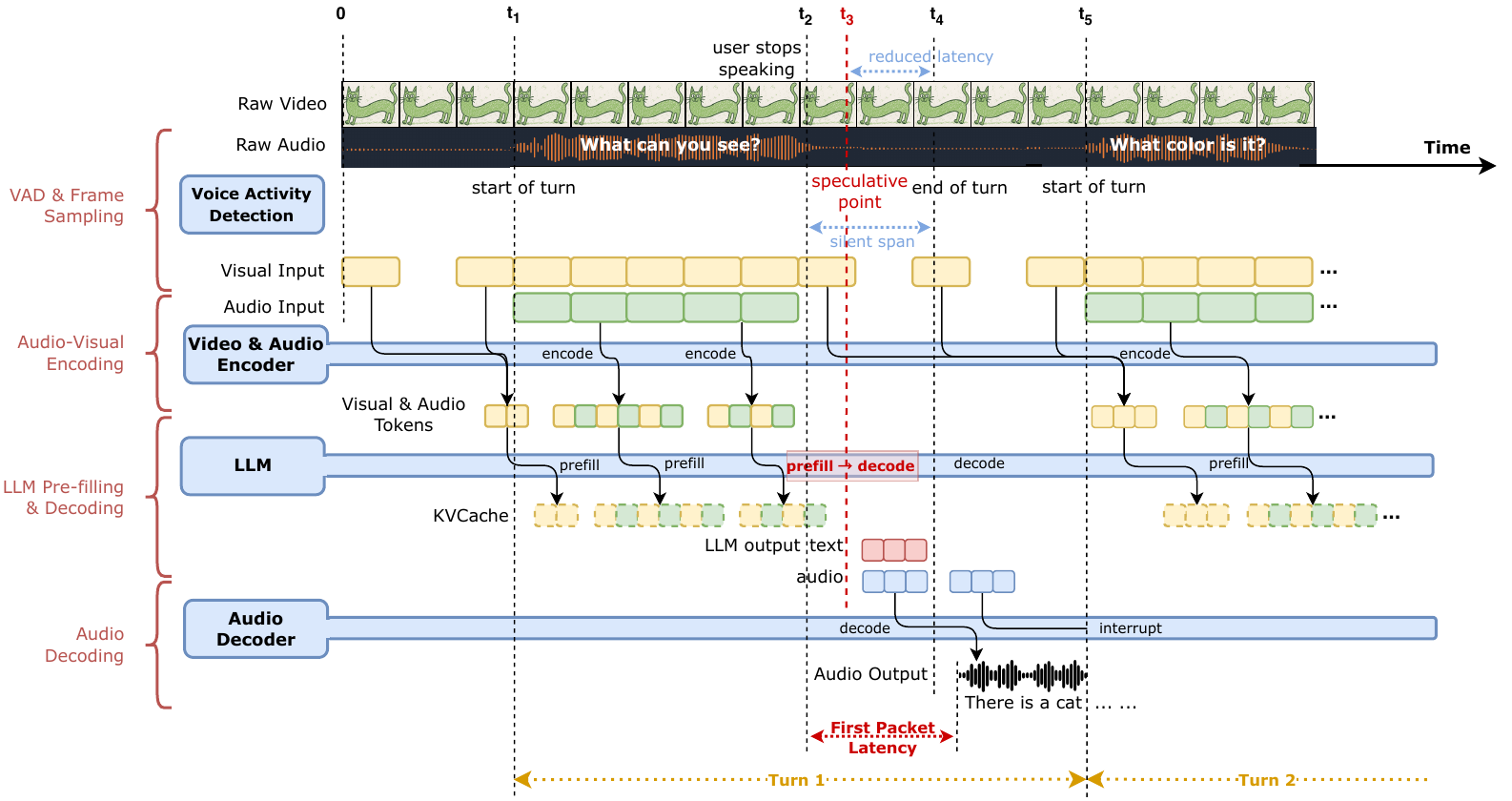}
    \caption{The asynchronous streaming pipeline for minimizing first packet latency.}
    \label{fig:streaming_pipeline}
\end{figure}

We design a highly efficient asynchronous streaming pipeline for model serving. As shown in Figure~\ref{fig:streaming_pipeline}, it consists of four sequentially linked and concurrently executed stages: VAD \& Frame Sampling, Audio-Visual Encoding, LLM Prefilling \& Decoding, and Audio Decoding. Each module supports incremental inference of streaming input and adaptive batching strategies, facilitating concurrent scheduling to reduce latency.

\textbf{Sparse-Dense Sampling Strategy} A voice activity detection (VAD) module is deployed to detect whether the user is speaking in real-time. When the user is speaking, densely sampled frames as well as the audio will be used as model input, facilitating a deeper analysis of the video content. Otherwise, only sparsely sampled frames are used to track the high-level overview of the video stream.

\textbf{Speculative Prefill-Decode Switching} Typically, the VAD model determines the end of a user turn by assessing whether the silence is longer than a predefined duration, referred to as the \textit{silent span} $[t_2, t_4]$. However, deferring the LLM state transition from prefill to decode until $t_4$ would incur substantial latency to the first packet. To overlap this latency with the latency of LLM decoding, we adopt a speculative prefill-decode switching strategy: The LLM starts decoding at an earlier speculative point $t_3$, reducing the latency from $t_3$ to $t_4$. If the user resumes speaking subsequently, a \textit{rollback} is triggered, wherein the LLM's generated content is discarded and the system reverts to the prefill state.

\textbf{Audio Delivery and Interruption} The audio will be delivered to the user once the VAD model explicitly detects the end of a turn, which is marked as $t_4$ in the figure, even if the first packet of audio response is generated earlier. However, should a user interrupt the audio generation at $t_5$, the process is immediately terminated. Consequently, the token sequence from the LLM is truncated at the nearest natural breakpoint, such as the most recent punctuation mark.


In our implementation, each request data packet from the user side consists of 1 second of audio and 2 corresponding video frames. Using the streaming prefill strategy, each request packet can be immediately prefilled, eliminating the need to wait until the user's turn to finish before initiating computation. By overlapping the VAD endpoint detection ($600\sim700 ms$) and the prefill process, users can receive the model response within $100ms$ after the endpoint detection. This asynchronous pipeline enables real-time omni-modal interaction.

\section{Evaluation} 
\label{sec5:eval}

In this section, we conduct comprehensive evaluations of \longcat, compare it with various proprietary and open-source multimodal models, including vision understanding, audio comprehension, text understanding and generation, cross-modality understanding, and audio-visual interaction.

\subsection{Vision Capability Evaluation}

\begin{table}[htbp]
    \caption{Evaluation of image understanding. Values marked with * are sourced from public reports. As GPT-4o does not support image grounding, we do not report its results on RefCOCO and ScreenSpot-v2.}
    \centering
    \resizebox{1.0\textwidth}{!}{
        \begin{tabular}{l|cccccc|cl}
            \toprule
            Benchmark & \begin{tabular}[c]{@{}c@{}}\textbf{LongCat-Flash-Omni} \\ Instruct\end{tabular} & \begin{tabular}[c]{@{}c@{}}Gemini-2.5-Pro \\ {\small ThinkingBudget128}\end{tabular} & \begin{tabular}[c]{@{}c@{}}Gemini-2.5-Flash \\ {\small non-thinking}\end{tabular} & \begin{tabular}[c]{@{}c@{}}Qwen3-Omni \\ Instruct\end{tabular} & Seed-1.6 & \begin{tabular}[c]{@{}c@{}}GPT-4o \\ 1120\end{tabular} & \begin{tabular}[c]{@{}c@{}}Qwen3-VL \\ {\small 235B-A22B-Instruct}\end{tabular} & \begin{tabular}[c]{@{}c@{}}Qwen2.5-VL \\ {\small 72B-Instruct}\end{tabular} \\ 
            \midrule
            \multicolumn{9}{c}{\textbf{General}} \\
            \midrule
            MMBench-EN$_{\text{test}}$ &87.5 & 89.8 & 89.3 & 86.8 & 88.5 & 83.7 &88.3 & 88.6* \\
            MMBench-ZH$_{\text{test}}$ &88.7 & 89.2 & 88.5 & 86.4 & 83.8  & 82.8 &89.8 & 87.9* \\
            RealWorldQA                & 74.8 & 76.0 & 73.9 & 72.9 & 74.5 & 74.1 &79.3* & 75.7* \\
            MMStar                     & 70.9 & 78.5* & 75.5 & 68.5* & 71.5 & 63.2 &78.4* & 68.2 \\
            \midrule
            \multicolumn{9}{c}{\textbf{STEM \& Reasoning}} \\
            \midrule
            MathVista$_{\text{mini}}$  & 77.9 & 77.7* & 77.1 & 75.9  & 78.7 & 62.8  &84.9* & 74.8* \\
            MMMU$_{\text{val}}$        & 70.7 & 80.9* & 76.3 & 69.1* & 74.9 & 69.4 &78.7* & 70.2* \\  
            MMVet                      & 69.0 & 80.7 & 79.5 & 68.9 & 74.4 & 76.6 &75.9 & 74.5 \\
            \midrule
            \multicolumn{9}{c}{\textbf{Multi-Image}} \\
            \midrule
            BLINK                      & 63.1 & 70.0* & 65.7 & 56.1 & 65.0 & 65.5 &70.7* & 60.1 \\
            MuirBench                  & 77.1 & 74.0* & 73.7 & 62.1 & 74.6 & 70.5 &72.8* & 70.7* \\  
            Mantis                     & 84.8 & 83.9  & 83.4 & 80.7 & 81.1 & 79.3 &79.7 & 82.0 \\
            \midrule
            \multicolumn{9}{c}{\textbf{Text Recognition \& Chart/Document Understanding}} \\
            \midrule
            ChartQA               & 87.6 & 71.7 & 77.6 & 86.8*  & 82.4 & 74.5  & 89.2 & 89.5* \\
            DocVQA                & 91.8 & 94.0* & 93.6*  & 95.7   & 94.3 & 80.9  & 94.6 & 96.4* \\
            OCRBench              & 84.9 & 87.2*  & 85.6 & 85.5   & 85.6 & 82.3  & 91.2 & 88.5 \\
            OmniDocBench$_{\text{EN/ZH}}$↓   & 22.8/29.0 & 31.9/24.5 & 22.8/32.9 & 28.4/40.5 & 22.0/27.6  & 25.9/37.7 & 13.6/17.5 & 22.6/32.4* \\
            \midrule
            \multicolumn{9}{c}{\textbf{Grounding \& Counting}} \\
            \midrule
            RefCOCO-avg        & 93.9 & 80.2 & 74.8  & 91.6  & 87.5  & -     & 88.2 & 92.3 \\
            CountBench             & 92.4 & 91.0* & 78.6  & 90.0* & 94.1  & 85.6* & 94.3 & 93.6* \\
            \midrule
            \multicolumn{9}{c}{\textbf{Graphical User Interface (GUI)}} \\
            \midrule
            VisualWebBench       & 78.7 & 81.1  & 73.5  & 79.3  & 81.1 & 77.1 & 80.8  & 82.3* \\
            ScreenSpot-v2        & 91.2 & 75.8  & 63.9  & 94.7  & 91.7 & -    & 93.4  & 92.9 \\
            AndroidControl   & 91.2 & 79.2  & 79.1  & 90.5  & 84.6 & 65.2 & 90.0  & 93.7* \\
            \bottomrule
        \end{tabular}
    }
    \label{tab:eval_image_und}
\end{table}

In this section, we compare LongCat-Flash-Omni with omni-models (Gemini-2.5-Pro\citep{comanici2025gemini}, GPT-4o\footnote{\url{https://openai.com/index/hello-gpt-4o}}, Seed-1.6\footnote{\url{https://seed.bytedance.com/en/seed1_6}}, and Qwen3-Omni\citep{xu2025qwen3}) and vision language models (Qwen3-VL\citep{yang2025qwen3} and Qwen2.5-VL-72B\citep{Qwen2.5-VL}) without thinking mode.

Note that GPT-4o and Seed-1.6 only provide a vision understanding API. Therefore, we evaluate them solely on visual capabilities. To ensure a fair comparison with instruct model, we follow the evaluation setting used in Qwen3-VL benchmark, limiting Gemini-2.5-Pro’s thinking budget to 128 tokens. All models are evaluated under their official configurations.

\subsubsection{Image-to-Text Evaluation}
We conduct a comprehensive evaluation of image understanding across six key dimensions, leveraging a diverse suite of benchmarks:
\begin{itemize}
    \item \textbf{General Domains}: MMBench~\citep{liu2024mmbench}, RealWorldQA~\citep{realworldqa}, and MMStar~\citep{mmstar}.
    
    \item \textbf{STEM \& Reasoning}: MathVista~\citep{lu2024mathvista}, MMMU\textsubscript{val}~\citep{yue2023mmmu}, and MMVet~\citep{yu2024mmvet}.
    
    \item \textbf{Multi-Image}: BLINK~\citep{fu2024blink}, MuirBench~\citep{wang2024muirbench}, and Mantis~\citep{jiang2024mantis}.
    
    \item \textbf{Text Recognition \& Document Understanding}: ChartQA~\citep{masry2022chartqa}, DocVQA~\citep{mathew2021docvqa}, and OCRBench~\citep{liu2024ocrbench}, OmniDocBench\textsubscript{EN/ZH}~\citep{ouyang2024omnidocbench}.
    
    \item \textbf{Grounding \& Counting}: RefCOCO, RefCOCO+~\citep{kazemzadeh-2014-refcoco} (averaged in RefCOCO-avg), and CountBench\citep{paiss2023countbench}.
    
    \item \textbf{Graphical User Interface (GUI)}: VisualWebBench~\citep{liu2024visualwebbench}, ScreenSpot-v2~\citep{wu2024atlas}, and AndroidControl~\citep{li2024androidcontrol}.
\end{itemize}

Results on image understanding benchmarks are summarized in Table~\ref{tab:eval_image_und}. Overall, LongCat-Flash-Omni achieves performance comparable to Gemini-2.5-Flash, while outperforming the open-sourced Qwen3-Omni. Its advantage is particularly pronounced on multi-image tasks, which benefit from the model’s training on high-quality interleaved image-text, multi-image data and video datasets.

\subsubsection{Video-to-Text Evaluation}
To assess the capability of video understanding, we consider three dimensions as below:
\begin{itemize}
    \item \textbf{Short Video}: MVBench~\citep{li2024mvbench}, NextQA~\citep{xiao2021nextqa}, and TempCompass~\citep{liu2024tempcompass}.
    \item \textbf{Long Video}: VideoMME~\citep{fu2024videomme} and LongVideoBench~\citep{wu2024longvideobench}.
    \item \textbf{STEM \& Reasoning}: MMVU~\citep{zhao2025mmvu}, and Video-MMMU~\citep{hu2025videommmu}.
\end{itemize}
We evaluate audio-visual understanding on VideoMME under the no-subtitle protocol (w/o sub), and report results for two conditions: with audio (“VideoMME w/ audio”) and without audio (“VideoMME w/o audio”).

As the Gemini models do not provide official evaluation entries for videos without audio, we report only the ``VideoMME w/ audio'' results.
All video benchmarks are evaluated using a maximum context length of 32K tokens to ensure consistent comparison.

As shown in Table~\ref{tab:eval_video_und}, LongCat-Flash-Omni achieves state-of-the-art performance on video-to-text tasks. Specifically, it surpasses all compared models by a significant margin on short video understanding, demonstrating superior video comprehension capabilities. On long video tasks, LongCat-Flash-Omni demonstrates performance on par with leading models such as Gemini-2.5-Pro and Qwen3-VL. Notably, in the VideoMME benchmark, it achieves the best performance among omni-modal models. This can be attributed to a combination of an advanced video processing strategy—using dynamic frame sampling and hierarchical token aggregation—and the strong long-context modeling capacity afforded by its efficient backbone.

\begin{table}[htbp]
    \caption{Evaluation of video understanding. Values marked with * are sourced from public reports.}
    \centering
    \resizebox{1.0\textwidth}{!}{
        \begin{tabular}{l|cccccc|cl}
            \toprule
            Benchmark & \begin{tabular}[c]{@{}c@{}}\textbf{LongCat-Flash-Omni} \\ Instruct\end{tabular} & \begin{tabular}[c]{@{}c@{}}Gemini-2.5-Pro \\ {\small ThinkingBudget128}\end{tabular} & \begin{tabular}[c]{@{}c@{}}Gemini-2.5-Flash \\ {\small non-thinking}\end{tabular} & \begin{tabular}[c]{@{}c@{}}Qwen3-Omni \\ Instruct\end{tabular} & Seed-1.6 & \begin{tabular}[c]{@{}c@{}}GPT-4o \\ 1120\end{tabular} & \begin{tabular}[c]{@{}c@{}}Qwen3-VL \\ {\small 235B-A22B-Instruct}\end{tabular} & \begin{tabular}[c]{@{}c@{}}Qwen2.5-VL \\ {\small 72B-Instruct}\end{tabular} \\ 
            \midrule
            \multicolumn{9}{c}{\textbf{Short Video}} \\
            \midrule
            MVBench           & 75.2  & 66.4  & 63.0 & 69.3* & 68.4 & 62.1  & 71.3 & 70.4* \\
            NextQA            & 86.2 & 84.2 & 81.4 & 82.4 & 84.1 & 79.7  & 81.3 & 82.3 \\
            TempCompass       & 82.2 & 80.8  & 80.2 & 73.5  & 79.4 & 76.4  & 80.5 & 74.8* \\
            \midrule
            \multicolumn{9}{c}{\textbf{Long Video}} \\
            \midrule
            VideoMME w/o audio & 76.2 & -     & -     & 70.5* & 75.2 & 73.2  & 79.2* & 73.3* \\
            VideoMME w/ audio & 78.2     & 80.6* & 78.5 & 73.0 & - & -  & - & - \\
            LongVideoBench    & 69.3 & 69.4  & 66.4 & 65.4 & 64.8 & 63.9  & - & 60.7* \\
            \midrule
            \multicolumn{9}{c}{\textbf{STEM \& Reasoning}} \\
            \midrule
            MMVU              & 67.1 & 75.6  & 72.4 & 62.4 & 67.3 & 67.4  & 69.3 & 62.9* \\
            Video-MMMU        & 67.5 & 79.4* & 76.6 & 60.3 & 75.4 & 68.0  & 73.7 & 59.3 \\
            \bottomrule
        \end{tabular}
    }
    \label{tab:eval_video_und}
\end{table}

\subsection{Audio Capability Evaluation}
\label{sec:audio-eval}

\subsubsection{Base Model Evaluation}

To systematically assess the audio capability of the base models produced in the pre-training stages (i.e., stage-1, stage-2, stage-3, and stage-4), we conduct evaluations with respect to the following three aspects: automatic speech recognition (ASR), text-to-speech (TTS), and speech continuation.
For ASR evaluation, we employ the SPEECHIO\_ASR\_ZH00002 (speechio02)~\footnote{\url{https://github.com/SpeechColab/Leaderboard}} to test Chinese speech recognition, while utilizing the LibriSpeech \citep{panayotov2015librispeech} for English speech recognition assessment. 
The evaluation results are presented in Table~\ref{tab:base_asr_tts_eval}. Remarkably, our model demonstrates competitive performance across these benchmark datasets, despite relying on discretized audio features.

\begin{table}[h!]
\centering
\caption{ASR and TTS performance of base models during the pre-training stages. Word error rate (WER) (for English) or character error rate (CER) (for Chinese) in percentage are reported.}
\resizebox{0.8\textwidth}{!}{
\begin{tabular}{lccccc}
\toprule
\textbf{Base Model} & \multicolumn{3}{c}{\textbf{ASR}} & \multicolumn{2}{c}{\textbf{TTS}} \\
\cmidrule(lr){2-4} \cmidrule(lr){5-6}
& \textbf{SpeechIO02↓}
 & \begin{tabular}[c]{@{}c@{}}\textbf{LibriSpeech} \\ \textbf{test-clean↓} \end{tabular} & \begin{tabular}[c]{@{}c@{}}\textbf{LibriSpeech} \\ \textbf{test-other↓} \end{tabular} & \textbf{SpeechIO02↓} & \textbf{LibriSpeech↓} \\
\midrule
Stage-1 & 2.93 & 1.98 & 3.96 & 4.12 & 4.72 \\
Stage-2 & 3.18 & 2.11 & 4.59 & 2.16 & 8.64 \\
Stage-3 & 3.01 & 1.93 & 3.74 & 2.53 & 3.68 \\
Stage-4 (32K) & 3.49 & 2.30 & 4.00 & 1.73 & 5.99 \\
Stage-4 (128K) & 3.46 & 2.12 & 4.15 & 2.62 & 7.38 \\
\bottomrule
\end{tabular}
}
\label{tab:base_asr_tts_eval}
\end{table}

The TTS evaluation framework uses text samples from the speechio02 and LibriSpeech~\citep{panayotov2015librispeech} test-clean/other datasets as input sources. The model uses discrete semantic-acoustic tokens as speech prompts, conducting speech token generation with text input as teacher-forcing guidance. The output speech tokens are reconstructed into waveform by the audio decoder. We compute the word error rate (WER) (for English) or character error rate (CER) (for Chinese) between transcription of the generated speech produced by a robust ASR system and the original input text as the measure for the evaluation. The results are shown in the right part of Table~\ref{tab:base_asr_tts_eval}. We observe that the base models are capable of generating robust speech from textual input, providing a solid foundation for speech output in spoken and audio-visual interaction modes.

To evaluate speech continuation capabilities, we develope a specialized test set comprising $1,200$ synthesized speech samples derived from text data from the CMMLU \citep{li2023cmmlu} benchmark.
The assessment requires the model to generate appropriate responses to multiple-choice questions based on provided speech input. For text outputs, we directly measure response accuracy, while speech outputs undergo ASR transcription before accuracy evaluation. The evaluation protocol employs a 1-shot learning approach, with detailed results presented in Table~\ref{tab:base_speech_cmmlu}. 
Our analysis reveals that the base models in different stage perform well in in-context speech continuation evaluation and there is only a negligible performance disparity between text and speech output.

\begin{table}[h!]
\centering
\caption{Speech continuation performance of base models across pre-training stages. Accuracy is reported as a percentage.}
\resizebox{0.5\textwidth}{!}{
\begin{tabular}{lcc}
\toprule
\textbf{Base Model} & \multicolumn{2}{c}{\textbf{Speech Continuation (CMMLU)↑}} \\
\cmidrule(lr){2-3}
& \textbf{Audio In Text Out}
 & \textbf{Audio In Audio Out} \\
\midrule
Stage-1 & 88.80 & 84.80 \\
Stage-2 & 89.60 & 84.80 \\
Stage-3 & 92.80 & 92.00 \\
Stage-4 (32K) & 91.20 & 91.20 \\
Stage-4 (128K) & 90.40 & 90.40 \\
\bottomrule
\end{tabular}
}
\label{tab:base_speech_cmmlu}
\end{table}

\subsubsection{Instruct Model Evaluation}

We conduct comprehensive evaluation of the audio capabilities of \longcat, covering speech recognition and translation, audio understanding and audio-to-text chat.

\textbf{Speech Recognition and Translation}

\begin{table}[t] 
    \caption{Automatic speech recognition (ASR) and speech-to-text translation (S2TT) evaluation results.}
    \centering
    \resizebox{1.0\textwidth}{!}{
\begin{tabular}{c|cccccc}
\toprule
Benchmark & \begin{tabular}[c]{@{}c@{}}\textbf{LongCat-Flash-Omni} \\ Instruct\end{tabular} & \begin{tabular}[c]{@{}c@{}}Gemini-2.5-Pro \\ {\small ThinkingBudget128}\end{tabular} & GPT-4o-Audio & \begin{tabular}[c]{@{}c@{}}Qwen3-Omni \\ Instruct\end{tabular} & Kimi-Audio & Step-Audio-2-mini  \\ 
\midrule
\multicolumn{7}{c}{\textbf{ASR}} \\
\midrule
\begin{tabular}[c]{@{}c@{}}\textbf{LibriSpeech} \\ test-clean | test-other\end{tabular} & 1.57 | 4.01 & 1.74 | 3.80 & 30.00 | 41.83 & 1.22 | 2.48 & 1.28 | 2.42 & 1.33 | 2.86 \\
\textbf{AISHELL-1} & 0.63 & 3.11 & 34.81  & 0.84  & 0.60 & 0.78 \\
\textbf{AISHELL-2} & 2.78 & 5.24 & 77.73 & 2.34 & 2.56 & 2.16 \\
\begin{tabular}[c]{@{}c@{}}\textbf{Fleurs} \\ zh | en \end{tabular} & 3.99 | 5.02 & 2.24 | 4.77 & 3.91 | 5.56 & 2.20 | 2.72 & 2.69 | 4.44 & 2.53 | 3.05 \\
\begin{tabular}[c]{@{}c@{}}\textbf{CommonVoice 15} \\ zh | en \end{tabular} & 4.98 | 13.59 & 47.30 | 49.86 & 42.83 | 23.88 & 4.31 | 6.05 & 8.46 | 7.92 & 5.00 | 6.75 \\
\begin{tabular}[c]{@{}c@{}}\textbf{WenetSpeech} \\ test-meeting | test-net \end{tabular} & 6.69 | 6.09 & 136.13 | 32.82 & 54.35 | 67.90 & 5.89 | 4.69 & 6.28 | 5.37 & 4.87 | 4.82 \\
\midrule
\multicolumn{7}{c}{\textbf{S2TT (BLEU)}} \\
\midrule
\textbf{CoVost2 en$\rightarrow$zh} & 47.23 & 41.94 & 29.32 & 48.72  & - & 49.12 \\
\textbf{CoVost2 zh$\rightarrow$en} & 27.32 & 25.38 & 16.01 & 21.51  & - & 29.47 \\

\bottomrule
\end{tabular}
}
\label{tab:eval_asr_ast}
\end{table}

\begin{table}[t]
    \caption{Evaluation results of audio understanding.}
    \centering
    \resizebox{1.0\textwidth}{!}{
\begin{tabular}{c|cccccc}
\toprule
Benchmark & \begin{tabular}[c]{@{}c@{}}\textbf{LongCat-Flash-Omni} \\ Instruct\end{tabular} & \begin{tabular}[c]{@{}c@{}}Gemini-2.5-Pro \\ {\small ThinkingBudget128}\end{tabular} & GPT-4o-Audio & \begin{tabular}[c]{@{}c@{}}Qwen3-Omni \\ Instruct\end{tabular} & Kimi-Audio & Step-Audio-2-mini  \\ 
\midrule
\textbf{MMAU} & 75.90 & 72.80 & 68.40 & 77.50  & 65.20 & 73.20 \\
\textbf{VocalSound} & 92.76 & 89.45 & 82.37  & 91.60  & 94.85 & 87.58 \\
\textbf{TUT2017} & 65.43 & 33.15 & 20.74 & 40.74  & 65.25 & 30.67 \\
\textbf{ClothoAQA} & 72.83 & 69.67 & 61.87  & 75.16  & 72.21 & 68.39 \\
\textbf{Nonspeech7k} & 93.79 & 87.59 & 72.28 & 80.83  & 93.93 & 73.24 \\
\textbf{CochlScene} & 70.02 & 45.34 & 34.94 & 43.03  & 80.42 & 44.58 \\
\textbf{MELD} & 54.60 & 46.74 & 39.00 & 50.80  & 59.13 & 31.44 \\

\bottomrule
\end{tabular}
}
\label{tab:eval_audio_understanding}
\end{table}

For ASR evaluation, as presented in Table~\ref{tab:eval_asr_ast}, we use LibriSpeech~\citep{panayotov2015librispeech}, AISHELL-1~\citep{bu2017aishell}, AISHELL-2~\citep{du2018aishell}, FLEURS~\citep{conneau2023fleurs}, CommonVoice15~\citep{ardila2019common}, and WenetSpeech~\citep{zhang2022wenetspeech} as benchmarks. We observe that \longcat consistently achieves superior performance compared to other competitors, including Gemini-2.5-Pro~\citep{comanici2025gemini}, GPT-4o-Audio, Qwen3-Omni-Instruct~\citep{xu2025qwen3}, Kimi-Audio~\citep{ding2025kimi}, and Step-Audio-2-mini~\citep{wu2025step}.
For speech-to-text translation (S2TT) evaluation, we adopt CoVost2~\citep{wang2021covost}. As shown in Table~\ref{tab:eval_asr_ast}, \longcat exhibits strong S2TT capabilities among all models\footnote{Kimi-audio defaults to ASR rather than executing the requested translation task.}. Taken together, the speech recognition and translation results demonstrate that LongCat-Flash-Omni possesses robust and comprehensive fundamental speech understanding capabilities.

\textbf{Audio Understanding}
As an omni-modal model, \longcat can effectively function as a native audio understanding model when vision input is not provided. We conduct a comprehensive evaluation of \longcat across a diverse set of audio comprehension tasks, including music, sound events, and speech understanding. Specifically, we use MMAU~\citep{sakshi2024mmau}, VocalSound~\citep{gong_vocalsound}, TUT2017~\citep{Mesaros2016_EUSIPCO}, ClothoAQA~\citep{lipping2022clotho}, Nonspeech7k~\citep{rashid2023nonspeech7k}, CochlScene~\citep{jeong2022cochlscene}, and MELD~\citep{poria2018meld} as benchmarks.
The results, presented in Table~\ref{tab:eval_audio_understanding}, show that \longcat consistently outperforms most competing models across all evaluated dimensions, and achieves state-of-the-art performance in several benchmarks. These findings highlight that \longcat possesses advanced capabilities in understanding a wide spectrum of general and complex acoustic information, going well beyond conventional speech recognition.

\textbf{Audio-to-Text Chat} We evaluate the ability of \longcat to engage in text-based conversations driven by audio instructions using the OpenAudioBench~\citep{li2025baichuan} and VoiceBench~\citep{chen2024voicebench} benchmarks. These benchmarks assess a wide range of capabilities, including world knowledge, domain-specific understanding, instruction following, and reasoning.
To ensure standardized and reproducible results, we employ the official evaluation code provided by the benchmark authors. The results, presented in Table~\ref{tab:eval_a2t_chat}, show that \longcat achieves strong performance across all benchmark subsets, reaching state-of-the-art results in several cases. Overall, these results demonstrate \longcat’s superior ability to conduct audio-driven conversations and perform complex reasoning tasks.

\begin{table}[t] 
    \caption{Evaluation results of audio-to-text chat.}
    \centering
    \resizebox{1.0\textwidth}{!}{
\begin{tabular}{c|cccccc}
\toprule
Benchmark & \begin{tabular}[c]{@{}c@{}}\textbf{LongCat-Flash-Omni} \\ Instruct\end{tabular} & \begin{tabular}[c]{@{}c@{}}Gemini-2.5-Pro \\ {\small ThinkingBudget128}\end{tabular} & GPT-4o-Audio & \begin{tabular}[c]{@{}c@{}}Qwen3-Omni \\ Instruct\end{tabular} & Kimi-Audio & Step-Audio-2-mini  \\ 
\midrule
\multicolumn{7}{c}{\textbf{OpenAudioBench}} \\
\midrule
\textbf{LlamaQuestions} & 83.33 & 83.00 & 86.30 & 83.30  & 79.33 & 69.70 \\
\textbf{ReasoningQA}    & 79.71 & 80.30 & 68.71 & 84.16  & 58.02 & 55.64 \\
\textbf{TriviaQA}       & 86.20 & 90.20 & 76.00 & 75.90  & 62.10 & 45.30 \\
\textbf{Webquestions}   & 76.00 & 80.90 & 81.20 & 75.20  & 70.20 & 54.40 \\
\textbf{AlpacaEval}     & 75.43 & 76.58 & 81.61 & 85.43  & 75.73 & 53.92 \\

\midrule
\multicolumn{7}{c}{\textbf{VoiceBench}} \\
\midrule
\textbf{AlpacaEval} & 4.94  & 4.70  & 4.73  & 4.74  & 4.46  & 3.84 \\
\textbf{CommonEval} & 4.32  & 4.11  & 4.37  & 4.54  & 3.97  & 3.19 \\
\textbf{OpenBookQA} & 93.41 & 95.16 & 87.90 & 89.70 & 83.52 & 72.97 \\
\textbf{SDQA}       & 82.46 & 83.54 & 90.10 & 76.90 & 63.12 & 44.85 \\
\textbf{MMSU}       & 81.95 & 88.32 & 78.90 & 69.00 & 62.17 & 52.00 \\
\textbf{AdvBench}   & 100   & 97.69 & 99.23 & 99.30 & 100   & 97.00 \\
\textbf{IFEval}     & 77.99 & 77.83 & 66.81 & 77.80 & 61.10 & 29.80 \\
\bottomrule
\end{tabular}
}
\label{tab:eval_a2t_chat}
\end{table}



\subsection{Text Capability Evaluation}
\subsubsection{Base Model Evaluation}
We compare the LongCat-Flash-Omni Base model (i.e., after pre-training stage 5) with other strong text-based models across the following core capabilities and corresponding benchmarks.: (1) \textbf{General Tasks:} MMLU~\citep{hendrycks2021measuringmassivemultitasklanguage}, MMLU-Pro~\citep{wang2024mmluprorobustchallengingmultitask}, C-Eval~\citep{huang2023ceval}, and CMMLU~\citep{li2023cmmlu}. (2) \textbf{Reasoning Tasks:} GPQA~\citep{rein2023gpqagraduatelevelgoogleproofqa}, SuperGPQA~\citep{pteam2025supergpqascalingllmevaluation}, BBH~\citep{BBH}, PIQA~\citep{piqa}, DROP~\citep{drop}, CLUEWSC~\citep{clue}, and WinoGrande~\citep{winogrande}. (3) \textbf{Math Tasks:} GSM8K~\citep{cobbe2021gsm8k}, MATH~\citep{mathbenchmark}. (4) \textbf{Coding Tasks:} MBPP+~\citep{humanevalmbppplus}, HumanEval+~\citep{humanevalmbppplus}, MultiPL-E~\citep{cassano2022multiplescalableextensibleapproach}, and CRUXEval~\citep{gu2024cruxevalbenchmarkcodereasoning}. We follow the same evaluation protocol as in \citep{meituan2025longcat_flash_chat} to ensure maximum fairness.

Table~\ref{tab:text_eval_base} presents the evaluation results across diverse benchmarks. 
LongCat-Flash-Omni Base model achieves performance on par with state-of-the-art base models despite its compact active/total parameter size. Furthermore, our model demonstrates no degradation in text capabilities after extensive training with multimodal data, indicating the effectiveness of our training strategy.

\begin{table}[htbp]
    \caption{Comparison between LongCat-Flash-Omni and other base models on text-only benchmarks. Values marked with * are sourced from public reports.}
    \centering
    \resizebox{0.8\textwidth}{!}{
    \begin{tabular}{l|c|ccc}
        \toprule
        Benchmark & \begin{tabular}[c]{@{}c@{}}\textbf{\longcat} \\ Base\end{tabular}  & \begin{tabular}[c]{@{}c@{}}LongCat-Flash \\ Base\end{tabular} & \begin{tabular}[c]{@{}c@{}}DeepSeek-V3.1 \\ Base\end{tabular} &  \begin{tabular}[c]{@{}c@{}}Kimi-K2 \\ Base\end{tabular} \\ 
        \midrule
        Architecture & MoE & MoE  & MoE  & MoE  \\
        \# Total Params & 560B & 560B & 671B & 1043B \\
        \# Activated Params & 27B & 27B & 37B & 32B \\
        \midrule
        \multicolumn{5}{c}{\textbf{General Domains}} \\
        \midrule
        MMLU $_{\text{(acc)}}$  & 86.81  & 87.05 & 87.46  & 87.47 \\
        MMLU-Pro $_{\text{(acc)}}$ & 69.05  & 70.32 & 59.29  & 68.36 \\
        CEval $_{\text{(acc)}}$ & 87.95  & 87.73 & 89.33  & 91.24 \\
        CMMLU $_{\text{(acc)}}$ & 87.14  & 87.19 & 88.21  & 90.35  \\
        \midrule
        \multicolumn{5}{c}{\textbf{General Reasoning}} \\
        \midrule
        GPQA $_{\text{(acc)}}$ & 51.76  & 51.09  & 47.16 & 45.89  \\
        SuperGPQA $_{\text{(acc)}}$ & 54.71  & 54.19 & -  & 44.70* \\
        BBH $_{\text{(acc)}}$ & 90.42 & 90.54 & 89.46 & 89.19 \\
        DROP $_{\text{(f1)}}$ & 80.75  & 78.39  & 80.74 & 69.81 \\
        PIQA $_{\text{(acc)}}$ & 92.27  & 92.33 & 93.00 & 95.10 \\
        WinoGrande $_{\text{(acc)}}$ & 85.95  & 85.08 & 83.50 & 82.87 \\
        CLUEWSC $_{\text{(acc)}}$ & 91.45  & 91.12 & 88.16 & 76.32 \\

        \midrule
        \multicolumn{5}{c}{\textbf{Mathematical Reasoning}} \\
        \midrule

        GSM8K $_{\text{(acc)}}$ & 93.10  & 92.19 & 92.22 & 92.27 \\
        MATH $_{\text{(acc)}}$ &  66.80 & 64.82 & 61.56 & 66.74 \\
        \midrule
        \multicolumn{5}{c}{\textbf{Coding}} \\
        \midrule
        MBPP+ $_{\text{(pass@1)}}$ & 76.46  & 77.25 & 59.26 & 80.49 \\
        HumanEval+ $_{\text{(pass@1)}}$ & 69.51  & 65.85  & 67.07  & 69.84 \\
        MultiPL-E $_{\text{(pass@1)}}$ & 70.76  & 69.25 & 62.00  & 59.22 \\
        CRUXEval-I $_{\text{(pass@1)}}$ & 71.88  & 71.63 & 65.87 & 65.87 \\
        CRUXEval-O $_{\text{(pass@1)}}$ & 73.50  & 75.88 & 71.25 & 68.75 \\
        \bottomrule
    \end{tabular}
    }
    \label{tab:text_eval_base}
\end{table}

\begin{table*}[t] 
     \centering 
     \caption{Evaluation results of frontier chat/instruct models. Values marked with * are sourced from other public reports. Note that DeepSeek-V3.1, Qwen3-235B-A22B, Gemini2.5-Flash, and Claude Sonnet-4 are evaluated under their non-thinking mode.}
     \label{tab:text_eval_results} 
     \resizebox{\textwidth}{!}{ 
     \begin{tabular}{l|c|c|ccc|ccc} 
         \toprule 
         Benchmark & \begin{tabular}[c]{@{}c@{}}\textbf{LongCat-Flash-Omni} \\ Instruct \end{tabular} & \begin{tabular}[c]{@{}c@{}}LongCat-Flash \\ \end{tabular} & \begin{tabular}[c]{@{}c@{}}DeepSeek \\ V3.1\end{tabular} & \begin{tabular}[c]{@{}c@{}}Qwen3 \\ MoE-2507\end{tabular} & \begin{tabular}[c]{@{}c@{}}Kimi-K2 \\ \end{tabular} & GPT-4.1 & \begin{tabular}[c]{@{}c@{}}Claude \\ Sonnet-4\end{tabular} & \begin{tabular}[c]{@{}c@{}}Gemini-2.5 \\ -Flash\end{tabular} \\ 
         \midrule 
         
Architecture & MoE & MoE & MoE & MoE & MoE & - & - & - \\ 
         \# Total Params & 560B & 560B & 671B & 235B & 1043B & - & - & - \\ 
         \# Activated Params & 27B & 27B & 37B & 22B & 32B & - & - & - \\ 
         \midrule 
         \multicolumn{9}{c}{\textbf{General Domains}} \\ 
  
        \midrule 
         MMLU $_{\text{(acc)}}$ & 90.30 & 89.71 & 90.96 & 90.23 & 89.86 & 89.64 & 91.75 & 86.33 \\ 
         MMLU-Pro $_{\text{(acc)}}$ & 82.73 & 82.68  & 84.45 & 84.83 & 82.06 & 81.72 & 83.74 & 81.95 \\ 
         CEval $_{\text{(acc)}}$ & 91.68 & 90.44 & 89.21 & 92.70 & 91.26 & 79.53 & 86.63 & 78.78 \\ 
         CMMLU $_{\text{(acc)}}$ & 89.39 & 84.34 & 88.04 & 88.14 & 89.66 & 77.65 & 86.51 & 78.30 \\ 
         \midrule 
         \multicolumn{9}{c}{\textbf{Instruction Following}} \\ 
        
 \midrule 
         IFEval $_{\text{(acc)}}$ & 82.44 & 89.65 & 86.69 & 88.54 & 88.91 & 85.58 & 88.35 & 83.92 \\ 
         COLLIE $_{\text{(acc)}}$ & 45.69 & 57.10 & 43.80 & 49.71 & 56.34 & 50.00 & 51.22 & 48.60 \\ 
         Meeseeks-zh $_{\text{(acc)}}$ & 39.05 & 43.03 & 33.83 & 35.32 & 42.79 & 41.54 & 35.07 & 34.84 \\ 
         \midrule 
       
  \multicolumn{9}{c}{\textbf{Mathematical Reasoning}} \\ 
         \midrule 
         MATH500 $_{\text{(acc)}}$ & 97.60 & 96.40   & 96.08 & 98.80 & 97.60 & 90.60 & 93.80 & 98.40\\ 
         AIME24 $_{\text{(avg@10)}}$ & 72.92 &  70.42  & 66.30* & 81.67 & 69.60* & 47.00 & 47.00 & 79.67 \\ 
     BeyondAIME $_{\text{(avg@10)}}$ & 47.40 & 43.00 & 36.50 & 57.60 & 36.60 & 22.10 & 20.50 & 44.20 \\ 
         \midrule 
         \multicolumn{9}{c}{\textbf{General Reasoning}} \\ 
         \midrule 
         GPQA-diamond $_{\text{(acc)}}$ & 74.41 & 73.23 & 74.90* & 77.43 & 75.76 & 67.68 & 70.71 & 80.30 \\ 
         DROP $_{\text{(f1)}}$ & 83.53 & 79.06  & 84.19 & 78.57 & 89.04 & 66.94 & 73.06 & 45.03 \\ 
         ZebraLogic $_{\text{(acc)}}$ & 86.00 & 89.30  & 85.30 & 94.22 & 89.11 & 56.30* & 80.10 & 57.00 \\ 
         GraphWalks-128K $_{\text{(precision)}}$ & 56.00 & 51.05  & 73.54 & 80.72 & 47.50 & 85.02 & 80.57 & 64.83 \\ 
         \midrule 
         \multicolumn{9}{c}{\textbf{Coding}} \\ 
         \midrule 
   
         LiveCodeBench $_{\text{(pass@1)}}$ & 52.64 & 48.02 & 56.40*  & 46.48 & 46.70 & 39.21 & 45.59 & 39.65 \\ 
         Humaneval+ $_{\text{(pass@1)}}$ & 90.85 & 88.41  & 92.68 & 94.51 & 85.98 & 93.29 & 94.51 & 87.80 \\ 
         MBPP+ $_{\text{(pass@1)}}$ & 80.16 & 79.63  & 79.89 & 79.89 & 81.75 & 79.37 & 80.16 & 76.19 \\ 
         \bottomrule 
     \end{tabular}
     } 
 \end{table*}

\subsubsection{Instruct Model Evaluation}

A comprehensive and rigorous text capability evaluation of the \longcat Instruct model is performed, covering diverse capability dimensions, including general domains, instruction following, mathematical reasoning,
general reasoning, and coding, using the following benchmarks:
\begin{itemize}
    \item \textbf{General Domains:} MMLU~\citep{hendrycks2021measuringmassivemultitasklanguage}, MMLU-Pro~\citep{wang2024mmluprorobustchallengingmultitask}, CEval~\citep{huang2023ceval}, and CMMLU~\citep{li2023cmmlu}.
    \item \textbf{Instruction Following:} IFEval~\citep{zhou2023ifeval}, COLLIE~\citep{yao2024collie}, and Meeseeks~\citep{meeseeks}, Meeseeks evaluates models' instruction-following capabilities in multi-turn scenarios through an iterative feedback framework that simulates realistic human-LLM interactions, enabling models to self-correct based on turn-specific failures and better reflect real-world usage patterns.
    \item \textbf{Mathematical Reasoning:}  MATH500~\citep{math500}, AIME24~\citep{AIME24}, and BeyondAIME~\citep{bytedance_seed_2025_beyondaime}.
    \item \textbf{General Reasoning:}  GPQA-diamond~\citep{rein2023gpqagraduatelevelgoogleproofqa}, DROP~\citep{drop}, ZebraLogic~\citep{lin2025zebralogic}, and GraphWalks~\citep{graphwalks}.
    \item \textbf{Coding:}  Humaneval+~\citep{humanevalmbppplus}, MBPP+~\citep{humanevalmbppplus}, and LiveCodeBench (2024.08-2025.05)~\citep{jain2025livecodebench}.
\end{itemize}

\begin{table*}
    \caption{Evaluation of cross-modality understanding. We use an internally corrected version of OmniBench, as the publicly released version contains scoring deficiencies.}
    \centering
    \resizebox{1.0\textwidth}{!}{
        \begin{tabular}{l|ccc|cc}
            \toprule    
            Benchmark & \begin{tabular}[c]{@{}c@{}} \textbf{LongCat-Flash-Omni} \\ Instruct\end{tabular} &  \begin{tabular}[c]{@{}c@{}} Gemini-2.5-Pro \\ {\small ThinkingBudget128}\end{tabular} & \begin{tabular}[c]{@{}c@{}} Gemini-2.5-Flash \\ non-thinking \end{tabular} & \begin{tabular}[c]{@{}c@{}}Qwen3-Omni \\ Instruct \end{tabular} & \begin{tabular}[c]{@{}c@{}} Qwen2.5-Omni \\ Instruct\end{tabular} \\ 
            \midrule    
            OmniBench  & 61.38  & 66.80 & 54.99       & 58.41 & 48.16  \\
            WorldSense & 60.89  & 63.96 & 58.72       & 52.01 & 46.69 \\
            DailyOmni  & 82.38  & 80.61 & 80.78       & 69.33 & 47.45 \\
            UNO-Bench  & 49.90  & 64.48 & 54.30       & 42.10 & 32.60 \\
            \bottomrule
        \end{tabular}
        }
    \label{tab:eval_cross_modal_und}
\end{table*}

We compare \longcat with some representative text chat models including DeepSeek-V3.1~\citep{deepseekai2025deepseekv3technicalreport}, Qwen3-235B-A22B (2507 version)~\citep{yang2025qwen3}, Kimi-K2~\citep{Kimi_K2_web_doc}, GPT-4.1~\citep{gpt4.1}, Claude Sonnet-4~\citep{claude4}, Gemini2.5-Flash~\citep{comanici2025gemini} and LongCat-Flash\citep{meituan2025longcat_flash_chat}. For closed-source models, we conduct evaluations through their official APIs. For models supporting both thinking and non-thinking modes (Qwen3-235B-A22B, Gemini-2.5-Flash, and Claude Sonnet-4), we explicitly configure these models to operate in non-thinking mode for a fair comparison.
The evaluation results are presented in Table~\ref{tab:text_eval_results}. the comprehensive evaluation demonstrates that \longcat maintains superior text capability, with consistently leading performance in different domains. Compared with LongCat-Flash, whose early base model serves as the foundation for LongCat-Flash-Omni, the latter not only exhibits no degradation in text capabilities but even achieves superior performance in certain domains. This highlights the effectiveness of our training strategy and underscores the potential synergy among different modalities in omni-modal model training.



\subsection{Cross-modality Evaluation}
In addition to the aforementioned capabilities, LongCat-Flash-Omni introduces advanced cross-modal understanding and human-like speech interaction with cross-modality inputs. To evaluate these abilities, we compare our model against several strong baselines, including Gemini-2.5-Pro, Gemini-2.5-Flash, Gemini-2.0-Flash, Qwen3-Omni, and Qwen2.5-Omni.
For a fair comparison between the LongCat-Flash-Omni instruct model and Gemini-2.5-Pro with thinking mode enabled, we follow the approach of Qwen3-VL by constraining Gemini’s thinking budget to 128 tokens, denoted as Gemini-2.5-Pro-ThinkingBudget128.
It's important to mention that we couldn't use APIs to test GPT-4o and Seed-1.6 because they don't have open ones for audio. Instead, we tested their performance on corresponding evaluations by interacting with their live applications in a real-time audio-visual scenario.


\subsubsection{Cross-modality Understanding Evaluation}
In this section, we evaluate cross-modal understanding using publicly available benchmarks, including OmniBench~\citep{omnibench}, WorldSense~\citep{worldsense}, and DailyOmni~\citep{zhou2025dailyomni}. We use an internally corrected version of OmniBench, as the publicly released version contains scoring deficiencies. 
To address the limited quality and coverage of existing benchmarks, we further introduce a new benchmark, UNO-Bench~\citep{chen2025unobench}, comprising 1,880 human-crafted questions spanning 44 task types, with 98\% of the questions requiring cross-modal reasoning.
We constructed this benchmark by manually annotating our in-house dataset. This approach prevents data contamination and ensures the benchmark is highly representative of real-world application scenarios.
In addition to conventional multiple-choice questions, the evaluation includes innovative multi-step open-ended questions, providing a more realistic and discriminative assessment of complex reasoning abilities.

As shown in Table~\ref{tab:eval_cross_modal_und}, LongCat-Flash-Omni outperforms Gemini-2.5-Flash-non-thinking and achieves performance comparable to Gemini-2.5-Pro-ThinkingBudget128. In particular, on WorldSense and DailyOmni, which emphasize real-world audio-video understanding, LongCat-Flash-Omni demonstrates superior performance, significantly surpassing other open-source omni-modal models. 
On UNO-Bench, which evaluates cross-modal perception and reasoning, LongCat-Flash-Omni also performs exceptionally well among open-source omni-modal models.
These results indicate that LongCat-Flash-Omni achieves highly effective multimodal integration, establishing it as the leading open-source omni-modal model.

\subsubsection{Real-time Audio-Visual Interaction Evaluation}
Existing cross-modal evaluation benchmarks still exhibit a noticeable gap from real-world user experiences, where real-time audio-visual interaction (audio-visual input with audio output) is essential. To the best of our knowledge, no prior work has systematically evaluated this form of real-time multimodal interaction.
To that end, we built a proprietary, end-to-end framework to measure the quality of a model's audio-visual interaction, specifically its ability to engage users naturally and fluently in real-world scenarios.

\begin{figure}[ht]
  \centering
  \includegraphics[width=0.9\textwidth]{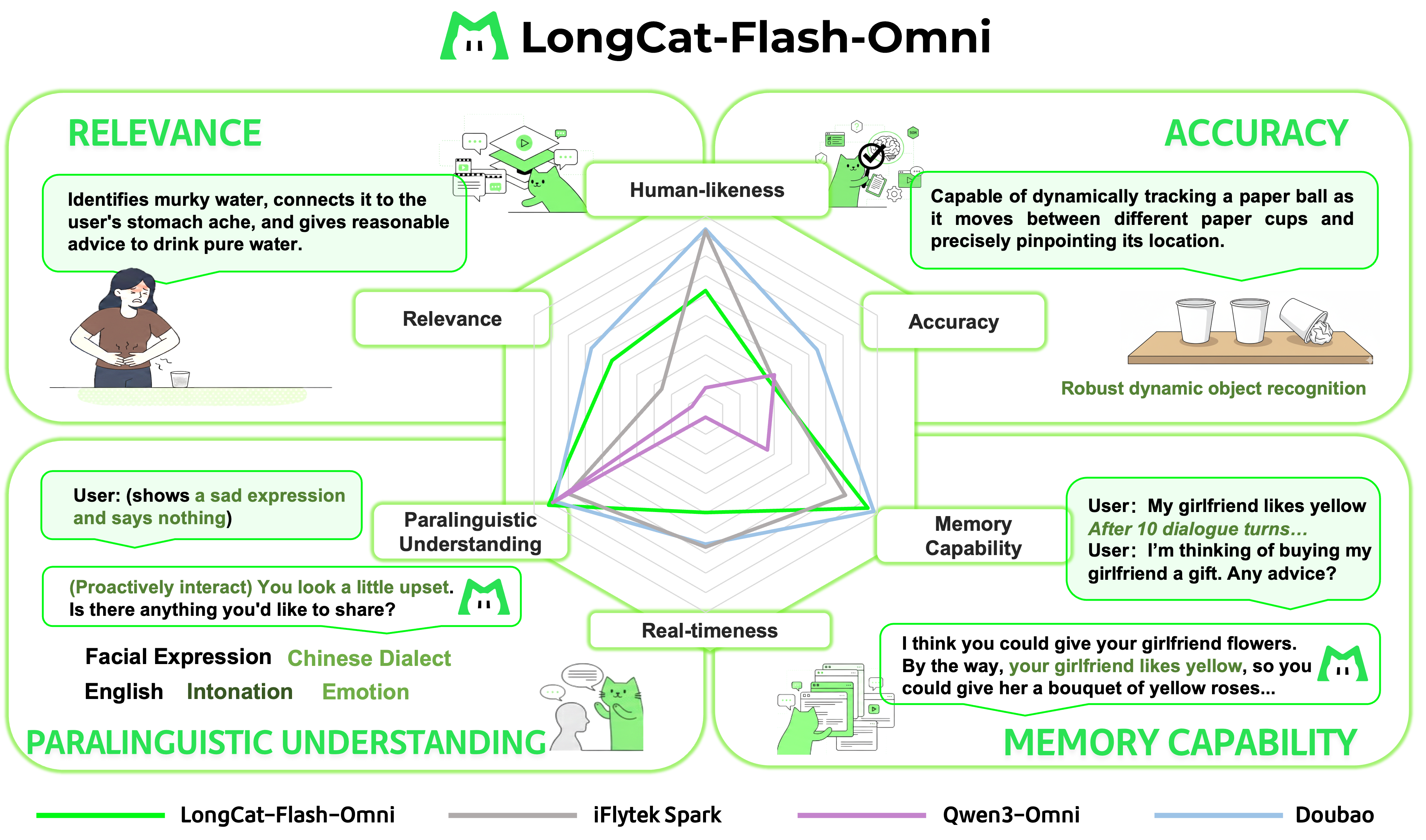}
  \caption{Qualitative analysis of real-time audio-visual interaction.}
  \label{fig:qualitative_analysis}
\end{figure}

\begin{table}[htbp]
    \caption{Real-time audio-visual interaction evaluation.}
    \centering
    \resizebox{1.0\textwidth}{!}{
        \begin{tabular}{l|ccccccc|c}
            \toprule    
            Metrics & Doubao & GPT-4o & iFlytek Spark & StepFun & ChatGLM & Qwen2.5-Omni & Qwen3-Omni & \textbf{LongCat-Flash-Omni}\\
            \midrule
            Score & 1.92 & 1.79 & 1.25 & 1.22 & 0.99 & 0.96 & 0.81 & 1.37\\
            95\% CI & [1.85, 1.98] & [1.72, 1.85] & [1.18, 1.32] & [1.15, 1.28] & [0.93, 1.05] & [0.89, 1.02] & [0.75, 0.87] & [1.30, 1.44]\\

            \bottomrule
        \end{tabular}
    }
    \label{tab:real_time_interaction}
\end{table}

\begin{table}[htbp]
    \caption{Percentage of good cases in qualitative analysis.}
    \centering
    \resizebox{1.0\textwidth}{!}{
        \begin{tabular}{l|ccccccc|c}
            \toprule    
            Metrics & Doubao & GPT-4o & iFlytek Spark & StepFun & ChatGLM & Qwen2.5-Omni & Qwen3-Omni & \textbf{LongCat-Flash-Omni}\\
            \midrule
            Real-timeness & 65.5 & 71.5 & 67.0 & 18.0 & 3.0 & 4.0 & 1.5 & 49.5 \\
            Human-likeness & 93.5 & 69.0 & 92.5 & 70.5 & 76.5 & 59.0 & 13.5 & 62.5 \\
            Paralinguistic Understanding & 88.0 & 87.5 & 80.0 & 87.0 & 84.5 & 89.0 & 89.0 & 91.5 \\
            Relevance & 66.5 & 60.0 & 25.5 & 31.5 & 25.0 & 16.0 & 8.0 & 54.5 \\
            Accuracy & 65.0 & 55.0 & 38.5 & 42.5 & 33.5 & 35.5 & 40.0 & 36.0 \\
            Memory Capability & 98.0 & 98.0 & 81.5 & 92.5 & 83.0 & 64.0 & 36.0 & 94.5 \\
            \bottomrule
        \end{tabular}
    }
    \label{tab:percentage_of_qualitative}
\end{table}

The data construction process involves 10 trained professional conversationalists who conducted multi-turn dialogue sessions (approximately three minutes each) with each model under evaluation. These interactions are carried out via real-time audio-visual interfaces on the official app or web platforms provided by the respective model developers.
A total of 200 dialogue sessions have been collected for each model, covering common real-world video call scenarios across four categories: problem solving, entertainment, self-improvement, and emotional support (50 samples per category).

The assessment process combines quantitative user ratings with qualitative expert analysis. 
For quantitative evaluation, 250 real users independently triple-annotated the complete interaction videos, rating naturalness and fluency on a four-point scale: 
0 for \textit{completely unnatural}, 
1 for \textit{partially unnatural and affecting interaction}, 
2 for \textit{partially unnatural but not affecting interaction}, 
and 3 for \textit{completely natural and fluent}.
For qualitative analysis, expert annotators conduct a dimensional breakdown to identify specific factors affecting naturalness and fluency across six key factors, including \textit{real-timeness}, \textit{human-likeness}, \textit{paralinguistic understanding}, \textit{relevance}, \textit{accuracy} and \textit{memory capability}.
This approach provides comprehensive qualitative insights into each model's performance.

We compare LongCat-Flash-Omni with five audio-visual interaction products including Doubao\footnote{\url{https://www.doubao.com/chat/}, the evaluation period was August 11-15, 2025.}, GPT-4o\footnote{\url{https://chat.chatbot.app/?model=gpt-4o}, the evaluation period was August 18-22, 2025.}, iFlytek Spark\footnote{\url{https://xinghuo.xfyun.cn/desk}, the evaluation period was August 25-29, 2025.}, StepFun\footnote{\url{https://www.stepfun.com/chats/new}, the evaluation period was September 8-12, 2025.} and ChatGLM\footnote{\url{https://chat.z.ai/}, the evaluation period was September 1-5, 2025.}, as well as two open-source models Qwen2.5-Omni and Qwen3-Omni.

The quantitative ratings are presented in Table~\ref{tab:real_time_interaction}. 
LongCat-Flash-Omni achieves the third-highest score for naturalness and fluency in end-to-end interaction. 
Comparing with audio-visual interaction products, LongCat-Flash-Omni ranks behind Doubao and GPT-4o but outperforms iFlytek Spark and StepFun. 
Notably, LongCat-Flash-Omni demonstrates a substantial advantage over open-source alternatives, scoring 0.56 points higher than the current SOTA open-source model, Qwen3-omni.

We present the qualitative analysis results in Table~\ref{tab:percentage_of_qualitative}, together with some case study in Figure~\ref{fig:qualitative_analysis}.
LongCat-Flash-Omni excels in paralinguistic understanding, relevance, and memory capability, performing on par with top-tier models. 
Notably, LongCat-Flash-Omni actively interprets user emotions from both facial expressions and vocal cues, demonstrating its superior paralinguistic understanding. 
Regarding relevance, LongCat-Flash-Omni exhibits strong comprehension capabilities by closely tracking dialogue topics and generating highly correlated responses. 
This comprehension capability, combined with robust memory retention, enables LongCat-Flash-Omni to recall information from many turns earlier, resulting in more fluent and natural interactions. 
However, certain gaps remain compared with leading models, particularly in real-timeness, human-likeness, and accuracy. Specifically, in terms of real-timeness, our model tends to be overly sensitive to user pauses, often initiating responses prematurely and interrupting users mid-conversation. Regarding human-likeness, it occasionally exhibits pronunciation errors, stuttering, and robotic or electronic audio artifacts. In terms of accuracy, while our model shows strong capability in recognizing dynamic objects, its recognition performance declines when processing text and numerical information. Additionally, we observe a tendency for our model to over-agree with users’ statements while overlooking relevant visual content.
These aspects will be further optimized in the future work.

\section{Conclusion} 
\label{sec8:conclusion}

In this report, we present \longcat, a next-generation open-source omni-modal model that unifies robust offline multimodal understanding with real-time audio-visual interaction in a single, cohesive framework. \longcat demonstrates that large-scale models can effectively perceive, integrate, and generate across diverse modalities, including text, audio, image, and video, without sacrificing performance in any individual domain.

We address major challenges in building such a system: cross-modal heterogeneity, unified offline and streaming capabilities, and low-latency interaction. Through a carefully designed multi-stage early-fusion pretraining pipeline, \longcat achieves deeply integrated representations that enable synergistic multimodal reasoning while preserving unimodality strength. Our introduction of human-in-the-loop data construction and a 128K-token context window further enhances multi-turn dialogue, temporal reasoning, and memory capabilities in dynamic interactive scenarios.
On the architectural side, the adoption of the ScMoE backbone with zero-computation experts, together with lightweight modality encoders and decoder, allows the model to support real-time audio-visual interaction.

Extensive evaluations show that \longcat not only achieves state-of-the-art performance on omni-modal benchmarks such as Omni-Bench and WorldSense, but also matches or exceeds closed-source systems in key unimodal tasks, including image and video understanding as well as audio comprehension. Moreover, subjective assessments confirm the model’s ability to deliver natural, low-latency, high-quality interactive experiences, highlighting its potential as a foundation for next-generation human-AI interfaces.

Looking forward, \longcat lays a strong foundation for the continued evolution of omni-modal intelligence. Future work will focus on expanding the diversity and scale of training data, integrating an adaptive thinking mode, refining streaming and generation capabilities, and exploring richer forms of embodied and interactive intelligence. We believe that the release of \longcat will not only accelerate research on multimodal understanding and generation but also inspire new applications and paradigms for building human-centered, AGI-oriented systems.

\clearpage

\section{Contributions}

The list of authors is in alphabetical order. Names marked with an asterisk ($\ast$) indicate people who have left our team.

\begin{CJK}{UTF8}{gkai}
\renewcommand{\arraystretch}{1.2}
\begin{tabular}{p{0.25\textwidth}p{0.25\textwidth}p{0.25\textwidth}p{0.25\textwidth}}
Bairui Wang & Jian Yang & Senbin Yang & Xuezhi Cao \\
Bayan & Jiaxing Liu & Shanbo Xu & Xunliang Cai \\
Bin Xiao & Jing Huang & Shanglin Lei & Yang Yang \\
Bo Zhang * & Jingang Wang & Shengze Ye & Yanli Tan \\
Bolin Rong & Jinrui Ding & Shimin Chen & Yao Yao \\
Borun Chen & Juchao Jiang & Shuaiqi Chen & Yerui Sun \\
Chang Wan & Jun Kuang & Shujie Hu & Yi Chen \\
Chao Zhang & Jun Wang & Shuo Li & Yifan Lu \\
Chen Chen (陈晨) & Junhui Mei & Siqi Yang & Yin Gong \\
Chen Chen (陈琛) & Ke Ding & Siyu Ren & Yining Zhang \\
Chen Huang & Kefeng Zhang & Siyu Xu & Yitian Chen \\
Chengxu Yang & Lei Chen & Song Li & Yiyang Gan \\
Chengzuo Yang & Liang Shi & Songxiang Liu & Yuchen Tang \\
Cong Han * & Limeng Qiao & Tianhao Bai & Yuchen Xie \\
Dandan Peng & Liming Zheng & Tianye Dai & Yueqian Wang \\
Delian Ruan & Lin Ma & Wei Hong & Yuewen Zheng \\
Detai Xin & Liuyang Guo & Wei Wang & Yufei Zhang \\
Disong Wang & Liya Ma & Weixiao Zhao & Yufeng Zhong \\
Dongchao Yang & Luying Sun & Wengang Cao & Yulei Qian \\
Fanfan Liu & Man Gao & Wenlong He & Yuqi Peng \\
Fengjiao Chen & Mengshen Zhu & Wenlong Zhu & Yuqian Li \\
Fengyu Yang & Miao Cao & Xi Nan & Yuwei Jiang \\
Gan Dong & Minliang Lin & Xi Su & Zeyang Hu \\
Gang Huang & Nuo Xu & Xiaohan Zhao & Zheng Zhang \\
Gang Xu & Peng Shi & Xiaohao Wang & Zhengkun Tian * \\
Guanglu Wan & Qi Zhang & Xiaoyu Li & Zhiqing Hong \\
Guoqiang Tan & Qian Fang & Xiaoyu Wang & Zhixiong Zeng \\
Guoqiao Yu & Qian Wang & Xiaoyu Zhao & Zhuqi Mi \\
Haibo Qiu & Qian Yang & Xin Chen & Ziran Li \\
Hao Lu & Quanxiu Wang & Xin Pan & Ziwen Wang \\
Hongbo Liu & Rongxiang Weng & Xiusong Sun & Ziyi Zhao \\
Hongyu Xiang & Rongxin Guo * & Xu Xiang & Ziyuan Zhuang \\
Jiaheng Wu & Ruoxuan Liang & Xudong Xing & Zizhe Zhao \\
\end{tabular}
\end{CJK}

\clearpage

\bibliographystyle{unsrtnat}
\bibliography{references}  

@inproceedings{kazemzadeh-2014-refcoco,
    title = "{R}efer{I}t{G}ame: Referring to Objects in Photographs of Natural Scenes",
    author = "Kazemzadeh, Sahar  and
      Ordonez, Vicente  and
      Matten, Mark  and
      Berg, Tamara",
    booktitle = "Proceedings of the 2014 Conference on Empirical Methods in Natural Language Processing ({EMNLP})",
    month = oct,
    year = "2014",
    address = "Doha, Qatar",
    publisher = "Association for Computational Linguistics",
    url = "https://aclanthology.org/D14-1086/",
    pages = "787--798"
}

@article{pratap2024scaling,
  title={Scaling speech technology to 1,000+ languages},
  author={Pratap, Vineel and Tjandra, Andros and Shi, Bowen and Tomasello, Paden and Babu, Arun and Kundu, Sayani and Elkahky, Ali and Ni, Zhaoheng and Vyas, Apoorv and Fazel-Zarandi, Maryam and others},
  journal={Journal of Machine Learning Research},
  volume={25},
  number={97},
  pages={1--52},
  year={2024}
}

@software{pyscenedetect,
  author    = {Breakthrough, Brandon},
  title     = {PySceneDetect: Video Scene Cut Detection Tool and Library},
  year      = {2025},
  version   = {v0.6.4},
  url       = {https://github.com/Breakthrough/PySceneDetect},
  urldate   = {2025-10-09},
  note      = {Available at \url{https://pyscenedetect.readthedocs.io/}}
}

@InProceedings{Goyal_2017_ICCV,
author = {Goyal, Raghav and Ebrahimi Kahou, Samira and Michalski, Vincent and Materzynska, Joanna and Westphal, Susanne and Kim, Heuna and Haenel, Valentin and Fruend, Ingo and Yianilos, Peter and Mueller-Freitag, Moritz and Hoppe, Florian and Thurau, Christian and Bax, Ingo and Memisevic, Roland},
title = {The "Something Something" Video Database for Learning and Evaluating Visual Common Sense},
booktitle = {Proceedings of the IEEE International Conference on Computer Vision (ICCV)},
month = {Oct},
year = {2017}
}

@article{qiao2025univitar,
  title={UniViTAR: Unified Vision Transformer with Native Resolution},
  author={Qiao, Limeng and Gan, Yiyang and Wang, Bairui and Qin, Jie and Xu, Shuang and Yang, Siqi and Ma, Lin},
  journal={arXiv preprint arXiv:2504.01792},
  year={2025}
}

@article{zeng2025uitron,
  title={UItron: Foundational GUI Agent with Advanced Perception and Planning},
  author={Zeng, Zhixiong and Huang, Jing and Zheng, Liming and Han, Wenkang and Zhong, Yufeng and Chen, Lei and Yang, Longrong and Chu, Yingjie and He, Yuzhi and Ma, Lin},
  journal={arXiv preprint arXiv:2508.21767},
  year={2025}
}

@article{abbas2024effective,
  title={Effective pruning of web-scale datasets based on complexity of concept clusters},
  author={Abbas, Amro and Rusak, Evgenia and Tirumala, Kushal and Brendel, Wieland and Chaudhuri, Kamalika and Morcos, Ari S},
  journal={arXiv preprint arXiv:2401.04578},
  year={2024}
}

@article{xu2023demystifying,
  title={Demystifying clip data},
  author={Xu, Hu and Xie, Saining and Tan, Xiaoqing Ellen and Huang, Po-Yao and Howes, Russell and Sharma, Vasu and Li, Shang-Wen and Ghosh, Gargi and Zettlemoyer, Luke and Feichtenhofer, Christoph},
  journal={arXiv preprint arXiv:2309.16671},
  year={2023}
}

@misc{AIME24,
    author = {MAA},
    title = {AIME 2024},
    url = {https://maa.org/math-competitions/american-invitational-mathematics-examination-aime},
    year = {2024}
}

@inproceedings{BBH,
    author = {Suzgun, Mirac  and
Scales, Nathan  and
Sch{\"a}rli, Nathanael  and
Gehrmann, Sebastian  and
Tay, Yi  and
Chung, Hyung Won  and
Chowdhery, Aakanksha  and
Le, Quoc  and
Chi, Ed  and
Zhou, Denny  and
Wei, Jason},
    booktitle = {Findings of the Association for Computational Linguistics: ACL 2023},
    title = {Challenging {BIG}-Bench Tasks and Whether Chain-of-Thought Can Solve Them},
    year = {2023}
}

@misc{bytedance_seed_2025_beyondaime,
    author = {ByteDance-Seed},
    howpublished = {\url{https://huggingface.co/datasets/ByteDance-Seed/BeyondAIME}},
    journal = {Hugging Face repository},
    publisher = {Hugging Face},
    title = {BeyondAIME: Advancing Math Reasoning Evaluation Beyond High School Olympiads},
    year = {2025}
}

@article{cai2024shortcut,
    author = {Cai, Weilin and Jiang, Juyong and Qin, Le and Cui, Junwei and Kim, Sunghun and Huang, Jiayi},
    journal = {arXiv preprint arXiv:2404.05019},
    title = {Shortcut-connected expert parallelism for accelerating mixture-of-experts},
    year = {2024}
}

@article{cassano2022multiplescalableextensibleapproach,
    archiveprefix = {arXiv},
    author = {Federico Cassano and John Gouwar and Daniel Nguyen and Sydney Nguyen and Luna Phipps-Costin and Donald Pinckney and Ming-Ho Yee and Yangtian Zi and Carolyn Jane Anderson and Molly Q Feldman and Arjun Guha and Michael Greenberg and Abhinav Jangda},
    eprint = {2208.08227},
    journal = {arXiv preprint arXiv:2208.08227},
    primaryclass = {cs.LG},
    title = {{MultiPL-E}: A Scalable and Extensible Approach to Benchmarking Neural Code Generation},
    year = {2022}
}

@misc{claude4,
    author = {Anthropic},
    month = {May},
    title = {Introducing Claude 4},
    url = {https://www.anthropic.com/news/claude-4},
    year = {2025}
}

@inproceedings{clue,
    author = {Xu, Liang  and
Hu, Hai and
Zhang, Xuanwei and
Li, Lu and
Cao, Chenjie and
Li, Yudong and
Xu, Yechen and
Sun, Kai and
Yu, Dian and
Yu, Cong and
Tian, Yin and
Dong, Qianqian and
Liu, Weitang and
Shi, Bo and
Cui, Yiming and
Li, Junyi and
Zeng, Jun and
Wang, Rongzhao and
Xie, Weijian and
Li, Yanting and
Patterson, Yina and
Tian, Zuoyu and
Zhang, Yiwen and
Zhou, He and
Liu, Shaoweihua and
Zhao, Zhe and
Zhao, Qipeng and
Yue, Cong and
Zhang, Xinrui and
Yang, Zhengliang and
Richardson, Kyle and
Lan, Zhenzhong },
    booktitle = {Proceedings of the 28th International Conference on Computational Linguistics},
    title = {{CLUE}: A {C}hinese Language Understanding Evaluation Benchmark},
    year = {2020}
}

@article{cobbe2021gsm8k,
    author = {Cobbe, Karl and Kosaraju, Vineet and Bavarian, Mohammad and Chen, Mark and Jun, Heewoo and Kaiser, Lukasz and Plappert, Matthias and Tworek, Jerry and Hilton, Jacob and Nakano, Reiichiro and Hesse, Christopher and Schulman, John},
    journal = {arXiv preprint arXiv:2110.14168},
    title = {Training Verifiers to Solve Math Word Problems},
    year = {2021}
}

@article{comanici2025gemini,
    author = {Comanici, Gheorghe and Bieber, Eric and Schaekermann, Mike and Pasupat, Ice and Sachdeva, Noveen and Dhillon, Inderjit and Blistein, Marcel and Ram, Ori and Zhang, Dan and Rosen, Evan and others},
    journal = {arXiv preprint arXiv:2507.06261},
    title = {Gemini 2.5: Pushing the Frontier with Advanced Reasoning, Multimodality, Long Context, and Next Generation Agentic Capabilities},
    year = {2025}
}

@article{deepseekai2025deepseekv3technicalreport,
    archiveprefix = {arXiv},
    author = {DeepSeek-AI and Liu, Aixin and Feng, Bei and Xue, Bing and Wang, Bingxuan and Wu, Bochao and Lu, Chengda and Zhao, Chenggang and Deng, Chengqi and Zhang, Chenyu and Ruan, Chong and others},
    eprint = {arXiv preprint arXiv:2412.19437},
    journal = {arXiv preprint arXiv:arXiv preprint arXiv:2412.19437},
    primaryclass = {cs.CL},
    title = {DeepSeek-V3 Technical Report},
    year = {2025}
}

@article{drop,
    archiveprefix = {arXiv},
    author = {Dheeru Dua and Yizhong Wang and Pradeep Dasigi and Gabriel Stanovsky and Sameer Singh and Matt Gardner},
    eprint = {1903.00161},
    journal = {arXiv preprint arXiv:1903.00161},
    primaryclass = {cs.CL},
    title = {{DROP}: A Reading Comprehension Benchmark Requiring Discrete Reasoning Over Paragraphs},
    year = {2019}
}

@misc{gpt4.1,
    author = {OpenAI},
    month = {April},
    title = {Introducing {GPT-4.1} in the API},
    url = {https://openai.com/index/gpt-4-1/},
    year = {2025}
}

@misc{gpt4o,
    author = {OpenAI},
    month = {May},
    title = {Hello GPT-4o},
    url = {https://openai.com/index/hello-gpt-4o/},
    year = {2024}
}

@article{xu2025qwen3,
  title={Qwen3-Omni Technical Report},
  author={Xu, Jin and Guo, Zhifang and Hu, Hangrui and Chu, Yunfei and Wang, Xiong and He, Jinzheng and Wang, Yuxuan and Shi, Xian and He, Ting and Zhu, Xinfa and others},
  journal={arXiv preprint arXiv:2509.17765},
  year={2025}
}

@article{guo2025m2,
  title={M2-omni: Advancing omni-mllm for comprehensive modality support with competitive performance},
  author={Guo, Qingpei and Song, Kaiyou and Feng, Zipeng and Ma, Ziping and Zhang, Qinglong and Gao, Sirui and Yu, Xuzheng and Sun, Yunxiao and Chang, Tai-Wei and Chen, Jingdong and others},
  journal={arXiv preprint arXiv:2502.18778},
  year={2025}
}

@article{liu2025ola,
  title={Ola: Pushing the frontiers of omni-modal language model},
  author={Liu, Zuyan and Dong, Yuhao and Wang, Jiahui and Liu, Ziwei and Hu, Winston and Lu, Jiwen and Rao, Yongming},
  journal={arXiv preprint arXiv:2502.04328},
  year={2025}
}

@article{fu2025vita,
  title={Vita-1.5: Towards gpt-4o level real-time vision and speech interaction},
  author={Fu, Chaoyou and Lin, Haojia and Wang, Xiong and Zhang, Yi-Fan and Shen, Yunhang and Liu, Xiaoyu and Cao, Haoyu and Long, Zuwei and Gao, Heting and Li, Ke and others},
  journal={arXiv preprint arXiv:2501.01957},
  year={2025}
}

@misc{graphwalks,
    author = {OpenAI},
    title = {GraphWalks Dataset},
    url = {https://huggingface.co/datasets/openai/graphwalks},
    year = {2025}
}

@article{gu2024cruxevalbenchmarkcodereasoning,
    archiveprefix = {arXiv},
    author = {Alex Gu and Baptiste Rozière and Hugh Leather and Armando Solar-Lezama and Gabriel Synnaeve and Sida I. Wang},
    eprint = {2401.03065},
    journal = {arXiv preprint arXiv:2401.03065},
    primaryclass = {cs.SE},
    title = {CRUXEval: A Benchmark for Code Reasoning, Understanding and Execution},
    year = {2024}
}

@article{hendrycks2021measuringmassivemultitasklanguage,
    archiveprefix = {arXiv},
    author = {Dan Hendrycks and Collin Burns and Steven Basart and Andy Zou and Mantas Mazeika and Dawn Song and Jacob Steinhardt},
    eprint = {2009.03300},
    journal = {arXiv preprint arXiv:2009.03300},
    primaryclass = {cs.CY},
    title = {Measuring Massive Multitask Language Understanding},
    year = {2021}
}

@inproceedings{huang2023ceval,
    author = {Huang, Yuzhen and Bai, Yuzhuo and Zhu, Zhihao and Zhang, Junlei and Zhang, Jinghan and Su, Tangjun and Liu, Junteng and Lv, Chuancheng and Zhang, Yikai and Lei, Jiayi and Fu, Yao and Sun, Maosong and He, Junxian},
    booktitle = {Advances in Neural Information Processing Systems},
    title = {{C-Eval}: A Multi-Level Multi-Discipline Chinese Evaluation Suite for Foundation Models},
    year = {2023}
}

@article{humanevalmbppplus,
    archiveprefix = {arXiv},
    author = {Jiawei Liu and Songrun Xie and Junhao Wang and Yuxiang Wei and Yifeng Ding and Lingming Zhang},
    eprint = {2408.06450},
    journal = {arXiv preprint arXiv:2408.06450},
    primaryclass = {cs.SE},
    title = {Evaluating Language Models for Efficient Code Generation},
    year = {2024}
}

@inproceedings{jain2025livecodebench,
    author = {Naman Jain and King Han and Alex Gu and Wen-Ding Li and Fanjia Yan and Tianjun Zhang and Sida Wang and Armando Solar-Lezama and Koushik Sen and Ion Stoica},
    booktitle = {The Thirteenth International Conference on Learning Representations},
    title = {{LiveCodeBench}: Holistic and Contamination Free Evaluation of Large Language Models for Code},
    year = {2025}
}

@misc{Kimi_K2_web_doc,
    author = {MoonshotAI},
    title = {{{Kimi-K2}} Documentation},
    url = {https://moonshotai.github.io/Kimi-K2/},
    year = {2025}
}

@article{li2023cmmlu,
    archiveprefix = {arXiv},
    author = {Haonan Li and Yixuan Zhang and Fajri Koto and Yifei Yang and Hai Zhao and Yeyun Gong and Nan Duan and Timothy Baldwin},
    eprint = {2306.09212},
    journal = {arXiv preprint arXiv:2306.09212},
    primaryclass = {cs.CL},
    title = {{CMMLU}: Measuring massive multitask language understanding in Chinese},
    year = {2023}
}

@inproceedings{lin2025zebralogic,
    author = {Bill Yuchen Lin and Ronan Le Bras and Kyle Richardson and Ashish Sabharwal and Radha Poovendran and Peter Clark and Yejin Choi},
    booktitle = {Forty-second International Conference on Machine Learning},
    title = {ZebraLogic: On the Scaling Limits of {LLM}s for Logical Reasoning},
    year = {2025}
}

@article{liu2024deepseek,
    author = {Liu, Aixin and Feng, Bei and Wang, Bin and Wang, Bingxuan and Liu, Bo and Zhao, Chenggang and Dengr, Chengqi and Ruan, Chong and Dai, Damai and Guo, Daya and others},
    journal = {arXiv preprint arXiv:2405.04434},
    title = {Deepseek-v2: A strong, economical, and efficient mixture-of-experts language model},
    year = {2024}
}

@inproceedings{math500,
    author = {Lightman, Hunter and Kosaraju, Vineet and Burda, Yuri and Edwards, Harrison and Baker, Bowen and Lee, Teddy and Leike, Jan and Schulman, John and Sutskever, Ilya and Cobbe, Karl},
    booktitle = {The Twelfth International Conference on Learning Representations},
    title = {Let's verify step by step},
    year = {2023}
}

@article{mathbenchmark,
    author = {Dan Hendrycks
and Collin Burns
and Saurav Kadavath
and Akul Arora
and Steven Basart
and Eric Tang
and Dawn Song
and Jacob Steinhardt},
    journal = {arXiv preprint arXiv:2103.03874},
    title = {Measuring Mathematical Problem Solving With the MATH Dataset},
    year = {2021}
}

@article{meeseeks,
    author = {Wang, Jiaming and Zhao, Yunke and Ding, Peng and Kuang, Jun and Wang, Zongyu and Cao, Xuezhi and Cai, Xunliang},
    journal = {arXiv preprint arXiv:2504.21625},
    title = {Ask, Fail, Repeat: Meeseeks, an Iterative Feedback Benchmark for LLMs' Multi-turn Instruction-Following Ability},
    year = {2025}
}

@article{piqa,
    archiveprefix = {arXiv},
    author = {Yonatan Bisk and Rowan Zellers and Ronan Le Bras and Jianfeng Gao and Yejin Choi},
    eprint = {1911.11641},
    journal = {arXiv preprint arXiv:1911.11641},
    primaryclass = {cs.CL},
    title = {{PIQA}: Reasoning about Physical Commonsense in Natural Language},
    year = {2019}
}

@article{pteam2025supergpqascalingllmevaluation,
    archiveprefix = {arXiv},
    author = {{M-A-P Team, ByteDance.}},
    eprint = {2502.14739},
    journal = {arXiv preprint arXiv:2502.14739},
    primaryclass = {cs.CL},
    title = {{{SuperGPQA}}: Scaling {{LLM}} Evaluation across 285 Graduate Disciplines},
    year = {2025}
}

@article{rein2023gpqagraduatelevelgoogleproofqa,
    archiveprefix = {arXiv},
    author = {David Rein and Betty Li Hou and Asa Cooper Stickland and Jackson Petty and Richard Yuanzhe Pang and Julien Dirani and Julian Michael and Samuel R. Bowman},
    eprint = {2311.12022},
    journal = {arXiv preprint arXiv:2311.12022},
    primaryclass = {cs.AI},
    title = {{GPQA}: A Graduate-Level Google-Proof Q\&A Benchmark},
    year = {2023}
}

@article{su2024roformer,
    author = {Su, Jianlin and Ahmed, Murtadha and Lu, Yu and Pan, Shengfeng and Bo, Wen and Liu, Yunfeng},
    journal = {Neurocomputing},
    pages = {127063},
    publisher = {Elsevier},
    title = {Roformer: Enhanced transformer with rotary position embedding},
    volume = {568},
    year = {2024}
}

@article{wang2024mmluprorobustchallengingmultitask,
    archiveprefix = {arXiv},
    author = {Yubo Wang and Xueguang Ma and Ge Zhang and Yuansheng Ni and Abhranil Chandra and Shiguang Guo and Weiming Ren and Aaran Arulraj and Xuan He and Ziyan Jiang and Tianle Li and Max Ku and Kai Wang and Alex Zhuang and Rongqi Fan and Xiang Yue and Wenhu Chen},
    eprint = {2406.01574},
    journal = {arXiv preprint arXiv:2406.01574},
    primaryclass = {cs.CL},
    title = {{MMLU-Pro}: A More Robust and Challenging Multi-Task Language Understanding Benchmark},
    year = {2024}
}

@article{winogrande,
    author = {Sakaguchi, Keisuke and Bras, Ronan Le and Bhagavatula, Chandra and Choi, Yejin},
    journal = {arXiv preprint arXiv:1907.10641},
    title = {WinoGrande: An Adversarial Winograd Schema Challenge at Scale},
    year = {2019}
}

@article{yang2025qwen3,
    author = {Yang, An and Li, Anfeng and Yang, Baosong and Zhang, Beichen and Hui, Binyuan and Zheng, Bo and Yu, Bowen and Gao, Chang and Huang, Chengen and Lv, Chenxu and others},
    journal = {arXiv preprint arXiv:2505.09388},
    title = {Qwen3 technical report},
    year = {2025}
}

@inproceedings{yao2024collie,
    author = {Shunyu Yao and Howard Chen and Austin W. Hanjie and Runzhe Yang and Karthik R Narasimhan},
    booktitle = {The Twelfth International Conference on Learning Representations},
    title = {{COLLIE}: Systematic Construction of Constrained Text Generation Tasks},
    year = {2024}
}

@article{zhou2023ifeval,
    author = {Zhou, Jeffrey and Lu, Tianjian and Mishra, Swaroop and Brahma, Siddhartha and Basu, Sujoy and Luan, Yi and Zhou, Denny and Hou, Le},
    journal = {arXiv preprint arXiv:2311.07911},
    title = {Instruction-Following Evaluation for Large Language Models},
    year = {2023}
}

@misc{longcat-codec2025zhao,
      title={LongCat-Audio-Codec: An Audio Tokenizer and Detokenizer Solution Designed for Speech Large Language Models}, 
      author={Xiaohan Zhao and Hongyu Xiang and Shengze Ye and Song Li and Zhengkun Tian and Guanyu Chen and Ke Ding and Guanglu Wan},
      year={2025},
      eprint={2510.15227},
      archivePrefix={arXiv},
      primaryClass={eess.AS},
      url={https://arxiv.org/abs/2510.15227}, 
}

@inproceedings{graves2006connectionist,
  title={Connectionist temporal classification: labelling unsegmented sequence data with recurrent neural networks},
  author={Graves, Alex and Fern{\'a}ndez, Santiago and Gomez, Faustino and Schmidhuber, J{\"u}rgen},
  booktitle={Proceedings of the 23rd international conference on Machine learning},
  pages={369--376},
  year={2006}
}

@article{meituan2025longcat_flash_chat,
    title={LongCat-Flash Technical Report},
    author={Meituan},
    journal={arXiv preprint arXiv:2509.01322},
    year={2025}
}

@misc{meituan2025longcat_flash_chat_thinking,
      title={LongCat-Flash-Thinking Technical Report}, 
      author={Meituan},
      year={2025},
      eprint={2509.18883},
      archivePrefix={arXiv},
      primaryClass={cs.AI},
      url={https://arxiv.org/abs/2509.18883}, 
}

@inproceedings{panayotov2015librispeech,
  title={Librispeech: an asr corpus based on public domain audio books},
  author={Panayotov, Vassil and Chen, Guoguo and Povey, Daniel and Khudanpur, Sanjeev},
  booktitle={2015 IEEE international conference on acoustics, speech and signal processing (ICASSP)},
  pages={5206--5210},
  year={2015},
  organization={IEEE}
}

@article{masry2022chartqa,
  title={Chartqa: A benchmark for question answering about charts with visual and logical reasoning},
  author={Masry, Ahmed and Long, Do Xuan and Tan, Jia Qing and Joty, Shafiq and Hoque, Enamul},
  journal={arXiv preprint arXiv:2203.10244},
  year={2022}
}

@article{omnibench,	author={Yizhi Li and Ge Zhang and Yinghao Ma and Ruibin Yuan and Kang Zhu and Hangyu Guo and Yiming Liang and Jiaheng Liu and Zekun Wang and Jian Yang and Siwei Wu and Xingwei Qu and Jinjie Shi and Xinyue Zhang and Zhenzhu Yang and Xiangzhou Wang and Zhaoxiang Zhang and Zachary Liu and Emmanouil Benetos and Wenhao Huang and Chenghua Lin},	title={OmniBench: Towards the Future of Universal Omni-Language Models},	volume={abs/2409.15272},	year=2024,}

@article{worldsense,	author={Jack Hong and Shilin Yan and Jiayin Cai and Xiaolong Jiang and Yao Hu and Weidi Xie},	title={WorldSense: Evaluating Real-world Omnimodal Understanding for Multimodal LLMs.},	volume={abs/2502.04326},	year=2025,}

@inproceedings{liu2024mmbench,
  title={Mmbench: Is your multi-modal model an all-around player?},
  author={Liu, Yuan and Duan, Haodong and Zhang, Yuanhan and Li, Bo and Zhang, Songyang and Zhao, Wangbo and Yuan, Yike and Wang, Jiaqi and He, Conghui and Liu, Ziwei and others},
  booktitle={European conference on computer vision},
  pages={216--233},
  year={2024},
  organization={Springer}
}

@misc{realworldqa,
  title        = {RealWorldQA},
  author       = {xAI},
  year         = {2023},
  howpublished = {\url{https://huggingface.co/datasets/xai-org/RealworldQA}},
  note         = {Version 1.0, Accessed: 2024}
}

@misc{mmstar,
      title={Are We on the Right Way for Evaluating Large Vision-Language Models?}, 
      author={Lin Chen and Jinsong Li and Xiaoyi Dong and Pan Zhang and Yuhang Zang and Zehui Chen and Haodong Duan and Jiaqi Wang and Yu Qiao and Dahua Lin and Feng Zhao},
      year={2024},
      eprint={2403.20330},
      archivePrefix={arXiv},
      primaryClass={cs.CV},
      url={https://arxiv.org/abs/2403.20330}, 
}

@inproceedings{lu2024mathvista,
  title={MathVista: Evaluating Mathematical Reasoning of Foundation Models in Visual Contexts},
  author={Lu, Pan and Bansal, Hritik and Xia, Tony and Liu, Jiacheng and Li, Chunyuan and Hajishirzi, Hannaneh and Cheng, Hao and Chang, Kai-Wei and Galley, Michel and Gao, Jianfeng},
  booktitle={International Conference on Learning Representations (ICLR)},
  year={2024}
}

@inproceedings{yue2023mmmu,
            title={MMMU: A Massive Multi-discipline Multimodal Understanding and Reasoning Benchmark for Expert AGI},
            author={Xiang Yue and Yuansheng Ni and Kai Zhang and Tianyu Zheng and Ruoqi Liu and Ge Zhang and Samuel Stevens and Dongfu Jiang and Weiming Ren and Yuxuan Sun and Cong Wei and Botao Yu and Ruibin Yuan and Renliang Sun and Ming Yin and Boyuan Zheng and Zhenzhu Yang and Yibo Liu and Wenhao Huang and Huan Sun and Yu Su and Wenhu Chen},
            booktitle={Proceedings of CVPR},
            year={2024},
          }

@inproceedings{yu2024mmvet,
  title={Mm-vet: Evaluating large multimodal models for integrated capabilities},
  author={Yu, Weihao and Yang, Zhengyuan and Li, Linjie and Wang, Jianfeng and Lin, Kevin and Liu, Zicheng and Wang, Xinchao and Wang, Lijuan},
  booktitle={International conference on machine learning},
  year={2024},
  organization={PMLR}
}

@article{fu2024blink,
  title={BLINK: Multimodal Large Language Models Can See but Not Perceive},
  author={Fu, Xingyu and Hu, Yushi and Li, Bangzheng and Feng, Yu and Wang, Haoyu and Lin, Xudong and Roth, Dan and Smith, Noah A and Ma, Wei-Chiu and Krishna, Ranjay},
  journal={arXiv preprint arXiv:2404.12390},
  year={2024}
}

@article{wang2024muirbench,
          title={MuirBench: A Comprehensive Benchmark for Robust Multi-image Understanding},
          author={Wang, Fei and Fu, Xingyu and Huang, James Y and Li, Zekun and Liu, Qin and Liu, Xiaogeng and Ma, Mingyu Derek and Xu, Nan and Zhou, Wenxuan and Zhang, Kai and others},
          journal={arXiv preprint arXiv:2406.09411},
          year={2024}
        }

@misc{mathew2021docvqa,
      title={DocVQA: A Dataset for VQA on Document Images}, 
      author={Minesh Mathew and Dimosthenis Karatzas and C. V. Jawahar},
      year={2021},
      eprint={2007.00398},
      archivePrefix={arXiv},
      primaryClass={cs.CV},
      url={https://arxiv.org/abs/2007.00398}, 
}

@article{liu2024ocrbench,
   title={OCRBench: on the hidden mystery of OCR in large multimodal models},
   volume={67},
   ISSN={1869-1919},
   url={http://dx.doi.org/10.1007/s11432-024-4235-6},
   DOI={10.1007/s11432-024-4235-6},
   number={12},
   journal={Science China Information Sciences},
   publisher={Springer Science and Business Media LLC},
   author={Liu, Yuliang and Li, Zhang and Huang, Mingxin and Yang, Biao and Yu, Wenwen and Li, Chunyuan and Yin, Xu-Cheng and Liu, Cheng-Lin and Jin, Lianwen and Bai, Xiang},
   year={2024},
   month=dec }

@misc{ouyang2024omnidocbench,
      title={OmniDocBench: Benchmarking Diverse PDF Document Parsing with Comprehensive Annotations}, 
      author={Linke Ouyang and Yuan Qu and Hongbin Zhou and Jiawei Zhu and Rui Zhang and Qunshu Lin and Bin Wang and Zhiyuan Zhao and Man Jiang and Xiaomeng Zhao and Jin Shi and Fan Wu and Pei Chu and Minghao Liu and Zhenxiang Li and Chao Xu and Bo Zhang and Botian Shi and Zhongying Tu and Conghui He},
      year={2024},
      eprint={2412.07626},
      archivePrefix={arXiv},
      primaryClass={cs.CV},
      url={https://arxiv.org/abs/2412.07626}, 
}

@misc{liu2024visualwebbench,
      title={VisualWebBench: How Far Have Multimodal LLMs Evolved in Web Page Understanding and Grounding?}, 
      author={Junpeng Liu and Yifan Song and Bill Yuchen Lin and Wai Lam and Graham Neubig and Yuanzhi Li and Xiang Yue},
      year={2024},
      eprint={2404.05955},
      archivePrefix={arXiv},
      primaryClass={cs.CL}
}

@article{li2024androidcontrol,
  title={On the Effects of Data Scale on Computer Control Agents},
  author={Li, Wei and Bishop, William and Li, Alice and Rawles, Chris and Campbell-Ajala, Folawiyo and Tyamagundlu, Divya and Riva, Oriana},
  journal={arXiv preprint arXiv:2406.03679},
  year={2024}
}

@misc{li2024mvbench,
      title={MVBench: A Comprehensive Multi-modal Video Understanding Benchmark}, 
      author={Kunchang Li and Yali Wang and Yinan He and Yizhuo Li and Yi Wang and Yi Liu and Zun Wang and Jilan Xu and Guo Chen and Ping Luo and Limin Wang and Yu Qiao},
      year={2024},
      eprint={2311.17005},
      archivePrefix={arXiv},
      primaryClass={cs.CV},
      url={https://arxiv.org/abs/2311.17005}, 
}

@InProceedings{xiao2021nextqa,
    author    = {Xiao, Junbin and Shang, Xindi and Yao, Angela and Chua, Tat-Seng},
    title     = {NExT-QA: Next Phase of Question-Answering to Explaining Temporal Actions},
    booktitle = {Proceedings of the IEEE/CVF Conference on Computer Vision and Pattern Recognition (CVPR)},
    month     = {June},
    year      = {2021},
    pages     = {9777-9786}
}

@article{liu2024tempcompass,
  title   = {TempCompass: Do Video LLMs Really Understand Videos?},
  author  = {Yuanxin Liu and Shicheng Li and Yi Liu and Yuxiang Wang and Shuhuai Ren and Lei Li and Sishuo Chen and Xu Sun and Lu Hou},
  year    = {2024},
  journal = {arXiv preprint arXiv: 2403.00476}
}

@article{fu2024videomme,
  title={Video-MME: The First-Ever Comprehensive Evaluation Benchmark of Multi-modal LLMs in Video Analysis},
  author={Fu, Chaoyou and Dai, Yuhan and Luo, Yondong and Li, Lei and Ren, Shuhuai and Zhang, Renrui and Wang, Zihan and Zhou, Chenyu and Shen, Yunhang and Zhang, Mengdan and others},
  journal={arXiv preprint arXiv:2405.21075},
  year={2024}
}

@misc{wu2024longvideobench,
      title={LongVideoBench: A Benchmark for Long-context Interleaved Video-Language Understanding}, 
      author={Haoning Wu and Dongxu Li and Bei Chen and Junnan Li},
      year={2024},
      eprint={2407.15754},
      archivePrefix={arXiv},
      primaryClass={cs.CV},
      url={https://arxiv.org/abs/2407.15754}, 
}

@misc{zhao2025mmvu,
      title={MMVU: Measuring Expert-Level Multi-Discipline Video Understanding}, 
      author={Yilun Zhao and Lujing Xie and Haowei Zhang and Guo Gan and Yitao Long and Zhiyuan Hu and Tongyan Hu and Weiyuan Chen and Chuhan Li and Junyang Song and Zhijian Xu and Chengye Wang and Weifeng Pan and Ziyao Shangguan and Xiangru Tang and Zhenwen Liang and Yixin Liu and Chen Zhao and Arman Cohan},
      year={2025},
      eprint={2501.12380},
      archivePrefix={arXiv},
      primaryClass={cs.CV},
      url={https://arxiv.org/abs/2501.12380}, 
}

@article{hu2025videommmu,
    title={Video-MMMU: Evaluating Knowledge Acquisition from Multi-Discipline Professional Videos},
    author={Kairui Hu and Penghao Wu and Fanyi Pu and Wang Xiao and Yuanhan Zhang and Xiang Yue and Bo Li and Ziwei Liu},
    booktitle={arXiv preprint arXiv:2501.13826},
    year={2025},
    url={https://arxiv.org/abs/2501.13826}
}

@inproceedings{DBLP:conf/iclr/MichelHN22,
  author       = {Paul Michel and
                  Tatsunori Hashimoto and
                  Graham Neubig},
  title        = {Distributionally Robust Models with Parametric Likelihood Ratios},
  booktitle    = {The Tenth International Conference on Learning Representations, {ICLR}
                  2022, Virtual Event, April 25-29, 2022},
  publisher    = {OpenReview.net},
  year         = {2022},
  url          = {https://openreview.net/forum?id=a34GrNaYEcS},
  bibsource    = {dblp computer science bibliography, https://dblp.org}
}

@inproceedings{DBLP:conf/nips/Xie0DDLLLL0Y23,
  author       = {Sang Michael Xie and
                  Hieu Pham and
                  Xuanyi Dong and
                  Nan Du and
                  Hanxiao Liu and
                  Yifeng Lu and
                  Percy Liang and
                  Quoc V. Le and
                  Tengyu Ma and
                  Adams Wei Yu},
  editor       = {Alice Oh and
                  Tristan Naumann and
                  Amir Globerson and
                  Kate Saenko and
                  Moritz Hardt and
                  Sergey Levine},
  title        = {DoReMi: Optimizing Data Mixtures Speeds Up Language Model Pretraining},
  booktitle    = {Advances in Neural Information Processing Systems 36: Annual Conference
                  on Neural Information Processing Systems 2023, NeurIPS 2023, New Orleans,
                  LA, USA, December 10 - 16, 2023},
  year         = {2023},
  url          = {http://papers.nips.cc/paper\_files/paper/2023/hash/dcba6be91359358c2355cd920da3fcbd-Abstract-Conference.html},
  bibsource    = {dblp computer science bibliography, https://dblp.org}
}

@article{rafailov2023direct,
  title={Direct preference optimization: Your language model is secretly a reward model},
  author={Rafailov, Rafael and Sharma, Archit and Mitchell, Eric and Manning, Christopher D and Ermon, Stefano and Finn, Chelsea},
  journal={Advances in neural information processing systems},
  volume={36},
  pages={53728--53741},
  year={2023}
}

@article{krishna2017visualgenome,
  title={Visual genome: Connecting language and vision using crowdsourced dense image annotations},
  author={Krishna, Ranjay and Zhu, Yuke and Groth, Oliver and Johnson, Justin and Hata, Kenji and Kravitz, Joshua and Chen, Stephanie and Kalantidis, Yannis and Li, Li-Jia and Shamma, David A and others},
  journal={International journal of computer vision},
  volume={123},
  number={1},
  pages={32--73},
  year={2017},
  publisher={Springer}
}

@inproceedings{mao2016refcocoplusg,
  title={Generation and comprehension of unambiguous object descriptions},
  author={Mao, Junhua and Huang, Jonathan and Toshev, Alexander and Camburu, Oana and Yuille, Alan L and Murphy, Kevin},
  booktitle={Proceedings of the IEEE conference on computer vision and pattern recognition},
  pages={11--20},
  year={2016}
}

@inproceedings{lin2014microsoftcoco,
  title={Microsoft coco: Common objects in context},
  author={Lin, Tsung-Yi and Maire, Michael and Belongie, Serge and Hays, James and Perona, Pietro and Ramanan, Deva and Doll{\'a}r, Piotr and Zitnick, C Lawrence},
  booktitle={European conference on computer vision},
  pages={740--755},
  year={2014},
  organization={Springer}
}

@inproceedings{shao2019objects365,
  title={Objects365: A large-scale, high-quality dataset for object detection},
  author={Shao, Shuai and Li, Zeming and Zhang, Tianyuan and Peng, Chao and Yu, Gang and Zhang, Xiangyu and Li, Jing and Sun, Jian},
  booktitle={Proceedings of the IEEE/CVF international conference on computer vision},
  pages={8430--8439},
  year={2019}
}

@article{deitke2024molmo,
  title={Molmo and pixmo: Open weights and open data for state-of-the-art multimodal models},
  author={Deitke, Matt and Clark, Christopher and Lee, Sangho and Tripathi, Rohun and Yang, Yue and Park, Jae Sung and Salehi, Mohammadreza and Muennighoff, Niklas and Lo, Kyle and Soldaini, Luca and others},
  journal={arXiv e-prints},
  pages={arXiv--2409},
  year={2024}
}

@inproceedings{paiss2023countbench,
  title={Teaching clip to count to ten},
  author={Paiss, Roni and Ephrat, Ariel and Tov, Omer and Zada, Shiran and Mosseri, Inbar and Irani, Michal and Dekel, Tali},
  booktitle={Proceedings of the IEEE/CVF International Conference on Computer Vision},
  pages={3170--3180},
  year={2023}
}

@article{Qwen2.5-VL,
  title={Qwen2.5-VL Technical Report},
  author={Bai, Shuai and Chen, Keqin and Liu, Xuejing and Wang, Jialin and Ge, Wenbin and Song, Sibo and Dang, Kai and Wang, Peng and Wang, Shijie and Tang, Jun and Zhong, Humen and Zhu, Yuanzhi and Yang, Mingkun and Li, Zhaohai and Wan, Jianqiang and Wang, Pengfei and Ding, Wei and Fu, Zheren and Xu, Yiheng and Ye, Jiabo and Zhang, Xi and Xie, Tianbao and Cheng, Zesen and Zhang, Hang and Yang, Zhibo and Xu, Haiyang and Lin, Junyang},
  journal={arXiv preprint arXiv:2502.13923},
  year={2025}
}

@inproceedings{bu2017aishell,
  title={Aishell-1: An open-source mandarin speech corpus and a speech recognition baseline},
  author={Bu, Hui and Du, Jiayu and Na, Xingyu and Wu, Bengu and Zheng, Hao},
  booktitle={2017 20th conference of the oriental chapter of the international coordinating committee on speech databases and speech I/O systems and assessment (O-COCOSDA)},
  pages={1--5},
  year={2017},
  organization={IEEE}
}

@article{du2018aishell,
  title={Aishell-2: Transforming mandarin asr research into industrial scale},
  author={Du, Jiayu and Na, Xingyu and Liu, Xuechen and Bu, Hui},
  journal={arXiv preprint arXiv:1808.10583},
  year={2018}
}

@inproceedings{zhang2022wenetspeech,
  title={Wenetspeech: A 10000+ hours multi-domain mandarin corpus for speech recognition},
  author={Zhang, Binbin and Lv, Hang and Guo, Pengcheng and Shao, Qijie and Yang, Chao and Xie, Lei and Xu, Xin and Bu, Hui and Chen, Xiaoyu and Zeng, Chenchen and others},
  booktitle={ICASSP 2022-2022 IEEE International Conference on Acoustics, Speech and Signal Processing (ICASSP)},
  pages={6182--6186},
  year={2022},
  organization={IEEE}
}

@article{li2025baichuan,
  title={Baichuan-audio: A unified framework for end-to-end speech interaction},
  author={Li, Tianpeng and Liu, Jun and Zhang, Tao and Fang, Yuanbo and Pan, Da and Wang, Mingrui and Liang, Zheng and Li, Zehuan and Lin, Mingan and Dong, Guosheng and others},
  journal={arXiv preprint arXiv:2502.17239},
  year={2025}
}

@inproceedings{conneau2023fleurs,
  title={Fleurs: Few-shot learning evaluation of universal representations of speech},
  author={Conneau, Alexis and Ma, Min and Khanuja, Simran and Zhang, Yu and Axelrod, Vera and Dalmia, Siddharth and Riesa, Jason and Rivera, Clara and Bapna, Ankur},
  booktitle={2022 IEEE Spoken Language Technology Workshop (SLT)},
  pages={798--805},
  year={2023},
  organization={IEEE}
}

@article{sakshi2024mmau,
  title={Mmau: A massive multi-task audio understanding and reasoning benchmark},
  author={Sakshi, S and Tyagi, Utkarsh and Kumar, Sonal and Seth, Ashish and Selvakumar, Ramaneswaran and Nieto, Oriol and Duraiswami, Ramani and Ghosh, Sreyan and Manocha, Dinesh},
  journal={arXiv preprint arXiv:2410.19168},
  year={2024}
}

@inproceedings{lipping2022clotho,
  title={Clotho-aqa: A crowdsourced dataset for audio question answering},
  author={Lipping, Samuel and Sudarsanam, Parthasaarathy and Drossos, Konstantinos and Virtanen, Tuomas},
  booktitle={2022 30th European Signal Processing Conference (EUSIPCO)},
  pages={1140--1144},
  year={2022},
  organization={IEEE}
}

@INPROCEEDINGS{gong_vocalsound,
  author={Gong, Yuan and Yu, Jin and Glass, James},
  booktitle={ICASSP 2022 - 2022 IEEE International Conference on Acoustics, Speech and Signal Processing (ICASSP)}, 
  title={Vocalsound: A Dataset for Improving Human Vocal Sounds Recognition}, 
  year={2022},
  pages={151-155},
  doi={10.1109/ICASSP43922.2022.9746828}
}

@article{rashid2023nonspeech7k,
  title={Nonspeech7k dataset: Classification and analysis of human non-speech sound},
  author={Rashid, Muhammad Mamunur and Li, Guiqing and Du, Chengrui},
  journal={IET Signal Processing},
  volume={17},
  number={6},
  pages={e12233},
  year={2023},
  publisher={Wiley Online Library}
}

@article{poria2018meld,
  title={Meld: A multimodal multi-party dataset for emotion recognition in conversations},
  author={Poria, Soujanya and Hazarika, Devamanyu and Majumder, Navonil and Naik, Gautam and Cambria, Erik and Mihalcea, Rada},
  journal={arXiv preprint arXiv:1810.02508},
  year={2018}
}

@inproceedings{Mesaros2016_EUSIPCO,
    author = "Mesaros, Annamaria and Heittola, Toni and Virtanen, Tuomas",
    title = "{TUT} Database for Acoustic Scene Classification and Sound Event Detection",
    year = "2016",
    address = "Budapest, Hungary",
    booktitle = "24th European Signal Processing Conference 2016 (EUSIPCO 2016)"
}

@inproceedings{jeong2022cochlscene,
  title={Cochlscene: Acquisition of acoustic scene data using crowdsourcing},
  author={Jeong, Il-Young and Park, Jeongsoo},
  booktitle={2022 Asia-Pacific Signal and Information Processing Association Annual Summit and Conference (APSIPA ASC)},
  pages={17--21},
  year={2022},
  organization={IEEE}
}

@article{ding2025kimi,
  title={Kimi-audio technical report},
  author={Ding, Ding and Ju, Zeqian and Leng, Yichong and Liu, Songxiang and Liu, Tong and Shang, Zeyu and Shen, Kai and Song, Wei and Tan, Xu and Tang, Heyi and others},
  journal={arXiv preprint arXiv:2504.18425},
  year={2025}
}

@article{wu2025step,
  title={Step-audio 2 technical report},
  author={Wu, Boyong and Yan, Chao and Hu, Chen and Yi, Cheng and Feng, Chengli and Tian, Fei and Shen, Feiyu and Yu, Gang and Zhang, Haoyang and Li, Jingbei and others},
  journal={arXiv preprint arXiv:2507.16632},
  year={2025}
}

@article{ardila2019common,
  title={Common voice: A massively-multilingual speech corpus},
  author={Ardila, Rosana and Branson, Megan and Davis, Kelly and Henretty, Michael and Kohler, Michael and Meyer, Josh and Morais, Reuben and Saunders, Lindsay and Tyers, Francis M and Weber, Gregor},
  journal={arXiv preprint arXiv:1912.06670},
  year={2019}
}

@inproceedings{wang2021covost,
  title={CoVoST 2 and massively multilingual speech translation.},
  author={Wang, Changhan and Wu, Anne and Gu, Jiatao and Pino, Juan},
  booktitle={Interspeech},
  volume={2021},
  pages={2247--2251},
  year={2021}
}

@article{chen2024voicebench,
  title={VoiceBench: Benchmarking LLM-Based Voice Assistants},
  author={Chen, Yiming and Yue, Xianghu and Zhang, Chen and Gao, Xiaoxue and Tan, Robby T. and Li, Haizhou},
  journal={arXiv preprint arXiv:2410.17196},
  year={2024}
}

@inproceedings{zhang2018deep,
  title={Deep-FSMN for large vocabulary continuous speech recognition},
  author={Zhang, Shiliang and Lei, Ming and Yan, Zhijie and Dai, Lirong},
  booktitle={2018 IEEE International Conference on Acoustics, Speech and Signal Processing (ICASSP)},
  pages={5869--5873},
  year={2018},
  organization={IEEE}
}

@inproceedings{zhai2023sigmoid,
  title={Sigmoid loss for language image pre-training},
  author={Zhai, Xiaohua and Mustafa, Basil and Kolesnikov, Alexander and Beyer, Lucas},
  booktitle={Proceedings of the IEEE/CVF International Conference on Computer Vision},
  pages={11975--11986},
  year={2023}
}

@inproceedings{radford2021clip,
	title={Learning transferable visual models from natural language supervision},
	author={Radford, Alec and Kim, Jong Wook and Hallacy, Chris and Ramesh, Aditya and Goh, Gabriel and Agarwal, Sandhini and Sastry, Girish and Askell, Amanda and Mishkin, Pamela and Clark, Jack and others},
	booktitle={ICML},
	pages={8748--8763},
	year={2021}
}

@article{zheng2025orchestrate,
  title={Orchestrate Multimodal Data with Batch Post-Balancing to Accelerate Multimodal Large Language Model Training},
  author={Zheng, Yijie and Xiao, Bangjun and Shi, Lei and Li, Xiaoyang and Wu, Faming and Li, Tianyu and Xiao, Xuefeng and Zhang, Yang and Wang, Yuxuan and Liu, Shouda},
  journal={arXiv preprint arXiv:2503.23830},
  year={2025}
}

@article{ma2025veomni,
  title={VeOmni: Scaling Any Modality Model Training with Model-Centric Distributed Recipe Zoo},
  author={Ma, Qianli and Zheng, Yaowei and Shi, Zhelun and Zhao, Zhongkai and Jia, Bin and Huang, Ziyue and Lin, Zhiqi and Li, Youjie and Yang, Jiacheng and Peng, Yanghua and others},
  journal={arXiv preprint arXiv:2508.02317},
  year={2025}
}

@inproceedings{zhang2025disttrain,
  title={Disttrain: Addressing model and data heterogeneity with disaggregated training for multimodal large language models},
  author={Zhang, Zili and Zhong, Yinmin and Jiang, Yimin and Hu, Hanpeng and Sun, Jianjian and Ge, Zheng and Zhu, Yibo and Jiang, Daxin and Jin, Xin},
  booktitle={Proceedings of the ACM SIGCOMM 2025 Conference},
  pages={24--38},
  year={2025}
}

@article{xue2025pipeweaver,
  title={PipeWeaver: Addressing Data Dynamicity in Large Multimodal Model Training with Dynamic Interleaved Pipeline},
  author={Xue, Zhenliang and Hu, Hanpeng and Chen, Xing and Jiang, Yimin and Song, Yixin and Mi, Zeyu and Zhu, Yibo and Jiang, Daxin and Xia, Yubin and Chen, Haibo},
  journal={arXiv preprint arXiv:2504.14145},
  year={2025}
}

@inproceedings{feng2025optimus,
  title={Optimus: Accelerating $\{$Large-Scale$\}$$\{$Multi-Modal$\}$$\{$LLM$\}$ Training by Bubble Exploitation},
  author={Feng, Weiqi and Chen, Yangrui and Wang, Shaoyu and Peng, Yanghua and Lin, Haibin and Yu, Minlan},
  booktitle={2025 USENIX Annual Technical Conference (USENIX ATC 25)},
  pages={161--177},
  year={2025}
}

@article{chen2024timemarker,
  title={Timemarker: A versatile video-llm for long and short video understanding with superior temporal localization ability},
  author={Chen, Shimin and Lan, Xiaohan and Yuan, Yitian and Jie, Zequn and Ma, Lin},
  journal={arXiv preprint arXiv:2411.18211},
  year={2024}
}

@misc{qi2024pipelineparallelismcontrollablememory,
  title={Pipeline Parallelism with Controllable Memory}, 
  author={Penghui Qi and Xinyi Wan and Nyamdavaa Amar and Min Lin},
  year={2024},
  eprint={2405.15362},
  archivePrefix={arXiv},
  primaryClass={cs.LG},
  url={https://arxiv.org/abs/2405.15362}, 
}

@article{megatron-lm,
  title={Megatron-LM: Training Multi-Billion Parameter Language Models Using Model Parallelism},
  author={Shoeybi, Mohammad and Patwary, Mostofa and Puri, Raul and LeGresley, Patrick and Casper, Jared and Catanzaro, Bryan},
  journal={arXiv preprint arXiv:1909.08053},
  year={2019}
}

@inproceedings{paszke2019pytorch,
  title        = {PyTorch: An Imperative Style, High-Performance Deep Learning Library},
  author       = {Paszke, Adam and Gross, Sam and Massa, Francisco and Lerer, Adam and Bradbury, James and Chanan, Gregory and Killeen, Trevor and Lin, Zeming and Gimelshein, Natalia and Antiga, Luca and Desmaison, Alban and Kopf, Andreas and Yang, Edward and DeVito, Zachary and Raison, Martin and Tejani, Alykhan and Chilamkurthy, Sasank and Steiner, Benoit and Fang, Lu and Bai, Junjie and Chintala, Soumith},
  booktitle    = {Advances in Neural Information Processing Systems 32},
  editor       = {Wallach, H. and Larochelle, H. and Beygelzimer, A. and d'Alch{\'e}-Buc, F. and Fox, E. and Garnett, R.},
  pages        = {8024--8035},
  year         = {2019},
  publisher    = {Curran Associates, Inc.},
  url          = {http://papers.nips.cc/paper/9015-pytorch-an-imperative-style-high-performance-deep-learning-library.pdf}
}

@inproceedings{wang2025koala,
  title={Koala-36m: A large-scale video dataset improving consistency between fine-grained conditions and video content},
  author={Wang, Qiuheng and Shi, Yukai and Ou, Jiarong and Chen, Rui and Lin, Ke and Wang, Jiahao and Jiang, Boyuan and Yang, Haotian and Zheng, Mingwu and Tao, Xin and others},
  booktitle={Proceedings of the Computer Vision and Pattern Recognition Conference},
  pages={8428--8437},
  year={2025}
}

@misc{zhao2023pytorchfsdpexperiencesscaling,
      title={PyTorch FSDP: Experiences on Scaling Fully Sharded Data Parallel}, 
      author={Yanli Zhao and Andrew Gu and Rohan Varma and Liang Luo and Chien-Chin Huang and Min Xu and Less Wright and Hamid Shojanazeri and Myle Ott and Sam Shleifer and Alban Desmaison and Can Balioglu and Pritam Damania and Bernard Nguyen and Geeta Chauhan and Yuchen Hao and Ajit Mathews and Shen Li},
      year={2023},
      eprint={2304.11277},
      archivePrefix={arXiv},
      primaryClass={cs.DC},
      url={https://arxiv.org/abs/2304.11277}, 
}

@article{carreira2019short,
  title={A short note on the kinetics-700 human action dataset},
  author={Carreira, Joao and Noland, Eric and Hillier, Chloe and Zisserman, Andrew},
  journal={arXiv preprint arXiv:1907.06987},
  year={2019}
}

@article{jiang2024mantis,
  title={Mantis: Interleaved multi-image instruction tuning},
  author={Jiang, Dongfu and He, Xuan and Zeng, Huaye and Wei, Cong and Ku, Max and Liu, Qian and Chen, Wenhu},
  journal={arXiv preprint arXiv:2405.01483},
  year={2024}
}

@article{wu2024atlas,
  title={Os-atlas: A foundation action model for generalist gui agents},
  author={Wu, Zhiyong and Wu, Zhenyu and Xu, Fangzhi and Wang, Yian and Sun, Qiushi and Jia, Chengyou and Cheng, Kanzhi and Ding, Zichen and Chen, Liheng and Liang, Paul Pu and others},
  journal={arXiv preprint arXiv:2410.23218},
  year={2024}
}

@misc{zhou2025dailyomni,
      title={Daily-Omni: Towards Audio-Visual Reasoning with Temporal Alignment across Modalities}, 
      author={Ziwei Zhou and Rui Wang and Zuxuan Wu},
      year={2025},
      eprint={2505.17862},
      archivePrefix={arXiv},
      primaryClass={cs.AI},
      url={https://arxiv.org/abs/2505.17862}, 
}

@misc{chen2025unobench,
      title={UNO-Bench: A Unified Benchmark for Exploring the Compositional Law Between Uni-modal and Omni-modal in OmniModels}, 
      author={Chen Chen and ZeYang Hu and Fengjiao Chen and Liya Ma and Jiaxing Liu and Xiaoyu Li and Xuezhi Cao},
      year={2025},
      eprint={2510.18915},
      archivePrefix={arXiv},
      primaryClass={cs.CL},
      url={https://github.com/meituan-longcat/UNO-Bench}, 
}

\clearpage


\end{document}